# MAKING SENSE OF THE
# ROBOTIZED
# PANDEMIC RESPONSE

A Comparison of Global and Canadian Robot Deployments and Success Factors

University of Toronto Robotics Institute

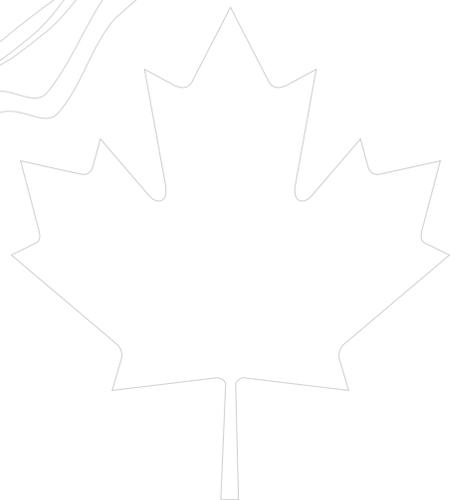

(alphabetical) T Barfoot, J Burgner-Kahrs, E Diller, A Garg, A Goldenberg, J Kelly, X Liu, HE Naguib, G Nejat, AP Schoellig, F Shkurti, H Siegel*, Y Sun, SL Waslander

*Corresponding Author (robotics@utoronto.ca)

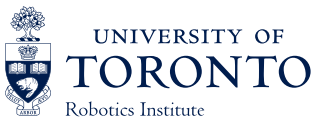

# Executive Summary

*From disinfection and remote triage, to logistics and delivery, countries around the world are making use of robots to address the unique challenges presented by the COVID-19 pandemic. Robots are being used to manage the pandemic here in Canada too, but relative to other regions, we have been more cautious in our adoption — this despite the important role that robots of Canadian origin are now playing on the global stage. This whitepaper discusses why this is the case, and argues that strategic investment and support for the Canadian robotics industry are urgently needed to bring the benefits of robotics home, where we have greater control in shaping the future of this game-changing technology. Such investments will not only serve to support Canada's current pandemic response and post-pandemic recovery, but will also prepare this country to weather future crises. Without such support, Canada risks falling behind other developed nations that are investing heavily in hardware automation at this time.*

Robots have been playing an ever-expanding role in contemporary society, enabling the automation of dangerous and repetitive tasks in a wide range of applications. Now the COVID-19 pandemic is putting robots to the test. Written in plain language and free of technical jargon, this paper aims to provide decision-makers with a primer on the use of robots and their potential in the fight against COVID. We survey how other countries around the world are making use of robots, describe the current state of affairs in Canada, and provide recommendations on how we can better leverage our robotics capabilities to benefit Canadians in this challenging time. To understand how to best make an impact with robots here in Canada, we conducted interviews with robotics industry professionals and stakeholders in areas as diverse as hospitals, public health, public transportation, manufacturing/retail industries, education, and government labs. We used these dialogues, along with our collective 130 years of experience working across the spectrum of the field to identify Canada's biggest strengths, weaknesses, and opportunities for robotics innovation.

It is clear from our study that countries around the world are actively adopting robots to improve frontline safety and efficiency during the pandemic. And yet we caution against approaches that put too much stock into demonstration systems and the ability to convert them to deployable robot systems in the short term. Disinfection robots

> Countries around the world are actively adopting robots to improve frontline safety and efficiency during the pandemic, for everything from disinfection to logistics.

in particular are being deployed extensively to service hospitals and medical facilities where COVID is an issue, and show great promise in reducing exposure risk and augmenting cleaning capacity when workers fall ill or cannot come into work. Yet details regarding their effectiveness are still unknown, in particular regarding their ability to adequately disinfect soft surfaces like bedding, seating, and personal protective equipment (PPE)[1]. Disinfection robots require continued validation and testing before they are broadly deployed, and more pilot studies in a variety of settings are urgently needed to determine proper disinfectant dosages, techniques, and protocols.

Beyond disinfection, the aggressive adoption of cleaning, triage, and diagnostics robots around the globe has led to increased demand and new large scale investments. Moreover, hospitals are not the only places where robots are being deployed; we also identified increased growth in robot applications that reduce the need for physical human interaction, including public safety drones and security monitoring robots, social interaction and companion bots, autonomous delivery and transport, and, especially, behind-the-scenes manufacturing and logistics automation. Across the board, it is those robots that were already widely deployed before the pandemic that appear to be expanding market share most rapidly, followed by existing robot platforms that can be most readily adapted to meet the most pressing current needs. New robot systems are also being developed to automate and standardize high-risk tasks such as swabbing and sample collection. While such platforms are unlikely to be ready in time for use in the current pandemic, investments made now will no doubt result in systems that are ready for future crises. Meanwhile, we observe that teleoperation can be used as an effective intermediate measure to keep front line workers safe while remotely performing tasks that are not yet fully automated.

The Canadian robotic industry is also responding to this global pandemic. While many new COVID-related robot deployments in Canada center around laboratory testing automation, with small scale trials in disinfection, tele-medicine and floor cleaning ongoing in the healthcare sector, behind the scenes it appears that logistics is where the appetite for robotics is growing most quickly here. Two Canadian robotics companies in particular stand out in their adaptation to the needs of the pandemic: OTTO Motors and Avidbots. These companies have seen dramatic growth in demand for their fully autonomous

> While Canada has significant core technical strengths in robotics, it is not exploiting its robotics assets as heavily as it could. More international experience is needed here if our robotics industry is to overcome Canada's "small market" disadvantage.

---

1 For a layperson's explanation of current issues in UV effectiveness, see: T. Irving, "Can UV light help hospitals disinfect masks and gowns? U of T researcher explains," University of Toronto News, Apr. 06, 2020. https://www.utoronto.ca/news/can-uv-light-help-hospitals-disinfect-masks-and-gowns-u-t-researcher-explains

warehouse transport and industrial cleaning robots, respectively. Both companies are expanding their product lines to include disinfection robots with significantly improved autonomous capabilities as compared to existing products, but large investments are needed to bring hardware solutions to market here in a short timeframe. The NGEN supercluster, which is focused on accelerating advanced manufacturing practices, is funding multiple COVID-19 related projects with an emphasis on disinfection, but more is needed.

> **The pandemic has become an inflection point for accelerating investment in robotics.**

The story of robots and COVID-19 is not just about which countries are using robots to help manage their pandemic response; it is also about how the pandemic has become an inflection point for accelerating investment in robotics more broadly. Both at the macro/country level and at the micro/organizational level, those who were already invested in robotics technology before the pandemic have been better able to make use of robots during the pandemic. They are also more likely to benefit from the sudden demand for these technologies in this moment, and, we believe, for the foreseeable future. Further, we expect much of this growth in robot adoption to persist as the pandemic subsides, due to the strong gains in safety, efficiency, and convenience that they afford.

Our bottom-line conclusion is that while Canada has significant core technical strengths in robotics, it is not exploiting its robotics assets as heavily as it could. As a result, Canada risks falling behind other nations in leveraging robotics technologies to navigate the pandemic, post-pandemic recovery, and the future beyond.

The reasons for this are complex, but generally point to 1) Canada's lack of infrastructure and pathways to bring robotics R&D to market, and 2) this country's relatively small market economy, which has tended to make Canada a place for robot sales offices rather than a centre for international industry-focused robotics R&D and investment. Ultimately these conditions have made it a challenge for Canadian robotics startups to scale and grow, as they must quickly prove themselves capable of exporting internationally. This need to internationalize requires access to technical talent and financing, but also — and perhaps more importantly — access to executive talent with in-depth knowledge of global markets, as well as product management, organizational design,

> Momentum for robotics was already building in Canada before the pandemic.

operational finance, and marketing skills[2,3]. With only a handful of Canadian robotics companies operating at scale on the global market, and with few global corporations with R&D headquarters in Canada, such executive-level robotics talent has been hard to come by in this country. More mentorship and international business experience is needed here if we are to overcome our "small market" disadvantage.

Despite these challenges, several promising efforts to bridge the gaps in Canada's innovation pipeline were already underway before the pandemic struck. These include the creation of the Government of Canada's Innovation and Skills Plan[4] and Economic Strategy Tables[5], and the launch of the Pan-Canadian Artificial Intelligence Strategy[6] in 2017 and the $950M Supercluster Initiative[7] in 2019. While these activities are not focused on robots per se, they do call out robotics as among the key advanced technologies that will enable many aspects of Canada's future economy. Beyond providing funding from government and Canadian firms, such flagship efforts have garnered Canada international attention and investment: since 2017, global corporations such as Google[8], Uber[9], NVIDIA[10], Samsung[11,12], LG[13], and others have all opened corporate R&D centres in this country seeking our AI & robotics talent. A number of Canadian robotics firms have also

---

received significant VC fundings or been acquired since the launch of these efforts as well[14]. In short: pre-COVID, momentum for robotics was already building in Canada.

But what of now? The COVID-19 crisis has undoubtedly challenged Canadian robotics firms. Like all businesses worldwide, they have had to cope with temporary workplace closures, physical distancing measures, and global economic uncertainty. Despite these immediate concerns, however, the robotics companies we interviewed generally reported a positive outlook, with one key caveat. The good news is that existing customers are leaning more heavily on automation to help meet productivity demands while at the same time managing worker shortages and more restricted operational environments. And there appears to be interest from new customers too, both in Canada and globally. Given the longer term economic uncertainties, however, it remains to be seen how many of these potential new customers will survive long enough and will have the access to the capital needed to invest in automation in the long run. In other words, our interviews suggest that a chief concern among Canadian robotics startups is the generation of new revenue streams[15] — not surprising given that scaling up is what most Canadian robotics businesses must do to survive. Despite this challenge, the current crisis has created an unprecedented need for government coordination to ensure access to essential medical supplies and goods to keep the economy moving. With this comes the opportunity for Canada to leverage its single-payer healthcare system to fund pilot programs that put robotics to the test, and eventually, into broader use. Likewise, business incentives could be created to enable Canadian manufacturers of key goods to upskill their workforce and modernize their warehouses with current-generation collaborative robots, making them more competitive on the global stage. If designed appropriately, such measures could help solve urgent pandemic needs, while also creating long term gains in innovation, productivity, demand for skilled jobs, and greater capacity to compete internationally.

> A chief concern among Canadian robotics startups is the generation of new revenue streams.

---

14  The number of Canadian robotics companies that have received VC funding since 2017 is too long to list here, however significant fundings (over $10M) since 2017 include: Titan Medical, Synaptive Medical, ATTAbotics, Avidbots, Kindred AI, OTTO Motors, Robotiq, Resson Aerospace, Kinova, and Drone Delivery Canada. Recent acquisitions include Aeryon Labs (acquired by FLIR Systems for $266M in 2019), and MDA (bought out by Northern Private Capital for CA$1B in 2019). Source: Crunchbase https://www.crunchbase.com/

15  These findings are echoed in a more general study on the impact of the pandemic on Canadian technology companies. See: JS. Denney and V. Vu, "The COVID-19 Crisis and Policy Preferences of Canadian Technology Scale-ups," Innovation Policy Lab at the Munk School of Global Affairs, p. 13, Mar. 2020. https://munkschool.utoronto.ca/ipl/files/2020/04/IPL-report-2020-02_COVID19-Policy-Preferences-Scale-ups-FINAL.pdf (accessed Sep. 01, 2020)

> When it comes to robotics, Canada lacks the strategic coordination of other nations. The time to act is now.

Canada's early and significant investments in space robotics helped put this country ahead of the curve in terms of robotics R&D know-how, and our current robotics research centres and startups — though relatively small in number — are known internationally. Recent data from Statistics Canada show that our high tech industry is weathering the pandemic better than other sectors, and is the fastest growing part of the Canadian economy (currently representing about 5% of GDP)[16]. Moreover, according to the Economist, Canada ranks fifth globally for "Automation Readiness" (behind robotics powerhouses South Korea, Germany, Singapore and Japan) due in large part to our strong public education system and progressive workplace policies[17]. And yet Canada lacks the focused robotics R&D funding and strategic stakeholder coordination of other nations such as China, Japan, the US, the EU, South Korea and Singapore, all which have significant multi-annual robotics programs that are typically linked to a larger digital innovation agenda.

Ultimately, it is our view that Canada needs a unified robotics roadmap, tying industry, government and academia together with a clear mandate for specialization and persistent investment in the robotics industry. With the right strategic investment focus going forward, Canada could reignite its robotics innovation pipeline as an important growth engine for the post-pandemic economic recovery. Given the urgent need, the exceptionally competitive global robotics landscape, and the volatility of the current economic climate, the time to act is now.

---

# TABLE OF CONTENTS

# 01
INTRODUCTION

# 1.1
# ABOUT THE AUTHORS

This white paper has been written collectively by faculty members and staff of the University of Toronto Robotics Institute, which represents the largest cluster of academic robotics researchers in Canada. Together we have domain expertise that spans from healthcare, to manufacturing, to the future of mobility, and technical expertise covering everything from dynamics and control, to machine learning, to robot vision and more.

The University of Toronto's rich history in robotics began in the early 1980s with several large robotics groups in the Department of Mechanical and Industrial Engineering (MIE) and the Institute for Aerospace Studies (UTIAS), many with close ties to Canadarm, Canada's internationally-known space robotics legacy. Subsequent decades saw UofT's expansion into the fields of medical robotics and advanced manufacturing. Reflecting the growing interdisciplinarity of the field, the UofT Robotics Institute was launched in 2019 to unite the many robotics research groups that have since emerged across the university's various Departments and Faculties, from Engineering, to Computer Science, to Architecture.

With the help of our international partners, our research network extends around the globe. Our extensive connections to Canada's largest health network and to academic, industry and government robotics partners from across the country contribute to our uniquely rich understanding of Canadian robotics in the global pandemic context and beyond.

# 1.2 ACKNOWLEDGEMENTS

We would like to extend thanks to the following organizations that provided critical background and advice in the preparation of this document:

ANVIV
Avidbots
Canadian Space Agency (CSA)
Clearpath Robotics / OTTO Motors
Defence R&D Canada (DRDC)
Drone Delivery Canada
ESI
FIRST Robotics Canada
Gatik
Hospital For Sick Children
Innovation Policy Lab, Munk School of Global Affairs, University of Toronto
Kinova
Loblaws
MDA Robotics
Metrolinx
Mount Sinai Hospital
National Research Council (NRC)
Responsible Autonomy & Intelligent System Ethics (RAISE) Lab, MGill University
RoboEye.ai
Robotiq
Savioke
Silicon Valley Robotics
Synaptive Medical
Toronto District School Board
Toronto Transit Commission

We also wish to acknowledge our many international robotics colleagues for their expert knowledge and thought leadership on the future of robotics and its application to the COVID pandemic. Robotics, like many sciences, is an international discipline and knows no borders. Research and opinion from these experts are cited throughout this document.

# 1.3 MOTIVATION

With its many overlapping technical, social, economic and political drivers, the field of robotics was complex and dynamic before the COVID-19 pandemic. Before the pandemic, robotics media hype was also the norm. Now, the pandemic appears to be accelerating investment and innovation in robotics, while at the same time exacerbating perennial robotics issues such as global competition, privacy, ethics, embedded bias, and job loss due to automation. Meanwhile COVID-19 misinformation and disinformation abounds.

> Non-experts must be able to distinguish between early stage robotics experiments and robots that are ready for real-world deployment.

It is our view that there has never been a greater need to help non-experts distinguish between early stage robotics experiments and robots that are ready for real-world deployment. Expert views are needed to counterbalance the plethora of news and social media accounts of robots being used (or pitched for use) in the pandemic, many of which obscure or overstate the actual capabilities of the machines in question. And it's not only the media that are guilty of using robots as "click-bait". Researchers have also been known to overstate the capabilities of their systems or underestimate research and development timelines, particularly when applying for grants. While sightings of novel use cases or innovative approaches can be an important signal of potential technologies to come, they are typically a better indicator of humankind's fascination with robots than of current technological capability. Unfortunately, when such stories receive attention that is disproportionate relative to the robot's actual capabilities, readers may be left with a distorted understanding of the technology, the timeline for its development, and thus its potential for impact.

Nor has there been a greater need for experts and non-experts alike to understand the larger economic, technical and social trends that are shaping this field as it responds to the coronavirus pandemic and the ensuing economic crisis. As one of the first countries in the world to have run a successful national robotics research program — the Canadarm Space Robotics program — Canada has a rich history in robotics and is recognized globally for its robotics expertise. And yet we have deployed relatively few robots in the COVID-19 pandemic. Why is this? Both internationally and here in Canada, there are tremendous opportunities for robotics to help navigate and recover from the pandemic.

Our aim with this whitepaper is to leverage expert robotics knowledge from research and industry to uncover useful boots-on-the-ground opportunities for Canadian robotics within the evolving context of the global pandemic and post-pandemic economic recovery.

# 1.4 RELATED WORK

Because the COVID-19 pandemic is new and we are still learning about the virus' transmission and its impacts, research on the actual application of robotics to COVID-specific contexts is only just now emerging[18]. In response to this information gap, calls for topics involving robots and COVID-19 are being issued by granting agencies and research publications alike[19 20 21 22]. We can therefore expect more science about robots and COVID-19 to emerge over the coming months and years.

Yet despite the current lack of COVID-specific robotics research, and though we are only a few months into the COVID-19 pandemic as we write this paper, a number of prominent robotics experts from both academia and industry have already published essays, opinion pieces, and surveys about robots and COVID-19. These works can be generally described as 1) documenting how and where robots are being used in the current pandemic, 2) exploring how the pandemic is accelerating the pace of robotics innovation and adoption, 3) explaining why robots are not being more widely used in the current crisis, and/or 4) making recommendations for future deployments. The paragraphs below describe such work by robotics experts in more detail.

---

18  See for example: A. A. Malik, T. Masood, and R. Kousar, "Repurposing factories with robotics in the face of COVID-19," Science Robotics, vol. 5, no. 43, Jun. 2020, doi: 10.1126/scirobotics.abc2782.
19  "Join the AI-ROBOTICS vs COVID-19 initiative of the European AI Alliance," Shaping Europe's digital future - European Commission, Mar. 25, 2020. https://ec.europa.eu/digital-single-market/en/news/join-ai-robotics-vs-covid-19-initiative-european-ai-alliance (accessed Sep. 01, 2020).
20  "Call for funding: robotic tech and solutions immediately deployable in hospitals," Shaping Europe's digital future - European Commission, Apr. 10, 2020. https://ec.europa.eu/digital-single-market/en/news/call-funding-robotic-tech-and-solutions-immediately-deployable-hospitals  (accessed Sep. 01, 2020).
21  "RAM Special Issue on Robotics Response for the COVID-19 Outbreak," IEEE Robotics and Automation Society. https://www.ieee-ras.org/about-ras/latest-news/1639-ram-special-issue-on-robotics-response-for-the-covid-19-outbreak  (accessed Sep. 01, 2020).
22  "Robotics, Autonomous Systems and AI for Nonurgent/Nonemergent Healthcare Delivery During and After the COVID-19 Pandemic," Frontiers Research Topic. https://www.frontiersin.org/research-topics/14462/robotics-autonomous-systems-and-ai-for-nonurgentnonemergent-healthcare-delivery-during-and-after-the  (accessed Sep. 01, 2020).

> "Robots used during an emergency are usually already in common use before the disaster. Technologists often rush out well-intentioned prototypes, but during an emergency, responders [...] are too busy and stressed to learn to use something new and unfamiliar. They typically can't absorb the unanticipated tasks and procedures, like having to frequently reboot or change batteries, that usually accompany new technology."
> – Murphy et al.[30]

In the early days of the pandemic, a prominent group of international robotics academics, Yang et al., published an editorial in Science Robotics that briefly summarizes many of the applications and opportunities for robotics in the COVID-19 pandemic context[23]. Tavakoli et al. subsequently published an essay in Advanced Intelligent Systems that provides examples of how autonomous robots and smart wearables can complement and support healthcare during the pandemic and beyond[24]. euRobotics lists active European robotics projects where robots are already helping, including disinfection, delivering medicine, and monitoring vital signs[25]. Articles such as these posit that robots can and will be a key aspect of dealing both with the current healthcare crisis and future pandemics.

Other expert articles discuss some of the critical challenges around implementing current robots during the pandemic. For example, Robohub, an online robotics research forum, highlighted a number of the COVID-19 robotics activities taking place around the world and discussed why we do not see more robots on active duty[26]. A group led by Prof. Robin Murphy (Texas A&M University), who is lauded for her research on robotics for disaster applications, has interviewed experts in public health, public safety, and emergency management[27] and documented over 100 press and social media reports from more than twenty countries about how robots are being used during the coronavirus pandemic[28]. Her group analyzed these findings and categorized reported robot use cases according to their frequency of deployment. Results from this analysis confirm a key finding from Murphy's prior research[29] in disaster robotics: namely that the robots which are most likely to be deployed during a disaster are the robots that were already in common use before the disaster. This is because first responders and front line workers do not have the time during an emergency to experiment with and adapt their procedures to

---

new technologies[30]. Despite this, Murphy's team observed that some COVID-19 responders are beginning to test robotics technologies that have already been proven in other application areas. For example, there have been reports of pesticide-spraying drones designed for agriculture being adapted into drones that spray disinfectant in public spaces[31].

Other groups have published reports and commentary exploring how the pandemic is accelerating the use of robotics more broadly, and what the economic, social or technical implications of this accelerated adoption might be. For example, researchers at Northwestern University's Retail Analytics Council recently published an article that documents how COVID-19 is accelerating demand for robots in grocery retail[32]. The authors observed that while retail sales generally declined in the early months of COVID-19, grocery sales remained strong. Given that grocery retailers already represented the largest retail segment invested in robotics technology before the pandemic, the authors concluded that the outbreak will drive grocery retailers to invest even more in robotics going forward. Another exploration of how COVID-19 may be accelerating the adoption of robotics is the COVID-19, Robots and Us[33] video series, co-produced by Silicon Valley Robotics and the CITRIS People and Robots Initiative headquartered at UC Berkeley. In this series, well-known robotics researcher Prof. Ken Goldberg suggests that our fear of the coronavirus is accelerating us towards a "low touch future" where autonomous robots help limit our physical interactions with people by performing tasks such as preparing our fast food and take-out coffee[34]. Another team exploring how pandemic psychology is accelerating robot adoption is Georgia Tech robotics expert Ayanna Howard and her ethics colleague Jason Borenstein. In their MIT Sloan Management Review article[35], they warn that in our rush to adopt robots in the emergency response, robots

---

30  J. Adams, R. Murphy, and V. Gandudi, "Robots are playing many roles in the coronavirus crisis – and offering lessons for future disasters," The Conversation, Apr. 22, 2020. http://theconversation.com/robots-are-playing-many-roles-in-the-coronavirus-crisis-and-offering-lessons-for-future-disasters-135527  (accessed Sep. 01, 2020).
31  J. Antunes, "Drones on the Front Lines of the COVID-19 Pandemic," Commercial UAV News, Mar. 26, 2020. https://www.commercialuavnews.com/public-safety/drones-on-the-front-lines-of-the-covid-19-pandemic  (accessed Sep. 01, 2020).
32  "Emerging Trends in Retail Robotics," Retail Analytics Council, Northwestern University, May 2020. https://s3.amazonaws.com/medill.wordpress.offload/WP%20Media%20Folder%20-%20imc-retail-analytics-council/wp-content/uploads/sites/3/2020/05/RAC-Research-Art.-Emerging-Trends-in-Retail-Robotics.pdf (accessed Sep. 01, 2020).
33  "COVID-19, Robots and Us: Weekly Online Discussions with Experts," Silicon Valley Robotics | CITRIS, People and Robots, 2020. https://www.youtube.com/watch?v=zsbmK5N5WTs&list=PLtjElujxzII5CJrgnm108SZ6itJcYPt5q&index=2&t=0s (accessed Sep. 01, 2020).
34  "COVID-19, Robots and Us: Weekly Online Discussions with Experts | April 7th 2020 Episode," Silicon Valley Robotics | CITRIS, People and Robots, 2020. https://www.youtube.com/watch?v=C6e5jQWdVY4 (accessed Sep. 01, 2020).
35  A. Howard and J. Borenstein, "AI, Robots, and Ethics in the Age of COVID-19," MIT Sloan Management Review, May 12, 2020. https://sloanreview.mit.edu/article/ai-robots-and-ethics-in-the-age-of-covid-19/ (accessed Sep. 01, 2020).

> Accelerated demand has exposed robots' technical limitations while at the same time making it difficult to see the real progress the field has made in recent years.

will be positioned as a technology that is necessary to protect our health, and we will be more likely to overlook concerns such as privacy or embedded bias as a result. As such, they advise that proactive planning and protections are required before robots are widely deployed for the pandemic.

Others still have argued that the accelerated demand has exposed robots' technical limitations while at the same time making it difficult to see the real progress the field has made in recent years. For example, The Robot Report, a prominent robotics industry news publication, released a special COVID-19 issue[36] declaring that fully autonomous robot solutions are still many years away, but that robot developers need not give up: they can rely on humans to help with tricky "edge cases" that robots can't handle on their own. To do so, they advise the robotics industry to develop better metrics for measuring system reliability and more transparent error-handling to make it easier to hand off control between humans and machines.

We expect that more expert works such as these will be published in the coming months as unforeseen use cases and challenges emerge, and as new research on the application of robotics to the coronavirus pandemic appears.

This whitepaper attempts to build off the works described above, leveraging our extensive Canadian and international robotics research networks, as well as interviews with end users from Canadian hospitals, long-term care homes, public transportation services and schools to bring a uniquely Canadian lens to the discussion.

---

# 1.5 SCOPE

We use this section to make clear what this whitepaper covers and what it does not.

The term "robot" means different things to different people. Our primary focus will be on systems that are physically embodied and have not only computing elements but also sensors to perceive the surrounding environment and, importantly, actuators (e.g., motors) to affect change in the environment. As such, while dis-embodied artificial intelligence algorithms ("software or sensing-only robots") will no doubt prove helpful in this pandemic, we will not cover them in this paper.

This definition of robotics is still general enough that it will be too difficult to cover all the ways robots can help us through the COVID-19 pandemic. Accordingly, we prioritize applications that are closer to the frontline such as in hospitals, long-term care facilities, and public health applications that are aimed to keep the general population safe. We also include essential services that are needed to keep the economy going, such as groceries and public transportation.

Finally, we also consider it in scope to discuss not only the technologies required to help but also the logistics of scaling solutions up. This requires that we analyze the capacity of industry to provide robots at scale, the operators and job skills needed to absorb and maintain technology solutions, the funding mechanisms needed to pay for these solutions, and the regulatory frameworks within which the technology must work. Given the relatively high cost and long timelines for robotics development (as compared to software solutions), it also requires us to consider the potential for post-pandemic demand, otherwise the investments in robotics technology that we make now risk becoming irrelevant after the pandemic.

To summarize, our three guiding principles in scoping this whitepaper are:

- Physically embodied robots
- Frontline and essential services applications
- Capacity to scale and longer term demand

# 1.6 HOW THIS DOCUMENT IS ORGANIZED

This document aims to provide decision-makers with information on the use of robots in the Canadian fight against the COVID-19 pandemic, including an assessment of the robotics industry (Canadian and international) and its capacity to deliver reliable and robust solutions that meet the needs of Canadian businesses and institutions. Thus far, the Introduction provided readers with information about the authors and our motivation for writing this paper. We also provided background information on how robot experts are thinking about the current pandemic situation. We then defined what we consider to be in scope and out of scope for this paper. The remainder of this whitepaper is organized as follows:

Section 2 provides essential background on robotics for non-experts, including a primer on robotics and an overview of the standard levels used to describe robot autonomy. This section also describes the Government of Canada's Technology Readiness Levels in the context of how robotics are developed from lab experiments through to deployment in real world applications. We end this section with a brief overview of the Canadian robotics landscape, giving a brief history of Canadian robotics, and an overview of the major robotics research centres and companies operating in this country. Readers who are already knowledgeable about robotics and the Canadian robotics landscape should feel free to skip this section.

Section 3 describes the main areas where robots are already being used to fight COVID-19 both globally and in Canada, and describes the various patterns of robot development during the crisis. Section 3.1 describes these use cases and includes specific examples of robots that have been deployed or are being developed, what function they are serving, who has produced them, and how much they cost (if available). Based on our knowledge and expertise developing robotics systems, we evaluate the maturity of these systems by grouping them according to the Technology Readiness Levels (TRLs) described in Section 2. In Section 3.2 we share our observations on how Canada is putting robots to use during the pandemic, covering both Canadian-based deployments, as well the deployments of made-in-Canada solutions. We finish with Section 3.3, which models the main pathways for robot development we have observed in the course of our research, including specific examples of each strategy in action.

Section 4 describes the various technological, social and systemic factors that are determining robot deployment success. Since most of these factors are common to any country developing robotics, we first describe them from a general perspective. We then follow up with a brief explanation of how the factors play out in a Canadian context, to highlight Canada's relative strengths and weaknesses.

Section 5 summarizes the main observations we made in the course of our research and evaluates Canada's overall ability to leverage robotics towards the pandemic response and future crises. We finish by proposing several steps Canada can take to better exploit its robotics assets.

Additional details on the Canadian robotics landscape are provided in the Appendix.

# ESSENTIAL BACKGROUND

This section provides background information on robotics, technology readiness levels, the usual robotics development approach, and a brief inventory of robotics assets in Canada. Those readers already familiar with this information can safely skip to Section 3.

# 2.1
## PRIMER ON ROBOTICS

We use this section to provide a basic summary of robotic concepts, and to explain (from a technological development standpoint) what is easy and what is challenging for robots to do.

Robots can have different types of "bodies". The three most common robot body types are the "manipulator", the "mobile robot", and the "drone", as depicted in Figure 1.

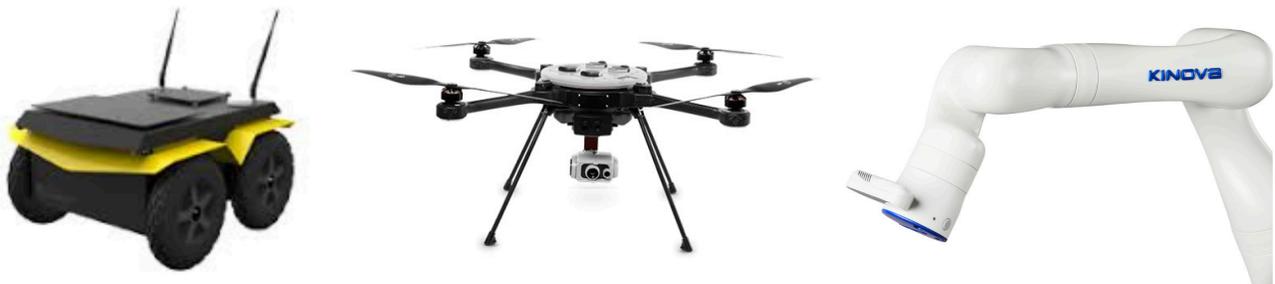

*Figure 1: Different robot body types, all developed in Canada: (left) Clearpath Robotics Jackal "mobile robot", (centre) FLIR Systems R80D SkyRaider "drone"; (right) Kinova Jaco for Gen3 robot "manipulator".*

Manipulators can be mounted on top of mobile robots to create "mobile manipulators". There are also many variations of robot bodies. For example, instead of wheels, mobile robots can have legs or tracks. Drones, which are also known as "unmanned aerial vehicles" (UAV) or "unmanned aerial systems" (UAS), are a form of mobile robot that can fly. Manipulators can also be specifically designed for high-performance tasks, such as robotic surgery. Researchers continue to experiment with different robot body types, such as soft robots, continuum (snake-like) robots, exoskeletons, and microrobots.

Robots can have different types of sensors mounted on them to perceive the surrounding environment. Some common sensors are optical cameras (similar to the camera on a phone), depth cameras (these can passively measure the distance between the camera and another object), laser rangefinders (these can actively measure distance to objects using laser light), and radar (these can actively measure distance to objects using radio-frequency radiation). Outdoor robots often use the Global Positioning System (GPS) to determine their position. Manipulators sometimes have sensors that help them measure how much force they are applying to objects. Individual motors usually have sensors to measure how fast and how far they have turned.

Robots can have different "control" approaches to make them behave as desired. The simplest method of control is "remote control" where an operator observes the robot from afar and provides low-level motion commands. An example of remote control can be found in the typical radio-controlled toy car, where the operator uses a joystick to tell the robots' motors to turn in a certain direction. If an operator cannot be physically close enough to see a robot directly, "teleoperation" can be used, where cameras mounted on the robot provide feedback to a remote display; the operator still provides low-level motion commands. In the case of manipulators (e.g., for surgery), other sensor information (such as force feedback) may also be provided to help the operator understand how hard the robot is pushing on something. Sometimes it is difficult to teleoperate a robot using video feedback due to time delays or the environment being cluttered; such situations may call for "assisted teleoperation" or "supervised autonomy", where the robot is given some basic ability to detect and avoid obstacles, or to drive on its own for a short distance for example. Finally, "full autonomy" usually refers to a robot that is able to operate without human supervision for long periods of time. Figure 2 shows these different levels of autonomy mapped against the industry-standard levels used to categorize self-driving cars. A similar system is used to describe autonomy levels in medical applications such as surgical robotics[37].

---

37  T. Haidegger, "Scaling the Autonomy of Surgical Robots," presented at the 2017 IEEE/RSJ International Conference on Intelligent Robots and Systems (IROS), Vancouver, 2017, Accessed: Sep. 01, 2020. Available: http://real.mtak.hu/86408/

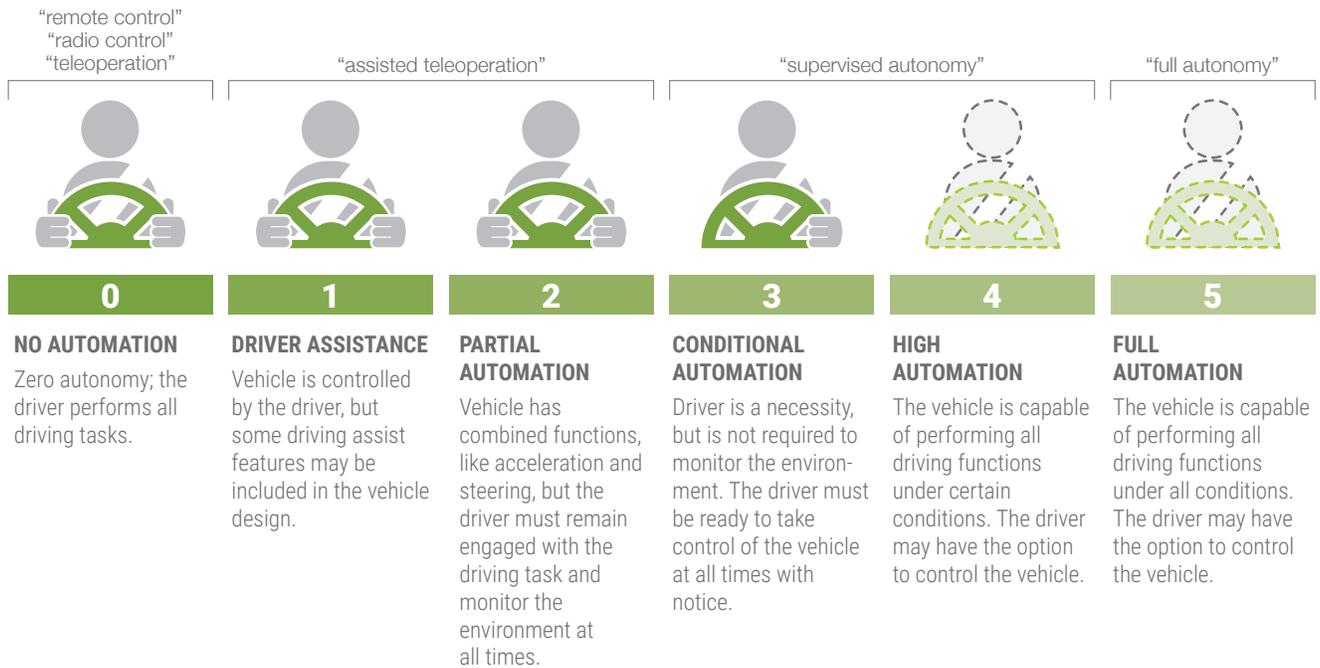

*Figure 2: Industry-standard autonomy levels for self-driving cars developed by the Society for Automotive Engineers (SAE) and mapped to our autonomy terminology. Image credit: Adapted from SAE..[38]*

There are two reasons to make a robot autonomous. The first reason is because the task that must be performed cannot be accomplished any other way. An example of this is a Mars rover, where the long communications time delay rules out teleoperation and so some operations must be autonomous in order to advance the mission in a timely manner. Another example is the quadrotor drone, where the dynamics are too complex for a human to control without assistance. The second reason to make a robot autonomous is to achieve some kind of productivity or safety improvement by reducing the human involvement required for operations. An example of this is the automotive assembly line, where many robots are able to carry out repetitive tasks, helping prevent repetitive stress injuries and freeing workers to focus on higher level operations. Full autonomy for highly unstructured environments (e.g., self-driving cars or a homecare robot) has not yet been achieved. Progress is being made, but generally speaking there is an inverse relationship between the structure of the task and the current state of the art in autonomy: the more unstructured the task, the less reliable are the current autonomous solutions.

At the risk of generalizing, mobile robots are fairly good at light-weight, low-speed applications where they move around and avoid static or slow-moving obstacles; current vacuum robots are an example of this. Scaling up to high-speed, heavy vehicles (i.e., self-driving cars) is still challenging unless the environment can be simplified, such as in warehouse and mining applications where human access to the environment is purposely restricted. Manipulator robots follow a similar trend. Tasks carried out in highly structured environments with humans removed are commonplace (e.g., assembly lines). Robots that can work alongside humans (known as collaborative robots[39] or "cobots") are becoming more

---

common. Picking up objects of known shape from a predetermined location using specialized tools is easy; picking up flexible or oddly shaped objects from unknown or cluttered locations using a generic "hand" is hard (e.g., getting an item from the back of the fridge).

In this whitepaper we provide an honest and realistic assessment of where robots are likely to be helpful in fighting COVID-19. The challenge is to find those applications where current technology can actually make a difference.

## 2.2 TECHNOLOGY READINESS LEVELS

> Technology readiness levels (TRLs) are used to help assess and compare the maturity of technologies across different stages of development.

Technology Readiness Levels (TRLs) are a scale developed in the 1970s and 1980s by the US National Aeronautics and Space Administration (NASA)[40] for assessing and comparing the maturity of technologies across different stages of development, and fostering mutual understanding among stakeholders from government, industry, and research. Variations of NASA's original TRLs have since been developed by government entities around the world, including the European Space Agency and the US Department of Defense (DoD)[41]. In the sections that follow, we will be assessing the maturity of different robotics technologies and their use in applications related to COVID-19 using the Government of Canada's Technology Readiness Levels (TRL)[42], summarized in Figure 3.

Naturally, a single TRL level is an oversimplification of the state of a given technology and it must be taken as an approximate assessment at best. Moreover, it is critical to understand the details of each specific application when making decisions about whether robots will actually be useful given the current state of the art. For example, "Telepresence" refers to teleoperating a mobile robot while also having a video conference with people at the remote location; this could be very useful for patient care and is quite reliable in certain situations, but also has its challenges such as handling the opening of doors, using elevators, and requiring a highly reliable communication connection. In this example, the TRL would be quite different depending on whether the robot must be able to open doors and operate elevators as part of its job. For this reason, a TRL level can only capture the true status of technology if the use case is very clearly defined.

---

40 A. Olechowski, S. D. Eppinger, and N. Joglekar, "Technology Readiness Levels at 40: A Study of State-of-the-Art Use, Challenges, and Opportunities," 2015 Proceedings of PICMET '15: Management of the Technology Age, p. 11, 2015. Available: http://web.mit.edu/eppinger/www/pdf/Eppinger_PICMET2015.pdf
41 M. Héder, "From NASA to EU: the evolution of the TRL scale in Public Sector Innovation," The Innovation Journal, vol. 22, p. 23, 2017. Available: https://core.ac.uk/download/pdf/94310086.pdf
42 "Technology readiness levels," Innovation, Science and Economic Development Canada, Jan. 23, 2018. https://www.ic.gc.ca/eic/site/080.nsf/eng/00002.html (accessed Sep. 01, 2020).

We also note that the language used in the TRL definitions of Figure 3 is not universal in its meaning. For example, in the robotics community "simulation" implies something quite different than it does in the medical community. In the context of the TRL definitions described below, we take simulation to mean something between a test in the lab and operation in the real world; it could be a pilot program in a mock physical environment, or field test of the component, subsystem, or system. Additionally, "qualified" must be taken to mean that the technology not only functions as planned but also adheres to any relevant regulatory framework for the particular application (e.g., from Health Canada, Transport Canada, etc.).

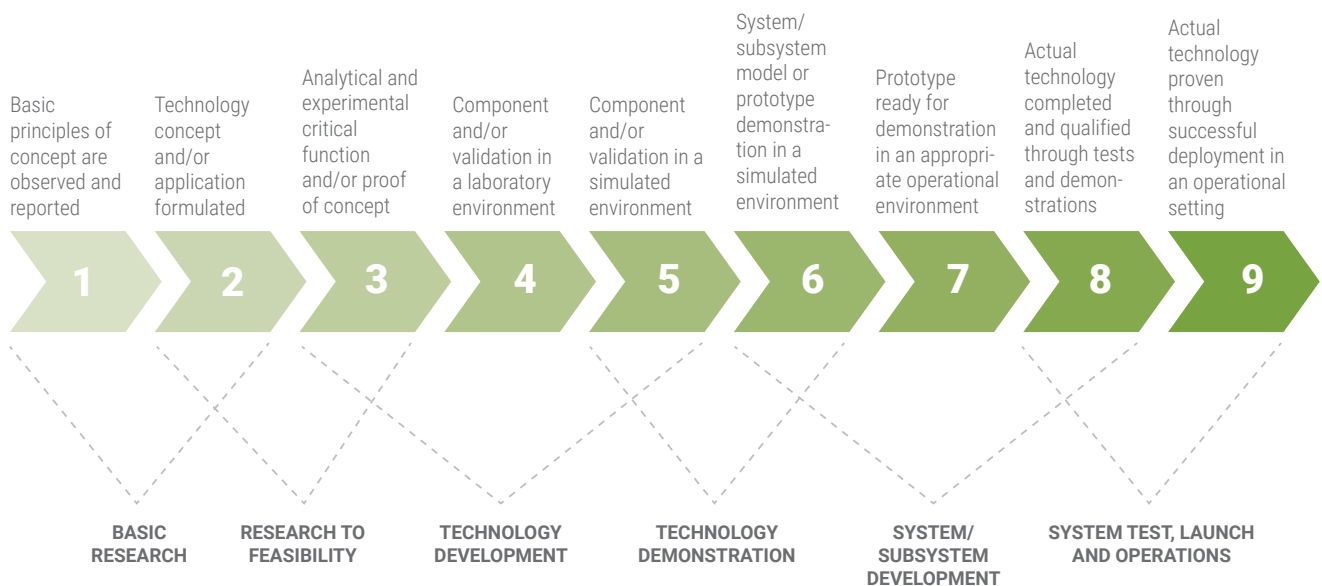

*Figure 3: Government of Canada Technology Readiness Levels (TRL) with typical development phases indicated below. Note that because Technology Readiness Levels stop short of considering market factors such as customer readiness, investment, or supply chains, TRLs alone cannot be used to measure capacity to scale or potential impact. Image credit: Adapted from Gamechangers[43].*

---

43 "Technology Readiness Levels," Game Changers Innovation. https://www.gamechangers.technology/technology-readiness-levels/ (accessed Sep. 01, 2020).

> Technology Readiness Levels stop short of considering market factors such as customer readiness, investment, or supply chains.

Though we will show some examples of emerging technology in this paper to help readers understand the breadth of robotics technologies being developed for the pandemic, we are specifically interested in robots that have been demonstrated or deployed in application environments, because these are the robots with the most realistic impact potential in the short to mid term. This means that our real interest is in robots with TRL7 or higher, as only these involve demonstrations or deployments in actual operational environments (TRL5 and TRL6 involve demonstrations of robots only in simulated environments). Because Technology Readiness Levels stop short of considering market factors such as customer readiness, investment or supply chains, TRLs alone cannot be used to measure capacity to scale or to assess potential impact. For this reason, while we use TRL levels to assess robot development, they alone are insufficient for determining whether a solution is ready for widespread deployment.

## 2.3 HOW ROBOTS ARE DEVELOPED

Below the TRL levels in Figure 3, we see typical labels applied to the development phases for technology. These labels map a progression from basic research, to proofs of concept (feasibility), to development (turning prototypes into robust products), to demonstrations (verification) and pilot deployments (validation), to product launch and operations. Real robotic systems are usually developed in this traditional way; the technology must climb the TRL ladder step by step. University research labs are typically involved in the low-TRL range (1-3), but may be involved in TRL4 when working with industry sponsors. University-based innovation labs and incubators may develop up to TRL6. Industrial robot developers then mature the technology (TRL3-7) whereupon the end users or 'systems integrators' build applications (TRL7-9)[44]. It is usually a multi-year process to design a new robot product from scratch due to the many hardware and software components that must be brought together in a functioning system that must be both safe and adhere to the relevant regulatory frameworks. See Appendix A.2 for more information on Canada's legal and regulatory landscape.

---

44  Note that due to the relative immaturity of the robot market, most robot developers also play the role of application developers.

> Developing robotics takes longer and is more expensive than developing than software solutions. Hardware must be prototyped, safety tested, and regulation-compliant before it is feasibly scaled up for manufacturing.

The development process for robotics is longer and costlier than a pure software product (such as an artificial intelligence algorithm or app) because hardware must not compromise safety, and yet it must be prototyped, tested, and regulation-compliant before it is feasibly scaled up for manufacturing. At the same time, robot development involves not only hardware (i.e. the robot's body and sensors) but also software, which must do everything from reasoning from the robots' sensor data and planning its motions, to controlling the robot's hardware and providing an interface for human control. The implication of this for COVID-19 is that if robots are to have an impact in the short term, only products that are well on their way to real applications (e.g., TRL7 and higher) can be ready in time.

## 2.4 BRIEF OVERVIEW OF CANADIAN ROBOTICS

> Canada's early investments through the 1970-80s helped establish our reputation as a key international player in robotics.

Canada has a rich history in robotics. Perhaps the most well-known accomplishments are the Canadarm and Canadarm2 space manipulators used on the Space Shuttles and International Space Station, respectively. Early and sustained investments in the Canada Space Arm (1975-2011)[45] and CIFAR's Artificial Intelligence, Robotics, and Society program (launched in 1983)[46] established Canada's reputation as a key international player in robotics. More recently, those early investments were renewed via a number of complementary initiatives. One of these is the $125 million Pan-Canadian Artificial Intelligence Strategy[47], launched in 2017, which established three national AI institutes: the Amii Institute in Alberta, the Mila Institute in Montreal, and the Vector Institute in Toronto. The $950 million

---

Supercluster Initiative launched its Scale.ai (artificial intelligence) and NGen (advanced manufacturing) programs[48] in 2018. Most recently, Innovation, Science and Economic Development Canada (ISED) launched its new Canadian Space Strategy (2019), which promises to invest $1.9 billion specifically to develop AI-enabled deep-space robotics systems[49].

Owing in part to this rich history, Canada has an excellent ecosystem of robotics experts distributed across the country in academia, industry, and government, and there are too many to cover in this paper. The Robot Report Global Map, which tracks robotics around the world, lists over 280 Canadian robotics organizations in Canada[50]. We briefly use this section to highlight some of Canada's key assets, depicted in Figure 4. More detail is provided in Appendix A.

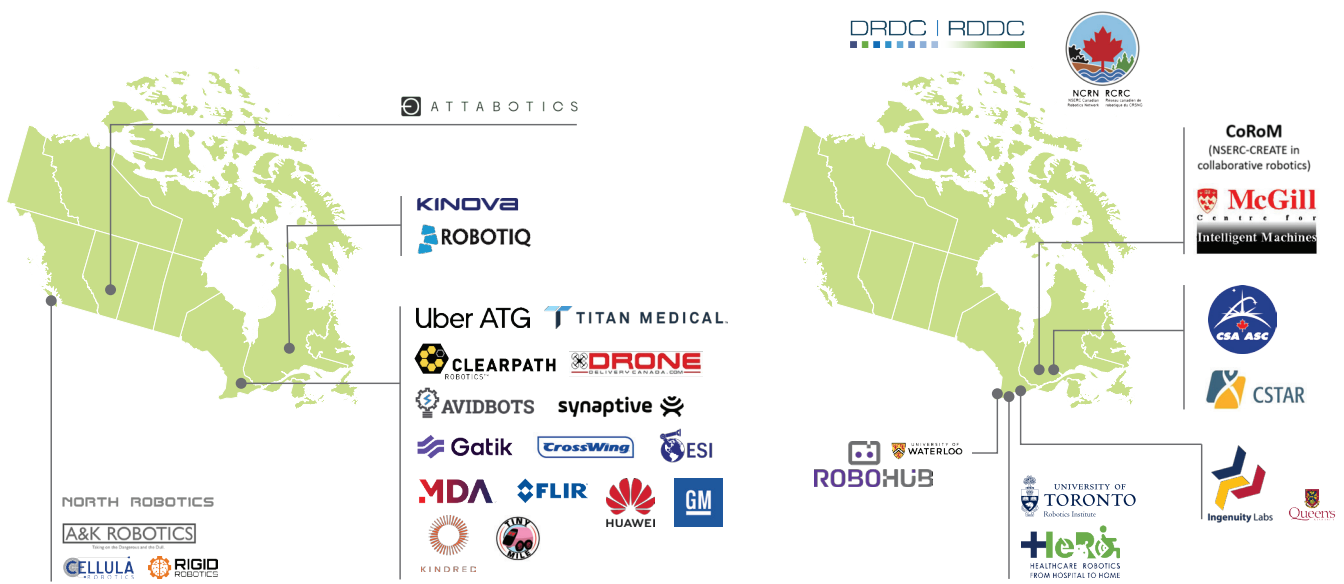

*Figure 4: Map of Canada's key robotics assets: (left) Distribution of select companies working directly in robotics in Canada; (right) Canada's largest academic and government robotics centers.*

Academically speaking, Canada's universities excel in fundamental and applied robotics research. For example, at the 2019 IEEE International Conference on Robotics and Automation, the University of Toronto ranked #6 in university conference contributions, but jumped to #1

---

> Canadian universities are on the whole following a global trend towards growing larger robotics centres involving multiple faculty from a variety of disciplines.

when it came to award nominations[51]. Since 2012, a Canada-wide network has existed called the NSERC[52] Canadian (Field) Robotics Network (NCRN)[53], led by McGill University. NCRN used to focus on field robotics, but more recently broadened its mandate. It now brings together many of the universities currently actively working on robotics, including McGill University, Queen's University, University of Alberta, University of Sherbrooke, University of Toronto, University of Waterloo, York University, University of Laval, University of Montreal, and Simon Fraser University. NCRN's mission is "to create the resilient and interacting robots of the future", making it a logical starting point for COVID-related robotics research contacts. Beyond NCRN, there are many robotics faculty and laboratories located in universities across the country. Some have significant expertise in areas that are relevant to the pandemic, such as medical robotics, for example the Canadian Surgical Technologies and Advanced Robotics (CSTAR) Engineering Group[54] at the University of Western Ontario and the NSERC HeRo[55] CREATE on healthcare robotics at the University of Toronto. Also relevant to a robotized pandemic response are our diverse research capabilities in the field of human-robot interaction (HRI), with expertise in everything from affective and social robots[56,57], to human-robot interfaces[58,59], collaborative industrial systems[60,61,62], HRI ethics[63], and rehabilitation[64,65]. Canadian universities are on the whole following a global trend towards growing larger robotics centres involving multiple faculty from a variety of disciplines. Examples of this trend include the University of Toronto Robotics Institute[66] (the authors of this white paper), and other large academic robotics hubs such as the Robohub[67] at Waterloo, the Ingenuity Center[68] at Queen's, and the Centre for Intelligent Machines[69] at McGill.

On the industry side, Canada is also very strong. MDA Robotics (formerly MD Robotics, formerly SPAR Aerospace) supplied the 1st Canadarm (called Space Shuttle Remote Manipulator System) in 1980 and subsequent versions of 2nd Canadarm manipulators, and continues to be an important original equipment manufacturer (OEM) supplier of both space and terrestrial robots for high-performance applications, including surgery. Clearpath Robotics has grown rapidly over the last decade to become a key global supplier of indoor and outdoor mobile robots for research and warehouse operations. Robotiq is a Quebec-based provider of tools and software for the collaborative robotics market, helping manufacturers start production faster. Also in Quebec is Kinova, a manufacturer of lightweight robot arms for testing, maintenance and inspection across a wide range of application industries. ESI is a leading provider of tactical robots for law enforcement, robot arms for industrial automation, medical robots for robot-assisted medical surgery, and personal robots for a range of applications. In terms of aerial robotics, FLIR (formerly Aeryon Labs) builds surveillance drones, and Drone Delivery Canada has quickly become a leader in aerial delivery, particularly to remote communities. Avidbots is a growing company providing autonomous floor-cleaning robots, and is currently developing a disinfection sprayer add-on. There are also many Canadian companies building robotics-related sensors including cameras (Lucid and FLIR, formerly Point Grey Research), laser rangefinders (Optech, LeddarTech, Neptec, Quanergy), and GPS positioning (Applanix, Novatel).

There are multiple Canadian government and industry groups with robotics expertise. Defence R&D Canada has been working in the robotics space for several decades at their various Centres, with expertise in teleoperation and autonomous operations of ground robots and drones. The Canadian Space Agency in Longueuil, Quebec has a strong in-house team of roboticists to help with operations of the Canadarms and to research future space exploration concepts including teleperation and autonomous operations of rovers. Hydro Quebec, Quebec's public hydro utility, has a research institute (IREQ) that has been developing robotics for hydro asset management for over twenty years[70]. The National Research Council also has in-house experts in robotics in their various centres. Unmanned Systems Canada is an industry association group focussed on autonomous vehicles, and the Regroupment des Enterprises en Automatisation Industrielle (REAI)[71] is an industry association of Quebec-based companies focused on industrial automation. Further details on the academic/government robotics landscape in Canada is given in the Appendix.

---

66 "University of Toronto Robotics Institute," University of Toronto. https://robotics.utoronto.ca/ (accessed Sep. 01, 2020).
67 "Waterloo Robohub," University of Waterloo. https://uwaterloo.ca/robohub/home (accessed Sep. 01, 2020).
68 "Ingenuity Labs Research Institute," Queen's University. https://www.queensu.ca/research/centres-institutes/ingenuity-labs-research-institute (accessed Sep. 01, 2020).
69 "Centre for Intelligent Machines," McGill University. https://www.cim.mcgill.ca/ (accessed Sep. 01, 2020).

We begin the next section by looking at where robots can be useful in this pandemic, and note that, despite the many robotics experts and assets across the country, at the time of writing, **our research revealed only a handful of COVID-specific deployments of robots in Canada**. Much of this paper is dedicated to analyzing why this is the case.

---

*70 "Advanced Robotics for Optimizing Asset Management," Hydro-Québec Research Institute (IREQ). http://www.hydroquebec.com/robotics/ (accessed Sep. 01, 2020).*
*71 "World-class automation expertise serving the Canadian eastern market," Regroupment des Enterprises en Automatisation Industrielle (REAI). https://reai.ca/en/ (accessed Sep. 01, 2020).*

# 03

# Deployments and Use Cases of Robots for COVID-19

This section describes the main areas where robots are already being used to address COVID-19 both globally and in Canada, and describes some novel use cases still in development. This is not intended to be an exhaustive list of all robots being experimented with in the COVID-19 pandemic; for a more comprehensive view of known use cases, please refer to the ongoing work of Robin Murphy and her team, described in Section 1.4. Instead, we focus on representing the breadth of robots being used to manage the pandemic crisis in order to describe general patterns in robot deployments. Our analysis is determined from our own experience deploying hardware systems, from media reports, as well as our own and others' interviews with experts from hospitals, public health, public transportation, manufacturing/retail industries, education, and government labs.

# 3.1
# WHAT ROBOTS CAN DO

This subsection describes the main use cases for robots during the pandemic, and includes examples. Each section begins with a brief summary of the need and how robots are being used to address it. The examples given are not intended to be an exhaustive list, but are instead meant to represent the range of jobs robots are doing in each use case category.

The examples are organized by Technology Readiness Levels (TRLs; see Section 2.2) so that the reader can understand each robot's relative maturity. Robots that are "ready to scale" are TRL9 robots that are tested and commercially available; here we indicate the robot cost where available. Robots that are "ready to pilot" are generally TRL7-8, though we include in this list the occasional TRL9 robot that has been repurposed from another application area but whose efficacy for this particular use case has yet to be studied (for example the Agras MG-1 agricultural spraying drone, which has been repurposed to spray disinfectants instead of pesticides)[72]. Robots that "need development" are generally TRL2-6 robots that require significant development but nonetheless offer a unique or potentially transformative capability.

Our goal with this section is to provide readers with an understanding of the relative maturity and capabilities of the technologies in question, such that readers can easily compare products and use cases.

---
[72] "Spain Is The First European Country To Use Agricultural Drones To Fight COVID-19," Quadcopter News, Apr. 04, 2020. https://www.quadcopternews.it/emergency/spain-is-the-first-european-country-to-use-agricultural-drones-to-fight-covid-19/ (accessed Sep. 01, 2020).

## 3.1.1 Disinfection & Cleaning

**NEED**

The need for disinfection and cleaning in hospitals and other patient care facilities during the pandemic is obvious. However, as economies reopen, public gathering hotspots such as airports, event venues, and public transportation vehicles, and transit stations, schools, malls and workplaces also play a vital role in preventing virus spread, as frequently-touched surfaces such as doorknobs, handrails, elevator buttons, faucet handles, seats, and tables, etc., are most likely to be contaminated. Surface and air contamination must be similarly managed in large workspaces such warehouses, retail stores, and factories — all without loss of productivity. Most countries have protocols and procedures in place for cleaning open access spaces and industrial facilities. Under pandemic conditions, however, cleaning capacity quickly reaches its limit and frequent cleaning cannot be ensured.

**HOW ROBOTS CAN HELP**

- Cleaning and disinfecting floors, large surfaces, personal protective equipment, and cluttered environments, and purifying air
- Identifying and targeting frequently-touched surfaces
- Reducing exposure risk and saving time by providing a coarse "first pass" of disinfection, even when manual cleaning is still needed
- Augmenting cleaning and capacity and frequency by supplementing cleaning staff
- Addressing labour shortages when cleaning staff fall ill or cannot come into work
- Assisting managers with oversight and planning by tracking, scheduling and reporting on cleaning routes and completion schedules
- Providing quality assurance for disinfection procedures by analysing (e.g. air) samples and alerting cleaning staff / managers when virus is present
- Reassuring customers, employees, and the public that hygiene measures are being taken

**EXAMPLES**

The robots described in this section use a range of disinfection modes, from UV-C light, to hydrogen peroxide fog, to electrostatic spray, and are designed to operate in a variety of locations, from hospitals to public spaces and even on airplanes. Some robots use multiple disinfection modes at once, and/or are designed to identify and target high-touch areas. UV-C is known to be effective for viral and bacterial disinfection in specific contexts, but is harmful to humans and may be ineffective if there is no direct line of sight. Disinfection may be less effective on certain surfaces (e.g. fabrics), and so correct dosage and technique are essential, and vary from context to context. For these reasons, further validation and testing of robotic disinfection devices is needed in a variety of pilot settings before they are widely deployed.

## TABLE 1: DISINFECTION & CLEANING ROBOTS

| | | | | |
|---|---|---|---|---|
| **READY TO SCALE (TRL9+)** | **UV-C room cleaning** 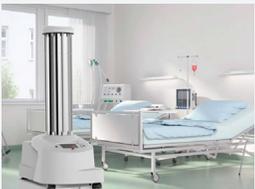 E.g. UVD Robot, Blue Ocean Robotics  Unit cost: ~$80K USD[73]  *Image: Blue Ocean Robotics* | **Compact UV-C cleaning for aircraft** 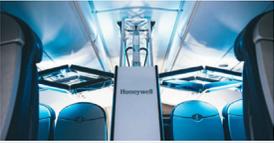 E.g. UV Cabin System, Honeywell  Unit cost: $10/use[74]  *Image: Honeywell* | **Floor cleaning - retail, office, hospital** 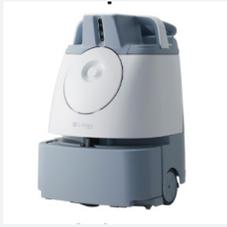 E.g. Whiz, Softbank Robotics  Unit cost: ~$500/mo USD[75]  *Image: Softbank Robotics* | **Floor cleaning - warehouse, factories**  For operating in dynamic environments with many moving obstacles 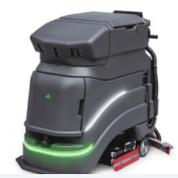 E.g. Neo, Avidbots  Unit cost: ~$50K[76] USD  *Image: Avidbots* |
| **READY TO PILOT (TRL7-8)** | **UV-C PPE disinfection**  Disinfect PPE for reuse [77] 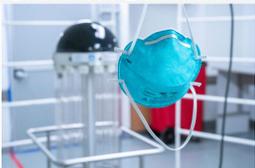 E.g. Tru-D Robot, Tru-D SmartUVC  *Image: University of Virginia* | **Targetted, multimodal disinfection**  UV/Dry-Fog/Air purification systems that can identify & target high-touch areas 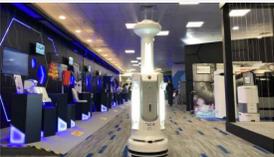 E.g. ISR, TMiROB  *Image: Time Medical Systems* | **Ultra low cost UV-C** 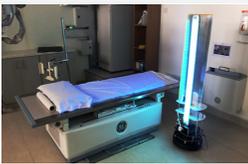 E.g. Violet UV Robot, Akara Robotics  *Image: Akara Robotics* | **Aerial disinfectant spraying** 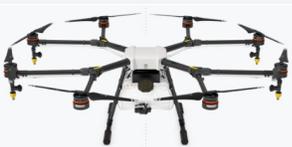 E.g. Agras MG-1 , DJI  Unit cost: ~$6K[78]  *Image: DJI (repurposed agricultural drone)* |
| **NEEDS DEVELOPING (TRL2-6)** | **Electrostatic spraying robots**  Distributes disinfectant over large areas and hidden surfaces 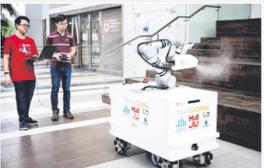 E.g. XDBot, Transforma Robotics & Nanyang Tech University  *Image: Transforma Robotics & Nanyang Tech University* | **All terrain disinfection robots**  Long-duration, high-speed, all-terrain UV + chemical disinfection 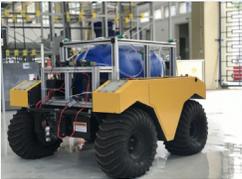 E.g. CIMATIC project, SENAI CIMATEC  *Image: SENAI CIMATEC* | **UV-C Drones**  For disinfecting large indoor spaces 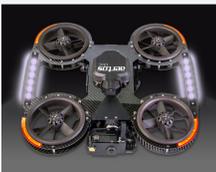 E.g. Aertos 120-UVC, Digital Aerolus  *Image: Digital Aerolus* | **UV-C Mobile Manipulator**  Open drawers/ manipulate objects 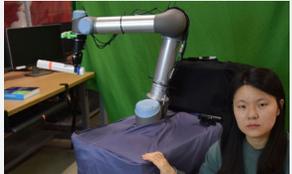 E.g. ADAMMS-UV, University of Southern California  *Image: University of Southern California* |

*Note: The examples listed on this page are meant to demonstrate the breadth of use cases and relative maturity of robotics for this application area, and are not meant to be exhaustive. Inclusion in this paper does not constitute endorsement on the part of the Authors.*

---

## 3.1.2 Medical Triage & Assistance

**NEED**

Under pandemic conditions, the number of people needing medical attention can rise rapidly. Patients in intensive care and elderly people locked down in nursing homes require regular monitoring, however every contact with a potential or confirmed infected patient requires a change of personal protective equipment (PPE), which takes time and creates significant waste. Additional PPE and safety protocols are necessary but can be exhausting for healthcare workers and may leave them with little time for their regular duties. The use of face coverings such as masks and shields can obstruct the view of healthcare workers' faces and make it difficult to communicate with patients. Healthcare workers are at added risk and may themselves become sick, be quarantined, and put their own families at risk.

**HOW ROBOTS CAN HELP**

- Preventing healthcare workers from getting sick by minimizing unnecessary physical contact with infected patients
- Enabling medical professionals to interview patients remotely or work from home while in quarantine
- Enabling remote symptom-checking, prescription, and referral in under-staffed hospitals and long-term care homes (tele-medicine)
- Enabling "face to face" communication without the need for masks
- Autonomously performing touchless temperature checks and conducting simple patient interviews
- Monitoring vital signs at the patients' bedside (e.g., temperature, blood oxygen level, respiration, and movement)
- Reducing the need for multiple changes of PPE, and therefore saving time and preventing waste
- Delivering medical supplies, food, medicine, linens, and lab samples
- Mobilizing coma-induced patients for improved recovery times

**EXAMPLES**

Telepresence — the ability to facilitate two-way communication between medical staff and patients — is a key feature of many medical deployments as it helps to keep frontline workers safe. While the robots described in this section are meant to perform a variety of tasks, those that perform simpler and more predictable tasks (e.g. Promobot Medical Assessor) tend to have higher TRL levels, and the multipurpose robots (e.g. the TRINA Nursing Assistant depicted in the table below) tend to be earlier in development. While a multipurpose nursing assistant like TRINA is unlikely to be ready in time for the COVID pandemic, it is being developed in response to the emergence of highly infectious diseases like Ebola and Zika[79], where robotics could perform essential tasks as part of a pandemic containment strategy.

| | TABLE 2: MEDICAL TRIAGE & ASSISTANCE ROBOTS | | |
|---|---|---|---|
| **READY TO SCALE (TRL9+)** | **Telepresence**<br>To enable medical professionals to interview patients remotely or work from home while in quarantine<br>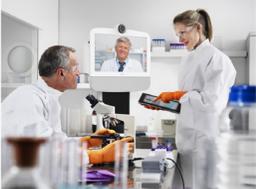<br>E.g. Ava 500, Ava Robotics<br>Unit cost: $1K/mo[80]<br>*Image: Ava Robotics* | **Vital sign assessment & patient intake**<br>To interview incoming patients, measure temperature, lung capacity measurement, blood sugar and oxygen saturation levels, and other health signs.<br>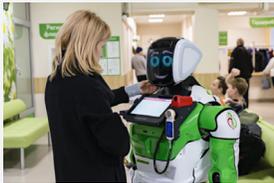<br>E.g. Medical Assessor, Promobot<br>*Image: Promobot* | **In-hospital delivery**<br>For secure delivery of medical supplies, food, medicine, linens, and lab samples<br>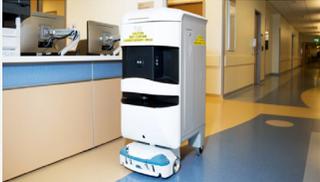<br>E.g. Tug, Aethon<br>Unit cost: $1500-2K/mo USD[81]<br>*Image: Aethon* |
| **READY TO PILOT (TRL7-8)** | **ICU Assistance**<br>To monitor heart rate and other vital signs; to enable doctors and nurses to communicate remotely with patients in isolation using two way communication<br>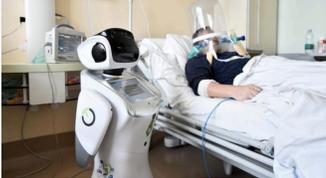<br>E.g. Elf, Sanbot<br>Unit cost: ~$16K USD[82]<br>*Image: Sanbot* | **Outdoor triage**<br>To enable medical staff to triage patients outdoors before they enter the hospital<br>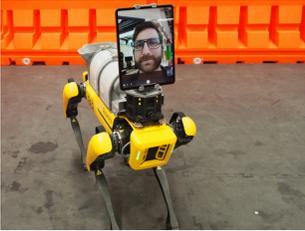<br>E.g. Spot with iPad, Boston Dynamics<br>Unit cost: >$74.5K USD[83]<br>*Image: Boston Dynamics* | **Physiotherapy assistance**<br>To help mobilize coma-induced patients and patients in recover<br>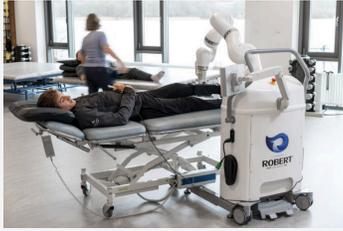<br>E.g. Robert Mobilization Assistant, Life Science Robotics<br>*Image: Life Science Robotics* |
| **NEEDS DEVELOPING (TRL2-6)** | **Multipurpose nursing assistance**<br>Mobile manipulator that can open drawers, doors, perform cleaning tasks, in quarantined areas etc.<br>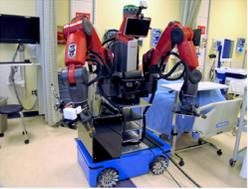<br>E.g. TRINA Nursing Assistant, University of Illinois Urbana-Champaign and Duke University<br>*Image: Intelligent Motion Lab (Duke)* | | |

*Note: The examples listed on this page are meant to demonstrate the breadth of use cases and relative maturity of robotics for this application area, and are not meant to be exhaustive. Inclusion in this paper does not constitute endorsement on the part of the Authors.*

---

79  "Tele-Robotic Intelligent Nursing Assistant (TRINA)," Human-Inspired Robotics Lab, Worcester Polytechnic Institute. http://labs.wpi.edu/hiro/project-portfolio/tele-robotic-intelligent-nursing-assistant-trina/ (accessed Sep. 02, 2020).
80  J. Vincent, "iRobot spinoff Ava Robotics launches autonomous telepresence bot," The Verge, Mar. 13, 2018. https://www.theverge.com/circuitbreaker/2018/3/13/17113494/telepresence-robot-ava-robotics-irobot (accessed Sep. 02, 2020).
81  "Blue-Collar Delivery Robots Making Gains in U.S. Hospitals," Robotics Business Review, Nov. 12, 2013. https://www.roboticsbusinessreview.com/health-medical/blue_collar_delivery_robots_making_gains_in_u-s-_hospitals (accessed Sep. 02, 2020).
82  "Sanbot Elf," Tribotix. https://tribotix.com/product/sanbot-elf/ (accessed Sep. 02, 2020).
83  E. Ackerman, "Boston Dynamics' Spot Robot Dog Now Available for $74,500," IEEE Spectrum: Technology, Engineering, and Science News, Jun. 16, 2020. https://spectrum.ieee.org/automaton/robotics/industrial-robots/boston-dynamics-spot-robot-dog-now-available (accessed Sep. 02, 2020).

## 3.1.3 Sampling & Diagnostics

**NEED**

In order to efficiently trace new infection cases and implement necessary measures against the spread of COVID-19, massive clinical testing needs to be performed on a daily basis. Manual sampling poses a transmission risk to frontline workers as they come in close contact with potentially infected people while swabbing nasal cavities or throats. Healthcare workers must also be trained in the proper sampling techniques because inconsistent manual sampling leads to viral load variations of the swab samples and thus low testing sensitivity and specificity. Incorrect swabbing procedures can also be painful or uncomfortable for patients. An added concern has been the shortage of swabs for collecting sample specimens.

Due to the multi-step nature and sophisticated reagent mixing procedures involved in COVID testing, streamlined and fully automated COVID-19 diagnostics have not been widely adopted in major hospitals[84].

**HOW ROBOTS CAN HELP**

- Enabling consistent, high throughput diagnostic testing
- Improving the consistency and accuracy of sample collection
- Enabling unskilled medical staff to perform swabbing procedures
- Reducing the risk of infection to healthcare workers
- Improving comfort to patients when swabbing
- Enabling population-level testing
- Enabling on-demand printing of swabs

**EXAMPLES**

The mature robots in this category are for lab automation, whereas swabbing robots still require significant development. Steerable robotic endoscopes and surgical tools could be adapted to nasal or throat swab sampling with improved operation consistency. However, less resource intensive and less invasive methods such as saliva sampling and sample pooling, if proven effective, may reduce the need to automate nasal and throat swabbing in the long run. Existing easy-to-use and cost-efficient sewage water sampling robots[85] show promise for identifying hotspots of COVID-19 infection in the population without requiring individual testing[86]. In the post-pandemic era, demands for serological antibody testing may also surge significantly, which could bring new market opportunities for blood-sampling robots. The robots for both swab and blood sampling could be further integrated with mobile diagnostic platforms for streamlined and automated COVID-19 diagnostics with minimal human intervention, enabling more accurate decentralized testing.

---

84   Based on interviews with clinicians from Toronto SickKids and UHN, May 2020.
85   K. Springer, "Meet Luigi: MIT's sewer-scouring robot," CNN Business, Sep. 30, 2016. https://money.cnn.com/2016/09/30/technology/mit-robots-sewers/index.html (accessed Sep. 02, 2020).
86   L. Osman, "Testing wastewater could give early warning of second wave of COVID-19," CTV News, May 22, 2020. https://www.ctvnews.ca/health/coronavirus/testing-wastewater-could-give-early-warning-of-second-wave-of-covid-19-1.4949874 (accessed Sep. 02, 2020).

## TABLE 3: SAMPLING & DIAGNOSTICS ROBOTS

**READY TO SCALE (TRL9+)**

**Liquid-handling**
For high throughput laboratory diagnostics

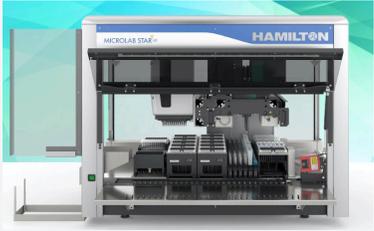

E.g. easyBlood STARlet for COVID-19, Hamilton
Unit cost: ~$100K
*Image: Hamilton*

**Opensource, low-cost liquid handling**
For large scale, high speed laboratory diagnostics

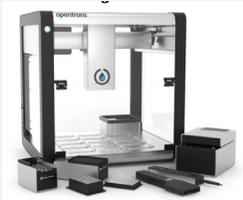

E.g. OpenTrons OT-2, OpenTrons
Unit cost: $5,000 USD
*Image: OpenTrons*

**READY TO PILOT (TRL7-8)**

**3d printing of nasal swabs**[87]

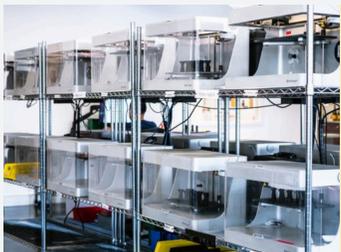

Markforged 3D printer
*Image: Markforged*

**Wastewater sampling**
Installed under manhole covers in target areas (e.g. nursing homes, hospitals, campuses, suspected hotspots) to enable population level testing

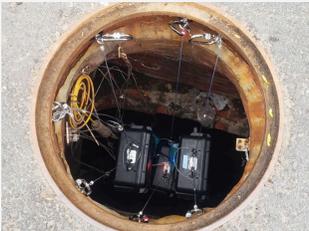

E.g. Biobot, Biobot Analytics (MIT Spinoff)
*Image: Biobot Analytics*

**Blood sampling**
For serological (antibody) testing

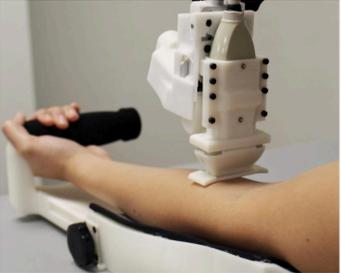

E.g. Blood drawing robot, Rutgers University
*Image: Rutgers University*

**NEEDS DEVELOPING (TRL2-6)**

**Autonomous throat swabbing**

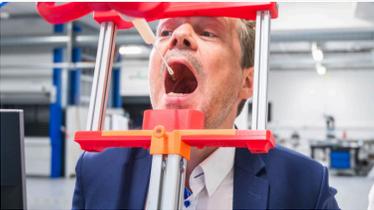

E.g. Automatic Swab Robot, Lifeline Robotics
*Image: Lifeline Robotics (Spin-off of Maersk Mc-Kinney Moller Institute, University of Southern Denmark (SDU)*

**Assisted throat swabbing**
Teleoperated throat swabbing to improve accuracy & comfort

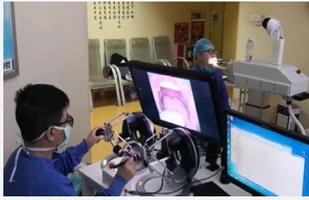

E.g. Throat Swabbing Assistant, Chinese Academy of Sciences
*Image: Shenyang Institute of Automation of the Chinese Academy of Science and First Affiliated Hospital of Guangzhou Medical University*

**Nasal swabbing**

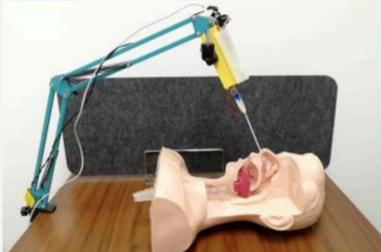

E.g. Low-cost Miniature Robot to Assist COVID-19 Nasopharyngeal Swab Sampling, Chinese Academy of Sciences Beihang University, and King's College London
*Image: Chinese Academy of Sciences Beihang University, and King's College London*

*Note: The examples listed on this page are meant to demonstrate the breadth of use cases and relative maturity of robotics for this application area, and are not meant to be exhaustive. Inclusion in this paper does not constitute endorsement on the part of the Authors.*

---

87  J. Camillo, "Manufacturers Shift to PPE Production to Fight COVID-19 Pandemic," Assembly Magazine, Jun. 17, 2020. https://www.assemblymag.com/articles/95741-manufacturers-shift-to-ppe-production-to-fight-covid-19-pandemic (accessed Sep. 02, 2020).

## 3.1.4 Public Safety & Security

**NEED**

To prevent the spread of this highly transmissible virus, most countries have adopted physical distancing policies and/or measures to restrict movement. Common restrictions include quarantining of (potentially) infected, physical distancing, limiting access to public spaces and workplaces, hand washing and mask wearing. There is a need to enforce these measures and guarantee a continued awareness of the infection risk in the population over long periods of time. However, enforcement capacity of a municipality or region can quickly reach their limits. Ensuring appropriate physical distancing and continued awareness of the infection risk in the population is a major challenge, as illustrated by recent examples of people crowding in city parks[88]. Moreover, similar enforcement measures are also required for individual healthcare facilities, workplaces, airports, public transportation, and shopping centers. Given the increased demand of enforcement personnel during a pandemic, physical distancing may not be consistently enforced.

**HOW ROBOTS CAN HELP**

- Alerting the public of safety policies
- Monitoring crowds in public spaces including airports, public transportation and shopping centers
- Performing temperature checks and identifying potentially infected individuals
- Alerting public health and/or law enforcement personnel to areas where physical distancing violations (including not wearing masks) or infection risks are detected
- Monitoring and/or controlling traffic and access to buildings, rooms, elevators etc, where restrictions are in place
- Enforcing quarantine rules, e.g. through reminders
- Securing vacant or poorly staffed facilities

**EXAMPLES**

Existing mobile robot platforms (driving, flying and walking vehicles) have been repurposed to be "mobile speakers" reminding people of important physical distancing guidelines both outdoors and indoors. Robots have also been used to monitor crowds in public spaces including airports, public transportation and shopping centers, and direct law enforcement personnel to areas where physical distancing violations (including not wearing masks) have been detected. Further, existing robots have been repurposed with relative ease to monitor traffic flow in areas where restrictions are in place (e.g., access to supermarkets or the subway). Privacy concerns must be appropriately addressed for the technology to obtain acceptance within a given context.

---

| | | | | |
|---|---|---|---|---|
| **READY TO SCALE (TRL9+)** | **Public health & mask/PPE detection** For non-contact temperature check, mask/PPE detection, alerts, etc 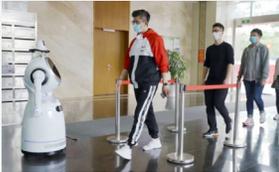 E.g. Cruzr, UBTech Robotics Unit cost: $35K USD[89] *Image: RobotLAB* | **Security patrol** For patrolling and monitoring vacant or poorly-staffed premises 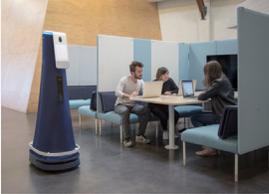 E.g. Cobalt, Cobalt Robotics Unit cost $6k/mo USD[90] *Image: Cobalt Robotics* | | |
| **READY TO PILOT (TRL7-8)** | **Employee health screening** For tracking employees/visitors as they enter & conducting temperature and health questionnaire screening 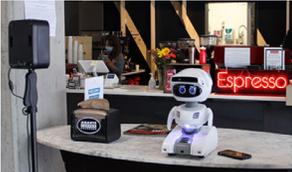 E.g. Misty II Temp Screening Assistant, Misty Robotics Unit cost: $3,000 + monthly fee *Image: Misty Robotics* | **Aerial crowd monitoring/enforcement** For detecting crowds, analyzing temperature & issuing physical distancing reminders from the air 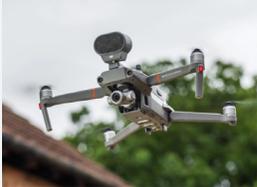 E.g. Mavic 2 with Loudspeaker, DJI Unit cost: ~$2K USD[91] *Image: DJI* | **Parkland patrol & enforcement** For detecting groups and issuing reminders to physically distance in outdoor areas such as parks 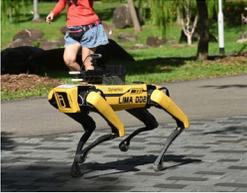 E.g. Spot, Boston Dynamics Unit cost: $75K USD[92] *Image: Boston Dynamics* | **Personal disinfection** Detect mask wearing, check temperature, sanitize personal items etc 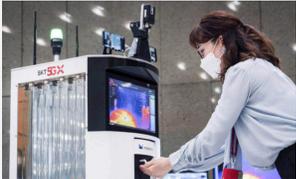 E.g. COVID-19 Quarantine Robot, Omron with SKT *Image: SKT* |
| **NEEDS DEVELOPING (TRL2-6)** | **Cloud-based multirobot monitoring systems** For integrating robots and smart devices, including security, temperature measurement, patient monitoring & more 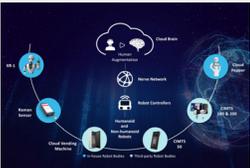 E.g. HARIX Cloud Brain, Cloudminds *Image: Cloudminds* | | | |

*Note: The examples listed on this page are meant to demonstrate the breadth of use cases and relative maturity of robotics for this application area, and are not meant to be exhaustive. Inclusion in this paper does not constitute endorsement on the part of the Authors.*

## 3.1.5 Social Interaction, Companionship & Healthy Living

**NEED**

Seniors have been particularly hard hit by the coronavirus, both because they are at higher risk of severe complications, and because they rely significantly on care staff and family members for assistance and encouragement with numerous activities of daily living. Seniors living at home alone find themselves isolated. Long-term care homes are the hardest hit facilities in this country, resulting in lockdowns and a large number of deaths. Long-term care homes in Canada have continuously had staff shortages and high turnover rates, so they were already hard-pressed before the pandemic hit. Some have required military personnel assistance during the COVID-19 outbreak, which is only a temporary solution.

Children have also been adversely impacted by the pandemic, with many forced to stay home and isolate at a time when social interaction is essential to their cognitive development. Some of the technologies described below may be equally useful to other vulnerable populations, such as those impacted by mental illness or who face mobility issues.

**HOW ROBOTS CAN HELP**

- Minimizing physical contact between patients/ nursing home residents, and caregivers and family members
- Coaching seniors through daily activities such as eating, drinking, dressing, getting out of bed, bathing, exercising, communication with others, companionship, and numerous recreational activities
- Providing a promising alternative to physical interaction when PPE is not available or in low supply, and transmission rates are high
- Connecting residents in locked down nursing homes and retirement homes with their family members and friends
- Supporting children's social and cognitive learning when physical interaction isn't possible

**EXAMPLES**

Robots that must navigate human-centred environments that contain frail and cognitively impaired individuals provide an interesting challenge to roboticists. As an emerging technology, the research and development of these robots has mainly been in research labs around the world, including at the University of Toronto[93]. The UofT team is developing intelligent and autonomous socially assistive robots to assist older adults including those living with dementia with activities of daily living, and cognitive and social interventions.

More recently a handful of international companies have developed relatively simple robotic platforms that can be used in applications ranging from senior care to supporting children with autism and other cognitive development concerns.

---

## TABLE 5: ROBOTS FOR SOCIAL INTERACTION, COMPANIONSHIP & HEALTHY LIVING

| | | | |
|---|---|---|---|
| **READY TO SCALE (TRL9+)** | **Greeting patients/visitors**[94]<br>Friendly welcome to hospital, airport or quarantine facility, to ease stress<br>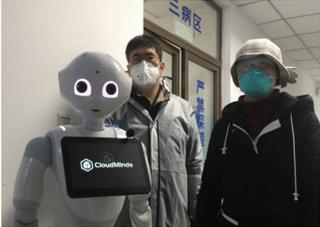<br>E.g. Pepper, Softbank Robotics<br>Unit cost: $1600 plus $360/mo[95]<br>*Image: Cloudminds* | **Telepresence**<br>Enable quarantined patients or senior in lockdown to communicate with family and friends<br>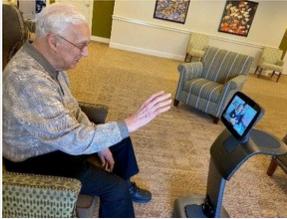<br>E.g. Temi, Robotemi<br>Unit cost: $4K USD<br>*Image: Connected Living + Temi* | **Pet therapy**<br>To help deal with social isolation and depression, by aiming to improve overall moods and quality of life<br>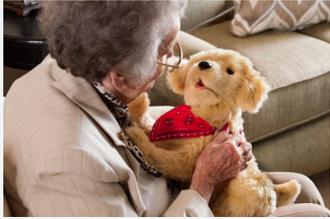<br>Golden Pup, Ageless Innovation<br>Unit cost: $130 USD[96]<br>*Image: Ageless Innovation* |
| **READY TO PILOT (TRL7-8)** | **Verbal coaching to assist independent living**<br>Cognitive stimulation, daily routine reminders, messaging and photo sharing with family & friends[97]<br>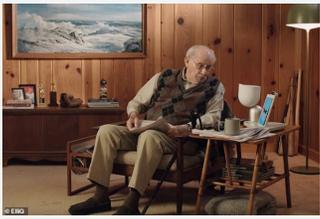<br>E.g. Elli-Q, Intuition Robotics<br>*Image: Intuition Robotics* | **Medication reminders**<br>For medication/treatment compliance, chronic illness management<br>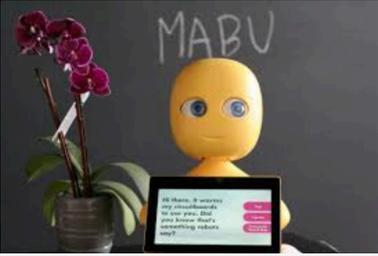<br>E.g. Mabu, Catalia Health<br>*Image: Catalia Health* | **Play-based learning & social interaction**<br>For supporting the cognitive development needs of isolated children<br>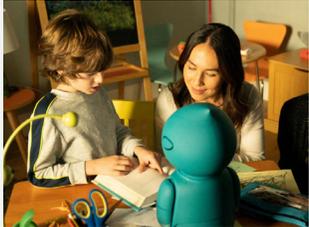<br>E.g. Moxie, Embodied<br>*Image: Embodied* |
| **NEEDS DEVELOPING (TRL2-6)** | **Socially assistive robots for everyday living**<br>Providing multisensory cues (verbal, physical, written) to provide assistance with a variety of daily activities such as eating, drinking, dressing, getting out of bed, bathing, exercising, etc.<br>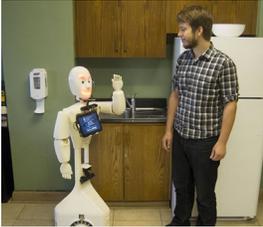<br>E.g. Casper Socially Assistive Robot, Autonomous Systems and Biomechatronics Lab, University of Toronto<br>*Image: Autonomous Systems and Biomechatronics Lab, University of Toronto* | | |

*Note: The examples listed on this page are meant to demonstrate the breadth of use cases and relative maturity of robotics for this application area, and are not meant to be exhaustive. Inclusion in this paper does not constitute endorsement on the part of the Authors.*

---

## 3.1.6 Manufacturing & Logistics

**NEED**

The pandemic has generated unprecedented demand for certain goods, in particular personal protective equipment (PPE), medical and grocery supplies. Demand for PPE is expected to rise as economies emerge from lockdown and PPE use becomes a requirement outside medical settings. Global supply chains have not always been able to deliver and so there is increasing demand for locally produced goods to ensure adequate supply and quality: we note that as of the writing of this paper, the Canadian government has received only a small fraction of the PPE it has ordered[98]. Flexible, modular and scalable fabrication processes that can be easily altered to accommodate change are desperately needed while the dynamics of supply and demand remain volatile. This applies not only to PPE production, but the production of other goods upon which Canada's economy depends. For example, food producers — who faced labour shortages even before the pandemic — are struggling to maintain production when employees on assembly lines fall sick, fewer migrant workers show up to harvest, and physical distancing rules mean fewer workers can be onsite at once. Meanwhile, retail grocery chains now face the logistical challenge of minimizing exposure to their employees and customers, while at the same time rapidly training new staff to replace sick and exhausted workers, and dealing with an increased demand for goods that has left some shelves empty. Grocery retailers are also going through a massive logistics shift as demand for online ordering and home delivery skyrocket.

**HOW ROBOTS CAN HELP**
- Scalable, localized production of critical supplies, including food
- Addressing labour shortages
- Touchless technology to reduce sick days and keep customers safe

**EXAMPLES**

Early on in the pandemic when PPE was in short supply, 3D printers were put to use manufacturing face shields and nasal swabs using open source designs; however this approach may become less cost-effective as assembly line production ramps up. Grocery and retail chains are beginning to make use of full-scale automated logistics systems, though most of these deployments were initiated before the pandemic. Farm automation was also an area for investment before the pandemic. One Danish slaughterhouse that had already heavily invested in automation before the pandemic reports that COVID has not slowed production, and that fewer than 10 of its 8000 employees have been infected[99]. Given the major outbreaks and stoppages at other slaughterhouse factories, some Canadian slaughterhouses are now looking to automate[100].

---

98   "Supplying Canada's response to COVID-19," Public Services and Procurement Canada. https://www.tpsgc-pwgsc.gc.ca/comm/aic-scr/provisions-supplies-eng.html (accessed Jun. 24, 2020).
99   M. Molteni, "Covid-19 Makes the Case for More Meatpacking Robots," WIRED, May 25, 2020. https://www.wired.com/story/covid-19-makes-the-case-for-more-meatpacking-robots/ (accessed Sep. 02, 2020).
100   H. Yourex-West, "Canadian meat-packing industry looks to make big changes following COVID-19," Global News, Jun. 17, 2020. https://globalnews.ca/news/7054288/meat-processing-changes-coronavirus/ (accessed Sep. 02, 2020).

# TABLE 6: MANUFACTURING & LOGISTICS ROBOTS

| | | | | |
|---|---|---|---|---|
| **READY TO SCALE (TRL9+)** | **Automated PPE assembly lines**<br>Large-scale PPE production<br>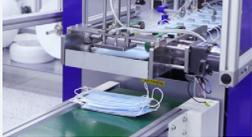<br>Mask Production Line, PIA Automation<br>*Image: PIA Automation* | **Grocery fulfillment & logistics**<br>End-to-end grocery delivery solutions reduces workers' contact with food<br>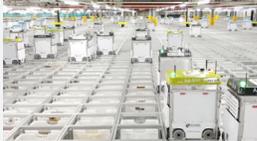<br>E.g. Ocado Smart Platform & Customer Fulfillment Centre, Ocado Group<br>*Image: Ocado Group* | **Retail shelf audit**<br>Anticipate inventory flow, support tired/untrained stockists<br>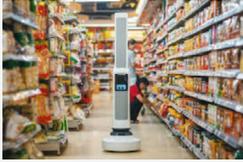<br>Tally, Simbe Robotics<br>Cost: Monthly based on store area/SKUs<br>*Image: Simbe Robotics* | **Automated food processing**<br>For maintaining throughput despite labour shortages; to reduce injury<br>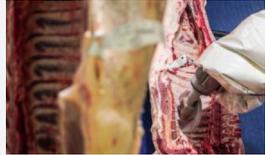<br>E.g. Beef Rib Cutter, Scott<br>*Image: Scott Automation* |
| **READY TO PILOT (TRL7-8)** | **3D printed medical supplies**<br>On-demand nasal swabs, face shields[101]<br>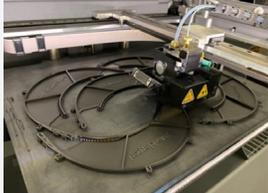<br>E.g. Markforged 3D printer<br>*Image: Markforged* | **Retail food preparation & service**<br>Help guests to physical distance & reduce workers' contact with food<br>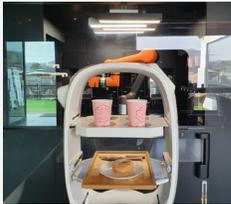<br>Storant Robot Cafe, Vision Semicon<br>*Image: Storant* | | |
| **NEEDS DEVELOPING (TRL2-6)** | **Fruit picking**<br>For harvesting soft and hard fruits despite farm labour shortages[102]<br>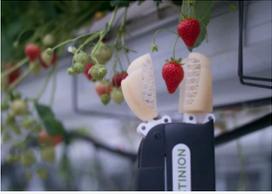<br>E.g. Rubion Strawberry Picker, Octinion<br>*Image: Octinion* | | | |

*Note: The examples listed on this page are meant to demonstrate the breadth of use cases and relative maturity of robotics for this application area, and are not meant to be exhaustive. Inclusion in this paper does not constitute endorsement on the part of the Authors.*

---

## 3.1.7 Delivery & Transportation

**NEED**

Delivery of supplies, both in the hospital setting and locally can help significantly in the reduction of social interaction and exposure of front-line workers to COVID-positive patients. In hospitals, deliveries of food, medicine and other supplies are often performed by different staff members, leading to significant mixing throughout the hospital setting. In urban environments, a large percentage of trips involve the pickup of prepared foods and pre-arranged purchases, and yet shoppers are still frequently entering stores and engaging in close interactions with staff. Last mile, local neighborhood, and indoor delivery can often become vectors for spreading the virus.

**HOW ROBOTS CAN HELP**

- Provide a valuable buffer between frontline workers and patients in hospitals, by reducing the number of daily interactions and the range of people that must come in contact with COVID patients
- Reduce the number of trips to grocery stores and restaurants for order pickup, ameliorating social distancing compliance and avoiding additional indoor interactions
- Reduce the amount of contact between people and food or other supplies
- Help prevent the spread of the virus to remote communities, and transport lab samples and medical supplies

**EXAMPLES**

Hospital-specific applications such as the secure delivery of medical supplies, food, medicine, linens, and lab samples are already deployed in numerous medical centers around the world, and save hospital staff thousands of kilometers of walking per year.

There is also a growing market for concierge robots that operate in the predictable hallways and elevators of hotels and highrise apartments. Such robots have been put to use during COVID as hotels become quarantine centres and high rise dwellers seek to limit their use of common elevators and hallways. Drone delivery may seem at first more challenging because it involves flying; however, in the air there are fewer obstacles for drones to contend with, which explains the relatively high TRL levels for aerial delivery.

Both on-road and sidewalk autonomous delivery solutions are now operational at low speeds, although this technology is still at the proof-of-concept phase as full safety certification remains a challenging open question. Regulatory issues aside, outdoor delivery robots have only become available in the last few years, and continue to mature and expand to new markets.

## TABLE 7: DELIVERY & TRANSPORTATION ROBOTS

| | | | | |
|---|---|---|---|---|
| **READY TO SCALE (TRL9+)** | **In-hospital delivery** <br> For secure delivery of medical supplies, food, medicine, linens, and lab samples <br> 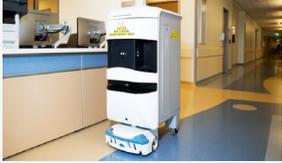 <br> E.g. Tug, Aethon <br> Unit cost: $1500-2K/mo USD[103] <br> *Image: Aethon* | **Concierge delivery in highrises, hotels, & quarantine facilities** <br> For contactless delivery of food, mail & personal supplies <br> 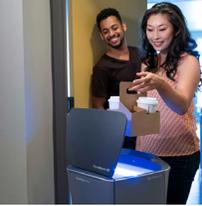 <br> E.g. Relay, Savioke <br> Unit cost: ~$2K/mo[104] <br> *Image: Savioke* | | |
| **READY TO PILOT (TRL7-8)** | **Drone delivery** <br> For delivery of goods, medical supplies, mail, and lab samples to and from remote communities <br> 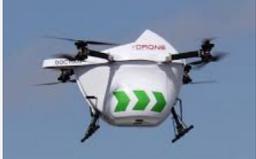 <br> E.g. Sparrow, Drone Delivery Canada. Delivery Cost: NA <br> *Image: Drone Delivery Canada* | **Last-mile grocery delivery** <br> For local goods transportation on low-speed roads <br> 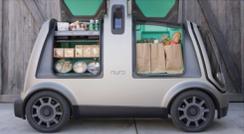 <br> E.g. Nuro Self Driving Vehicle, Nuro <br> *Image: Nuro* | **Takeout food delivery** <br> Short range sidewalk delivery for take-out food and small items <br> 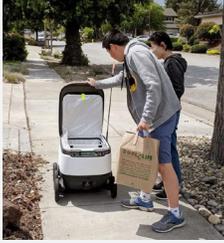 <br> E.g. Self-Driving Delivery Robot, Starship Technologies <br> *Image: Starship Technologies* | **Follow-me retail grocery carts** <br> For touchless shopping <br> 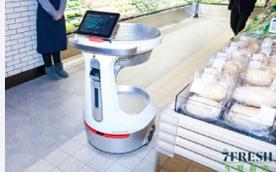 <br> E.g. 7Fresh Robot Grocery Cart, JD[105] <br> *Image: JD* |
| **NEEDS DEVELOPING (TRL2-6)** | **Warehouse to store delivery** <br> To transport goods between warehouse and retail facilities <br> 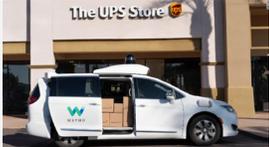 <br> *E.g. Waymo Self Driving Van, Waymo* <br> Current pilot deployments have test drivers on board[106] <br> *Image: Waymo* | | | |

*Note: The examples listed on this page are meant to demonstrate the breadth of use cases and relative maturity of robotics for this application area, and are not meant to be exhaustive. Inclusion in this paper does not constitute endorsement on the part of the Authors.*

---

## 3.1.8 Summary of Robot Deployments & Use Cases

In the course of this research, we reviewed over 60 robot platforms that are being used around the world to help organizations, businesses, and individuals cope with the COVID-19 pandemic. A number of these are described and categorized by their Technology Readiness Level in Section 3.

We found that some of these robots are still early-stage prototypes, many are ready for pilot tests, and still others have a high Technology Readiness Levels and are ready to be more widely deployed.

We describe some key observations below:

1. **COVID-19 is driving the proliferation of new disinfection robots, but not all are made equal, and further testing and validation is needed:**
   A quick web search for "COVID and robot" makes it obvious that disinfection is an important use case for robotics during the pandemic. A review of available disinfection robots reveals that these devices vary greatly in their level of autonomy, however, ranging from the fully autonomous UVD robot, to remotely controlled or teleoperated examples. Some, such as the XENEX Lightstrike Robot[107], are in fact stationary carts and are not "robotic" in the traditional sense. What may be less obvious to the non-expert is that prior to the pandemic there were relatively few disinfection robots available on the commercial market as compared to today. One notable exception is Blue Ocean Robotics' UVD robot. Blue Ocean was one of the early pioneers of UV disinfection technology, has taken their products through independent clinical testing[108], and has spent considerable time developing user-friendly interfaces and efficient workflows. Once the pandemic hit, demand for UV disinfection surged and many untested and lower-cost alternatives appeared in countries around the globe. It's important to note that while UV and other modes of robotic disinfection are promising, specific details regarding their effectiveness (including dosage and technique across a range of surface types) is still unknown[109]. Disinfection robots require further validation and testing before they are broadly deployed, and more pilot studies in a variety of settings are needed.

---

107   "LightStrikeTM Germ-ZappingTM Robots," XENEX Disinfection Systems. https://www.xenex.com/our-solution/lightstrike/ (accessed Sep. 04, 2020).
108   UVD Robots. http://www.uvd-robots.com/home/ (accessed Sep. 02, 2020).
109   For a layperson's explanation of current issues in UV effectiveness, see: T. Irving, "Can UV light help hospitals disinfect masks and gowns? U of T researcher explains," University of Toronto News, Apr. 06, 2020. https://www.utoronto.ca/news/can-uv-light-help-hospitals-disinfect-masks-and-gowns-u-t-researcher-explains

2. **The majority of new COVID-focused robots use pre-existing platforms that have been adapted for the pandemic:**
We observed that many of the deployments saw the same mobile bases being adapted for different COVID-19 use cases by upgrading them with new features or behaviours such as temperature monitoring, mask detection, or physical distancing reminders. This can also help explain the fast development of so many new UV disinfections robots. These tend to be produced by either robotics startups that have significant investor funding, or by established companies.

3. **Sample collection is the use case driving development of new platform types and robot designs:**
This is not surprising since the most recent widespread pandemic — the Spanish Flu — occurred between 1918 and 1920 before the first robots were invented. While most other use cases make use of existing platforms, several novel and unique robot platforms are being explored for collecting nasal, throat, blood, and wastewater samples. Given the long timelines involved in developing medical devices, swabbing robots are unlikely to be of use for this current pandemic. However blood sampling robots have already been trialed, and though they are still relatively early-stage, they may be pressed into service if serological testing becomes an important means of tracking the virus. Robotic wastewater sampling may have broader post-pandemic value if it can prove itself effective for population-level surveillance.

4. **Robot-as-a-Service models may reduce the cost risk for customers trialing robots during the pandemic:**
Many of the robotics solutions that are "ready to scale" are offered to customers for a fixed monthly fee. For example, refer to Softbanks' Whiz (floor cleaning) and Pepper (retail service), Ava Robotics' Ava 500 (telepresence), Aethon's Tug and Savioke's Relay (delivery) described earlier in this Section. In addition to not having to pay a large sum up front and having a predictable monthly expense, Robot-as-a-Service business models free customers from having to maintain and service their own robots, reducing the cost risk of trialing a robot during pandemic times. This can be particularly beneficial for undercapitalized businesses that wish to increase their use of automation. This business model also enables robot developers to improve their products over time, adapt to new customer needs, and push updates to their customers' fleets. Some robot developers have reported significant increased fleet use among existing customers since the start of the pandemic[110]. We discuss fee-for-service models in more detail in Section 4.4 under the heading "Long term outlook".

5. **Teleoperation is beginning to fill the autonomy gap in less structured environments:**
   For mobile robotics operating in less structured environments, mature deployments tend to have relatively low autonomy levels and may rely on teleoperators. In some cases the low autonomy levels may be obscured when Robot-as-a-Service models include teleoperators that take over when robots encounter troublesome edge cases, often without the end-user or customer being aware. Despite this promising trend, validation is still needed. We refer readers to Section 4.1 for a more detailed explanation of teleoperation, its present-day challenges, and its longer-term outlook.

6. **Some demonstrators are being pushed prematurely into service:**
   The pandemic may have accelerated the deployment of some "ready to pilot" demonstrator robots, but it remains to be seen whether these prove effective, useful, and safe enough to stimulate long term demand. Further testing and quality assurance are needed.

7. **Robots for security, manufacturing, and logistics, which were in common use before COVID-19, are seeing a pandemic-related surge:**
   While hospital disinfection robots and robot doctors are getting most of the media attention, robots working behind the scenes in manufacturing, warehouse and logistics are experiencing significant increased demand as the global shipping and logistics industry is faced with processing more orders with fewer staff on hand. Likewise, robots designed for security and patrol are also being put to greater use as organizations are challenged to keep track of the whereabouts and health of their employees for contact tracing purposes, and to monitor poorly staffed facilities while employees self-isolate.

---

110   For example Cobalt Robotics has launched COVID upgrades and reports a 50% increase in hours of coverage since COVID. See "Complete return to work solutions," Cobalt Robotics. https://cobaltrobotics.com/return-to-work/ (accessed Sep. 02, 2020).

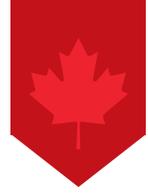

# 3.2
## HOW IS CANADA PUTTING ROBOTS TO USE DURING COVID-19?

Canada lags somewhat behind other industrialized nations in its robot usage and most Canadians will not encounter a robot in their day-to-day activities. Nonetheless, Canada installs a significant number of robots across the country each year, mostly in factories and warehouses. As described briefly in Section 2.4, it is also home to numerous homegrown robotics companies. The following figures provide a general overview of how Canada compares to other countries in its robot usage, as well as an overview of robot manufacturing in Canada.

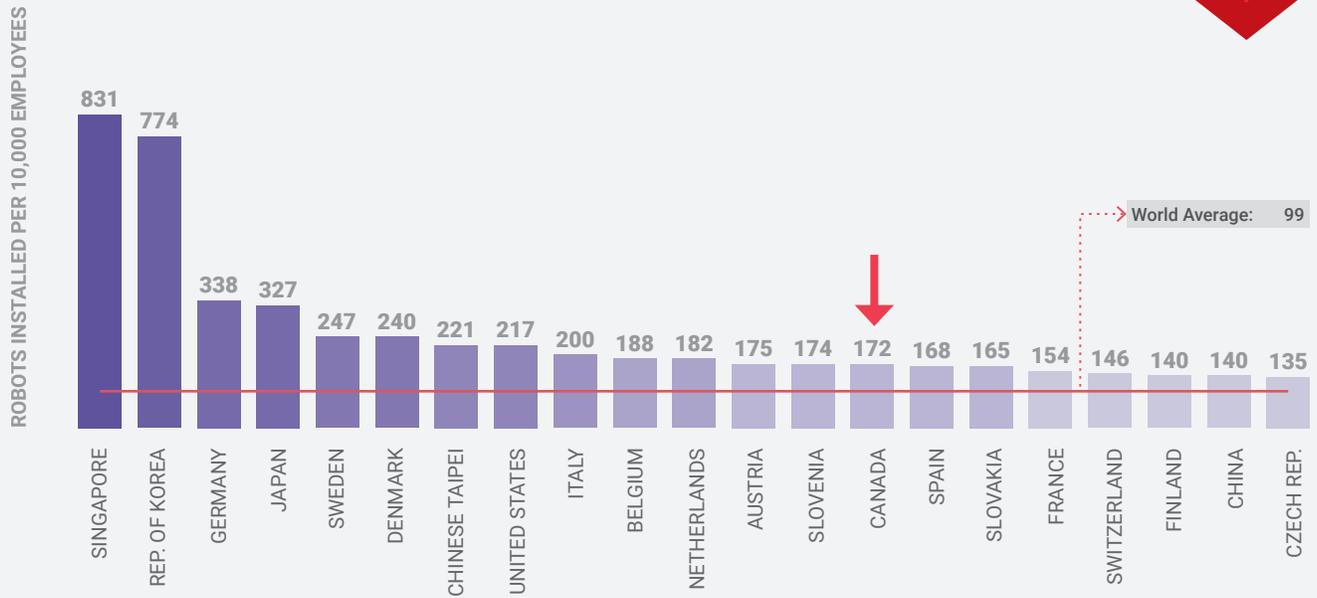

*Figure 5: Robot density in the manufacturing industry by country (2018).
Source: International Federation of Robotics*[111]

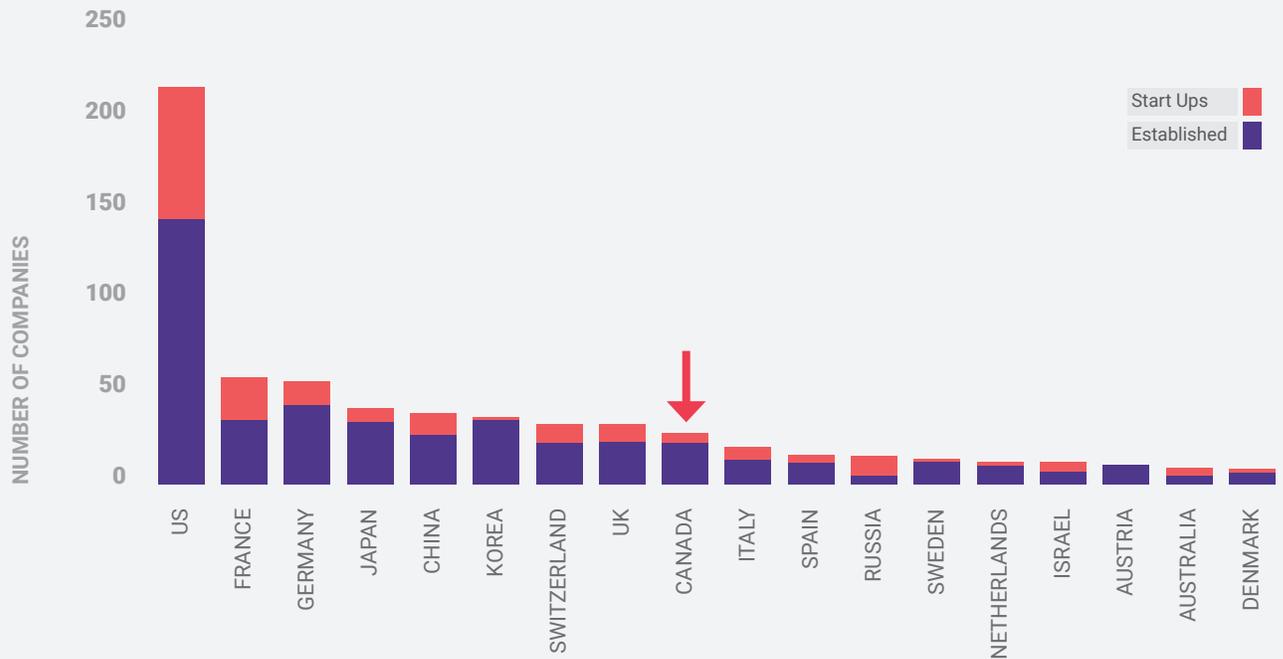

*Figure 6: Number of service robot manufacturers by country of origin (2017).
Source: International Federation of Robotics*[112]

---

111 "World Robotics 2018," *International Federation of Robotics*, 2018.
112 "World Robotics 2017," *International Federation of Robotics*, 2017.

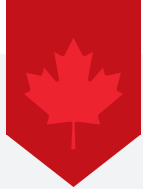
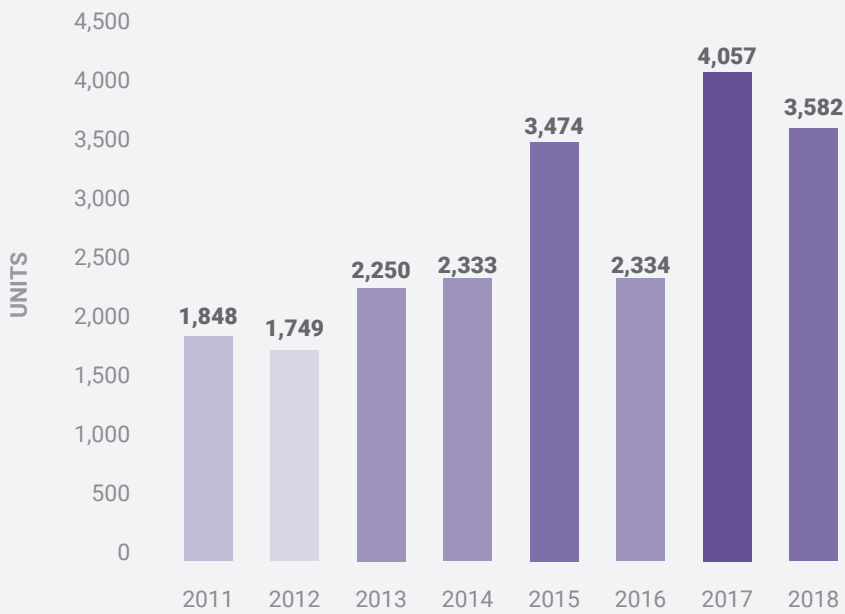

*Figure 7: Annual installations of industrial robots in Canada (2011-2018)*
*Source: International Federation of Robotics[113]*

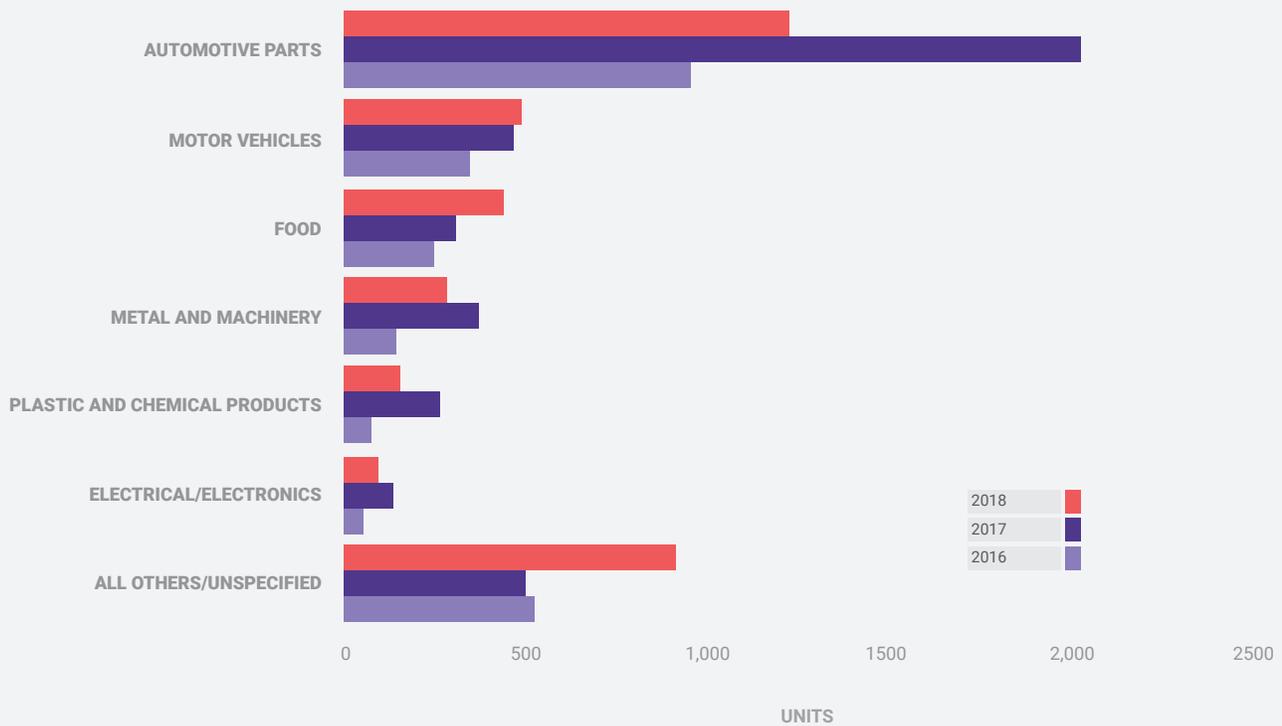

*Figure 8: Annual installations of industrial robots in Canada by industry (2011-2018)*
*Source: International Federation of Robotics[114]*

---

113  *"World Robotics 2019," International Federation of Robotics, 2019.*
114  *"World Robotics 2019," International Federation of Robotics, 2019.*

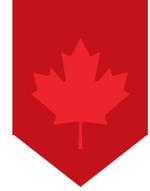

Canada's robot usage may seem small compared to power users like Singapore and South Korea, and its number of homegrown robotics companies may seem small compared to the US. However its "robot density" (the number of robots installed per 10,000 employees) is middle of the pack compared to many developed countries, and its number of homegrown companies puts it in the Top Ten of robot-producing nations. No doubt the numerous automotive, food, and machinery producers across Canada that have already installed robots are making good use of them to manage any labour shortages and increased demand due to the pandemic.

However, it is not our intention to focus on Canada's pre-existing base of robots in this paper. Rather, we are interested in understanding which new use cases Canada is finding for robots, and which traditional areas are seeing accelerated usage or new deployments due to the pandemic. In this section we give a (July 2020) snapshot that shows: 1) how robots (regardless of origin) are being used specifically in the COVID-19 response here in Canada, and 2) how made-in-Canada robots are being put to use during the pandemic globally.

## 3.2.1 Robots Used in Canada for Pandemic Management

Below we describe a variety of robots that Canadian organizations have deployed specifically for frontline pandemic management.

45    Use Cases   | Making Sense of the Robotized Pandemic Response  |  University of Toronto Robotics Institute

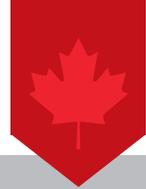

## TABLE 8: ROBOTS USED IN CANADA FOR PANDEMIC MANAGEMENT

| ROBOT | TRL | DESCRIPTION |
|---|---|---|
| **HIGH THROUGHPUT LIQUID HANDLING** | | |
| 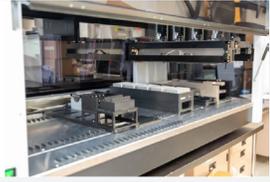 Hamilton robot liquid handling system (US origin) testing coronavirus samples at St. Joe's Hamilton. *Image: The Research Institute of St. Joe's Hamilton* <br><br> 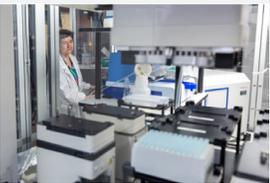 Robot liquid handling system for processing serological tests at Mount Sinai Hospital. *Image: Sinai Health* <br><br> 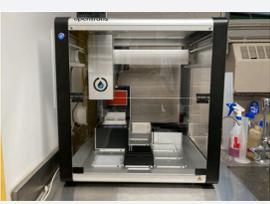 OpenTrons OT-2 liquid handling robot (US origin) used to develop portable test kits at the University of Toronto. *Image: BioMEMS Lab* | 9+ | High throughput testing is critical if the Canadian economy is to reopen before a vaccine is found.[115] Canada's main COVID testing facilities are making use of commercial robotic liquid handling platforms to ramp up testing capacity. <br><br> For example, researchers at **McMaster University and St. Joseph's Healthcare Research Institute**[116] in Hamilton are using a **robotic liquid handling** system developed by the US-based Hamilton Company[117] to speed up COVID-19 tests and enhance biosafety. <br><br> Researchers at **UofT** and **Mount Sinai Hospital** in Toronto are also using a robotic system to process serological tests to measure immunity in recovered COVID patients.[118] <br><br> Decentralized testing is also important for Canada because it can reduce backlogs at central testing facilities while also serving the needs of this country's remote and Indigenous communities. Sixteen diagnostics-related research projects were funded under CIHR's COVID Rapid Research program, many of which are focused on developing portable biosensor kits for viral and/or serological testing.[119] Liquid handling robots are used in the development of such kits to help standardize lab tests and benchmark assays, for example the **OpenTrons** (costing about $5K USD) used at the Microfluidics and **BioMEMS Lab at the University of Toronto.**[120] |
| **FLOORCLEANING - AIRPORT** | | |
| 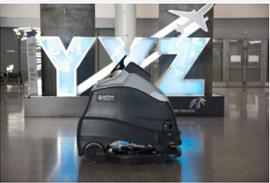 Nilfisk Liberty SC50 robot (Danish origin) at the Toronto International Airport. *Image: Greater Toronto Airports Authority* | 9+ | For the travel industry, cleanliness (both real and perceived) are essential to their continued survival. The **Toronto Pearson International Airport** recently announced that it is installing six robotic floor cleaners as part of a comprehensive COVID safety response.[121] Pearson is using the Liberty SC50 autonomous floor cleaning robot from **Nilfisk**, a supplier of professional cleaning equipment headquartered in Denmark. |
| **UV-C DISINFECTION** | | |
| 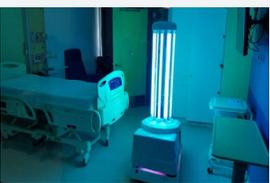 Blue Ocean UVD Robot (Danish origin) being piloted at the McGill University Health Centre. *Image: McGill University Health Centre* | 9+ | The **McGill University Health Centre** was the first in Canada to import a **Blue Ocean Robotics UVD Disinfection Robot** from Denmark.[122] The university is testing the safety and effectiveness of the robot against existing disinfection technologies currently used in Canadian hospitals, including its effectiveness in disinfecting stretchers and N-95 masks. Given the relatively high cost of the device ($120K CDN) the robot will also be tested at the Centre hospitalier de l'Université de Montreal and in long-term care homes before a decision is made on whether to purchase more. |

---

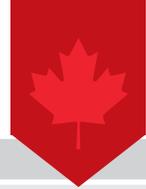

| ROBOT | | TRL | DESCRIPTION |
|---|---|---|---|
| **MANUFACTURE OF BOTTLED DISINFECTANTS** | | | |
| 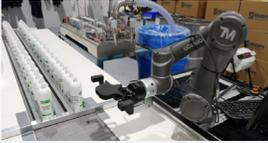 | Omron TM Collaborative Robot (Japanese origin) with a Robotiq Hand-E Adaptive Gripper (Canadian origin) manufacturing disinfectant in Quebec. *Image: Robotiq* | 9+ | A Canadian company dedicated to the contract manufacturing of colour cosmetics is using **Omron TM collaborative robots** combined with the **Robotiq Hand-E Adaptive Gripper** to manufacture bottled disinfectants. The Omron arm and Robotiq gripper are both mature TRL9 products, allowing the company to quickly set up and test the automated manufacturing workflow. The project is funded by the Quebec Government and they are looking at replicating the prototype in the upcoming weeks. |
| **TELEMEDICINE** | | | |
| 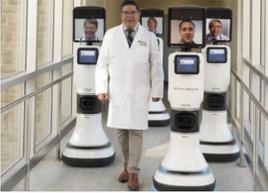 | Fleet of InTouch Health Vita Robots (US origin) at the University of Saskatchewan College of Medicine. *Image: University of Saskatchewan College of Medicine* | 9 | Pediatric critical care specialists at the **University of Saskatchewan College of Medicine**[123] are deploying 15 telepresence robots to various rural regions of the province in order to deliver remote healthcare. The team had already been experimenting with using these robots to deliver remote healthcare for five years prior to the pandemic. The team is using the **Vita robot**, developed by the US-based telemedicine provider **InTouch Health**. Each robot costs ~$80K USD or ~$6K USD / month.[124] |

In addition to these examples, we note the following major Canadian-based robot installations that have seen accelerated uptake as a result of the pandemic:

| ROBOT | | TRL | DESCRIPTION |
|---|---|---|---|
| **GROCERY LOGISTICS** | | | |
| 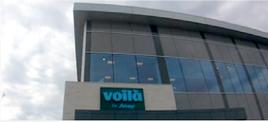 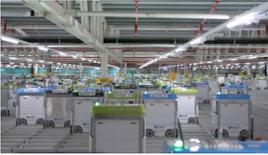 | Deployment at Sobeys Voilà Distribution Centre in Vaughan *Image: Sobeys* <br><br> *Ocado Customer Fulfilment Centre (CFC), Ocado Group (UK) Image: Ocado Group* | 9+ | In June 2020 **Sobeys**, a major Canadian grocery retailer, launched an online grocery delivery platform in **Vaughan**, Ontario to serve the greater Toronto area[125]. The delivery platform, called "Voilà", is powered by an automated warehouse developed by the UK-based **Ocado Group**. Ocado, which specializes in end-to-end grocery technology solutions, reports that its grid system and swarms of proprietary robots are able to pick a 50-item grocery order in minutes[126]. Though plans for this warehouse began in 2019, Sobeys reports exponential growth in their online sales in the first three months of Canada's pandemic lockdown and report that Ocado's ability to quickly deliver from a massive selection of both fresh and packaged goods with minimal handling are important features to customers in the pandemic. A second Ocado facility is currently being built in **Montreal** and is set to open in 2021. |
| **RETAIL LOGISTICS** | | | |
| 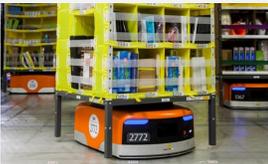 | Deployment at Amazon warehouse facility in Brampton Amazon Robotics (US) *Image: Amazon Robotics* | 9+ | **Amazon** installed its first Canadian robotics fulfillment centre in **Brampton**, Ontario in 2016[127]. Early in the pandemic, Amazon announced it would temporarily stop processing orders for non-essential items in order to prioritize delivery of medical supplies and household staples.[128] Despite these measures, employees in the Brampton facility report they are regularly working more than 50 hours a week in crowded conditions in order to meet the drastic increase in online orders.[129] Amazon claims to have hired an additional 100,000 workers since the start of the pandemic. While it's unclear how many of these have been hired in Canada, the company has increased the hourly pay for Canadian employees by $2, up from the previous median hourly wage of $18.39. Amazon plans to install a second robotic fulfillment centre in **Scarborough** but no date has been announced[130]. |

---

123  A. Hill, "Sask. ready to deploy medical robots to help fight against COVID-19," The Star Phoenix, Saskatoon, Apr. 10, 2020. https://thestarphoenix.com/news/local-news/sask-ready-to-deploy-medical-robots-to-help-fight-against-covid-19/ (accessed Sep. 02, 2020).
124  "Intouch Vita Telepresence Robot," Telepresence Robots. https://telepresencerobots.com/robots/intouch-health-rp-vita (accessed Sep. 02, 2020).
125  R. Redman, "Sobeys begins Voilà online grocery delivery in Toronto," Supermarket News, Jun. 22, 2020. https://www.supermarketnews.com/online-retail/sobeys-begins-voil-online-grocery-delivery-toronto (accessed Sep. 02, 2020).
126  "Fulfilment," Ocado Group. https://www.ocadogroup.com/our-solutions/fulfilment (accessed Sep. 02, 2020).
127  A. Gagnon, "Prime Minister Justin Trudeau opens first robotics fulfillment center in Canada," Day One: The Amazon Blog, Oct. 20, 2016. https://blog.aboutamazon.com/operations/canadian-prime-minister-justin-trudeau-gives-all-hands-at-amazon-fulfillment-center (accessed Sep. 03, 2020).
128  M. Nickelsburg, "Amazon warehouses will stop accepting non-essential items amid COVID-19 outbreak," GeekWire, Mar. 17, 2020. https://www.geekwire.com/2020/amazon-warehouses-will-stop-accepting-non-essential-items-company-triages-amid-covid-19 (accessed Sep. 03, 2020).
129  S. Mojtehedzadeh, "As online orders surge, what about the Amazon workers?," The Star, Mar. 19, 2020. https://www.thestar.com/news/canada/2020/03/19/as-online-orders-surge-what-about-the-amazon-workers.html (accessed Sep. 03, 2020).
130  "Amazon Expands in Ontario with New Fulfillment Centre in Scarborough," Business Wire, Seattle, Sep. 23, 2019. https://www.businesswire.com/news/home/20190923005639/en/Amazon-Expands-Ontario-New-Fulfillment-Centre-Scarborough (accessed Sep. 03, 2020).

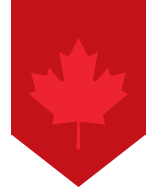

## 3.2.2 Made-in-Canada Robots

Several Canadian robotics developers with ready-to-scale or ready-to-pilot products are being used and tested for various aspects of pandemic management. These include:

| TABLE 9: MADE-IN-CANADA ROBOTS USED FOR PANDEMIC MANAGEMENT | | |
|---|---|---|
| **ROBOT** | **TRL** | **DESCRIPTION** |
| **LOGISTICS** | | |
| 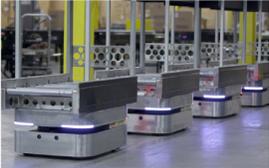 OTTO AMR, OTTO Motors *Image: OTTO Motors* | 9+ | **OTTO Motors**[131], the industrial division of Clearpath Robotics, has seen a surge in demand for its autonomous mobile **material-handling** robot as their customers (which include some of the world's largest manufacturing and warehouse facilities) face increased operational challenges during the pandemic. In June 2020, OTTO Motors announced they had raised $29M USD in Series C funding to meet this increased demand.[132] |
| **FLOOR CLEANING** | | |
| 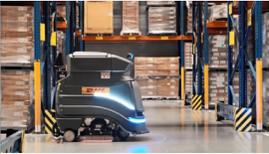 DHL Floorcleaner, Avidbots *Image: Avidbots* | 9+ | **Avidbots**[133], a Kitchener-based company, is partnering up with the Germany-based global shipping provider DHL to deploy their **floor cleaning robots** in DHL-run warehouses in more than 200 countries, including Canada[134]. Demand for the Neo has grown dramatically in the public sector, due to the increased need for cleaning of public spaces, counteracting increased caution on the retail and manufacturing fronts during the pandemic. In Q1 2021, Avidbots is planning to release a disinfection add-on to its robots that will allow them to chemically disinfect high touch points such as doors and slight switches. |
| **MANUFACTURE OF MEDICAL EQUIPMENT** | | |
| 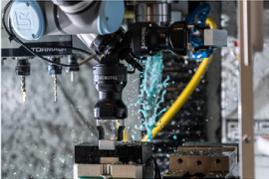 CNC Machine Tending Kit, Robotiq *Image: Robotiq* | 9+ | A **Robotiq CNC Machine Tending Kit** is being used in combination with a **UR 10E arm** by a US manufacturer in Louisiana to **mass-produce oxygen compressors and ventilator equipment**. The manufacturer provides the healthcare industry with a number of vacuum and blower solutions that are used both as components within medical equipment and for the supply of suction from a centralized vacuum system. |
| **DELIVERY TO REMOTE COMMUNITIES** | | |
| 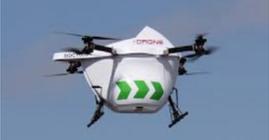 Drone Delivery Canada *Image: Drone Delivery Canada* | 7-8 | **Drone Delivery Canada**[135], a Toronto-based drone delivery company, has announced a partnership with GlobalMedic and Air Canada to **transport COVID-related cargo** such as PPE, test kids and swabs to the Beausoleil First Nation Community in Ontario[136]. The hope is that drone delivery will reduce the likelihood of the virus being transmitted to this community. Drone Delivery Canada has plans to expand internationally and recently began the process for commercial entry into the US market.[137] The Beausoleil pilot program is expected to launch in the third quarter of this year. |

---

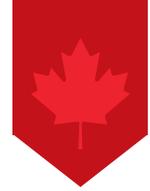

In addition, five disinfection robot projects were recently funded under the NGen Advanced Manufacturing[138] Supercluster COVID-19 response program, with a total of $10M CDN ($5M from industry and $5 from NGen). These products are at various stages of development, and not all are in use at the time of this writing:

| TABLE 10: DISINFECTION ROBOTS FUNDED BY NGEN COVID-19 RESPONSE PROGRAM | | | |
|---|---|---|---|
| **ROBOT** | | **TRL** | **DESCRIPTION** |
| **UV-C DISINFECTION** | | | |
| 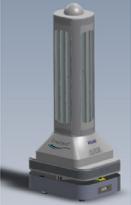 | "Violet" UV-C Robot, Prescientx (with OTTO Motors AMR base) *Image: OTTO Motors* | 8.5 | **Prescientx**[139] — a Cambridge, Ontario manufacturer of UV-C disinfection technology — is partnering with **OTTO Motors** (described in the table above) to develop an autonomous UV disinfection robot. Prescientx already has a number of mature UV-C disinfection solutions, including a portable, tethered UV-C base. The robot will be built on OTTO's TRL9 autonomous mobile robot base and will use the OTTO Fleet Manager[140] to coordinate the movement of the robots and track sanitization, and for this reason gets a high TRL. |
| **MULTIMODAL DISINFECTION** | | | |
| 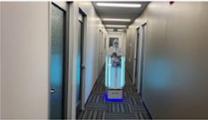 | UV-C Disinfection Service Robot (DSR), Global DWS. *Image: Global DWS* | 8 | **GlobalDWS**[141] — a Toronto-based systems integrator — is developing a multi-modal, voice-enabled UV/spray disinfection robot, which is currently being tested in various Toronto-area eldercare facilities[142] including the **Mon Sheong Home for the Aged**.[143] |
| **UV-C DISINFECTION** | | | |
| 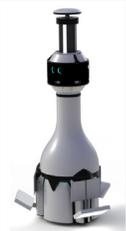 | Bishop Cleanbot UV-C Disinfecting Robot, CrossWing *Image: CrossWing* | 6-7 | **CrossWing**[144] — an Aurora, Ontario robotics OEM — is developing two disinfection robots, one that uses UV-C and a second that uses misting technology. The UV-C robot will be built on their existing Bishop platform[145], and will feature a telescoping payload and articulating UV-C LED panels to enable disinfection coverage in hard to reach places. |
| [no image available] | | 6-7 | **Advanced Intelligence Systems (AIS)**[146] — a Burnaby, BC robotics company — will develop a UV-C disinfection robot capable of mapping large scale environments. This company currently serves the agricultural and greenhouse market. |
| [no image available] | | 6-7 | **A&K Robotics**[147] — a Vancouver-based robotics company — will create a disinfection robot for large floor areas and high-touch surfaces. |

---

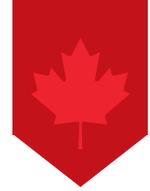

## 3.2.3 Observations in Canadian Deployments & Use Cases

We summarize our observations on Canadian deployments and use cases below:

1.  **Without much prior experience with robots, and lacking proof of their cost-effectiveness, Canadian organizations have been cautious in adopting them during the pandemic.**

    In her book on disaster robotics[148], Robin Murphy argues that most people do not have the capacity to experiment with new technologies when facing a crisis, and therefore the robots which are most likely to be deployed during a disaster are the robots that were already in common use before the disaster. Our examination of robot use in Canada during the COVID-19 pandemic confirms this premise.

    Based on our research, the majority of Canada's COVID-specific robot deployments appear to involve robots that were already in use or planned before the pandemic. The liquid handling robots being used to scale testing in Canadian hospitals were familiar devices in clinical facilities around the country prior to the pandemic. The 15 telepresence robots being used to triage patients in remote communities in Saskatchewan had been tested by doctors in clinical settings for five years. Also, the many mobile robots in Amazon, Ocado and DHL warehouses were either fully installed or planned long prior to the pandemic start.

    New deployments in Canada are being (correctly) treated as pilot projects. Drone Delivery Canada's Beausoleil deployment is a pilot program that is meant to validate previous demonstration trials. McGill University purchased a single UVD Disinfection Robot to study its effectiveness in hospitals and long term care facilities; given the robots' high cost and Canada's publicly-funded health system, these robots are being evaluated not just for their functionality and safety, but also for their cost-effectiveness as compared to manual cleaning.

    It appears that most Canadian organizations are taking a "wait and see" approach to robots. Few are eager to spend funds on unproven technology in uncertain economic conditions.

---

148   *Murphy, R. (2014) Disaster Robotics. MIT Press. ISBN: 9780262027359*

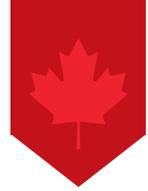

*Early adopters are trialing robots across many applications and use cases during the pandemic. Provided these deployments continue to prove their use, effectiveness, and return on investment (ROI), both in other countries and within the Canadian context, we anticipate more Canadian organizations will begin to adopt robots in the near future.*

2. **The majority of robots being deployed in Canada during COVID-19 are operating behind the scenes and out of the public's view.**

   Unlike the attention-grabbing robots being deployed in parks, shopping malls, on public transit, and in hospital waiting areas in Japan, China, US and the EU, most of Canada's robots are being more quietly deployed behind the scenes: Canada's liquid handling robots are tucked away in clinical testing facilities; for safety reasons, McGill's UVD robot is being tested behind closed doors at hospitals and long term care homes in quarantine; and the Canadian-based Ocado, Amazon and DHL warehouse deployments are not open to the public — only warehouse workers will see these in action.

   And yet this is changing. Drone Delivery Canada's pilot Beausoleil program, though not yet formally launched, will see airtime in public skies starting this fall. Once the program starts, Beausoleil community members will be catching glimpses of the drones on a regular basis. And at various health centres in northern Saskatchewan, community members are being triaged by one of 15 Vita telepresence robots that have been deployed to rural communities throughout the province. Finally, at Pearson Airport in Toronto, travellers are being reassured that hygiene measures are being kept up by the six Nilfisk floor cleaning robots that are roaming the hallways and waiting areas.

   *Public-facing robot deployments in Canada are few in number, and occur in rural areas or in places where, at present, few people are there to see them in action. If robot deployments become an increasingly common sight in Canada, Canadians' attitudes towards them are likely to shift.*

3. **Two established Canadian robotics companies are seeing significant increased exports as a result of the pandemic, and both serve customers in warehouse logistics.**

   While the robots that most readily come to mind as serving pandemic needs are doing frontline jobs like disinfecting, taking temperatures, and monitoring crowds, it's the behind-the-scenes robots working in the warehouse and logistics space that seem to have seen the largest increase in uptake.

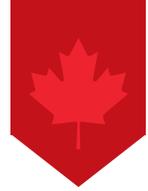

In Canada, since the pandemic took hold, Avidbots started deploying its floor cleaning robots to 200 DHL warehouses worldwide, and OTTO Motors has seen increased demand for its autonomous material handling robot both domestically and abroad, and has attracted significant investment.

*From a home-grown industry perspective, logistics and manufacturing is one of Canada's core robotics strengths. Increased global investment in warehouse automation is likely to benefit established Canadian robotics companies operating in this space.*

4. **OTTO Motors is well positioned to produce and export Canada's first homegrown disinfection robot to its warehouse logistics customers.**

Prior to COVID-19, Canada had no homegrown disinfection robot. Yet demand for these devices has surged since the pandemic. High-quality imported devices are expensive and many other countries have begun to develop their own home-grown solutions. Through NGen's COVID-19 response program, Canada is funding the development of several home-grown UV-C disinfection robots.

*OTTO Motors' new partnership with established Canadian UV disinfection supplier Prescientx, combined with its established customer base, puts the company in a strong position to meet the new disinfection demands of its warehouse and logistics customers, while also providing a local supply of disinfection robots for Canada.*

In conclusion, Canada has some key homegrown robotics companies that are scaling up due to pandemic demand, but these are focused on logistics rather than frontline work. Canada also has comparatively few robots deployed in-country. We will use the rest of the document to discuss why this is the case, and how this situation might be ameliorated.

# 3.3

# POTENTIAL PATHWAYS TO A ROBOTIZED PANDEMIC RESPONSE

Thus far, we have made some sense of the numerous COVID-related robot sightings in the media, and have arrived at the conclusion that the robots we are seeing represent a wide range of technology development stages. Our research also suggests that a variety of strategies are being employed to bring robots to the COVID-19 response, some of which are seemingly more effective than others. In the final part of Section 3, we attempt to model these strategies in order to describe the potential pathways for success.

Though no model can capture the complex reality of bringing a robot to market — especially in a crisis — we nonetheless find it useful to generalize the basic approaches, and the timelines and conditions of success for each. While the Technology Readiness Levels (TRLs) described in Section 2.2 are useful for understanding the technical feasibility of a robot, the TRL schema falls short of considering market factors such as customer desirability and readiness, investment, or supply chains. For this reason, we conclude that TRLs alone cannot be used to measure capacity to scale or give a realistic assessment of potential impact. Instead we propose the Venn diagram in Figure 9, which is based on a standard Desirability/Feasibility/Viability model, and provides a convenient visualization of the four potential pathways for robot deployment during COVID-19, their timelines, and how they relate:

1. **Deploy** proven robotics solutions in real pandemic use cases at a scale that is impactful within a very short time frame (immediate)
2. **Transfer** pre-existing commercially scaled robotics applications to pandemic applications (short term)
3. **Roboticize** by automating existing pandemic solutions that are already commonplace in frontline/essential service use cases with (short term)
4. **Develop** current prototypes into applications that can operate at scale (mid-long term)

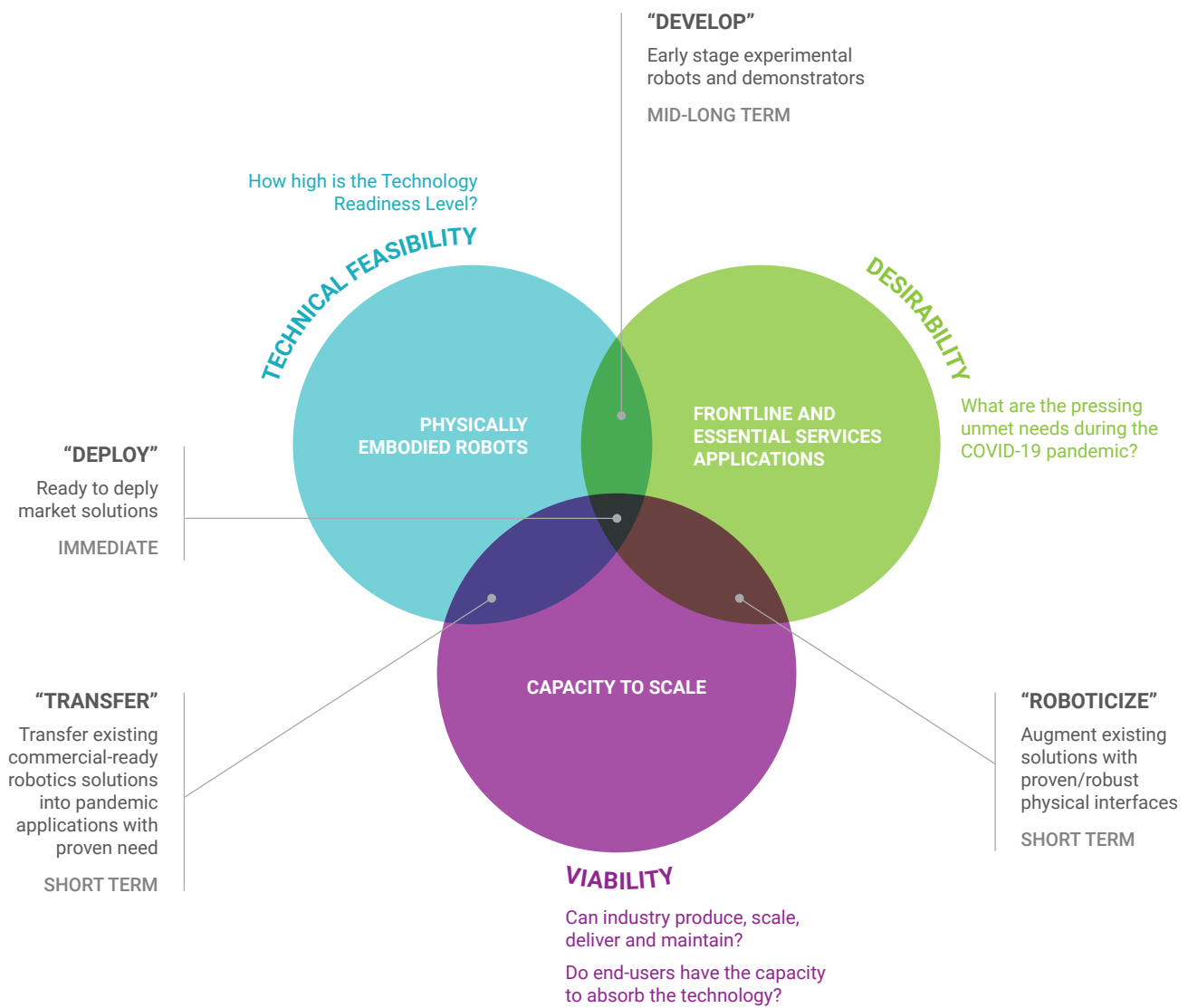

*Figure 9: Venn diagram depicting the four pathways to deployment-ready robots for COVID-related use cases and their relative timelines. The centre of the diagram where the three circles overlap represents robots that are ready to "deploy". There are three "paths" to get robots to the center where they are ready to deploy: "Roboticize", "Transfer" and "Develop".*

The *Develop* path has a mid to long term time horizon, and covers everything from promising prototypes to radically new and pioneering solutions. The *Roboticize* and *Transfer* paths are has a shorter time line, and are very similar to each other, with their chief difference being one of perspective:

- **If you are a robotics developer** and have a commercial-ready platform that you think can be helpful in the pandemic, then you are taking the **transfer** path, and need to find a partner with frontline expertise who has the capacity to co-develop a pandemic solution with you.
- **If you are not a robotics developer** but are looking to automate another product or process that you know already works, then you are taking the **roboticize** path, and need a reliable robotics provider who can help you integrate your solution into their hardware.

We provide some examples to illustrate these concepts below.

# DEPLOY (IMMEDIATE)

The *deploy* path is characterized by existing commercially-available robotics technology that can realistically be deployed to address pandemic needs right away. It assumes that:

- The robot is at the top of the Technology Readiness scale (TRL9)
- The solution is addressing a pressing unmet need
- End-users have the previous experience, or learning capacity and training support needed to adopt the technology
- The robot provider is able to consult, install, deliver, train and/or maintain the system as required
- The system has been validated and tested

**EXAMPLE**

## Putting a tested telepresence fleet into remote triage service

A ready-to-deploy example here in Canada is the small fleet of 15 telepresence robots at the University of Saskatchewan (described in Section 3.2.1). The team is using a commercially available Vita Robot from US-based telemedicine provider InTouch Health in order to triage patients in remote areas of the province. Because the team had already been field testing the robots for several years prior to the pandemic, they were not overwhelmed at the prospect of putting their fleet to use once the pandemic arrived. Vita uses a Robot-as-a-Service cost model; the monthly fee makes it easy to predict and manage costs. Should the need arise, the health team could easily scale this deployment up by ordering additional robots.

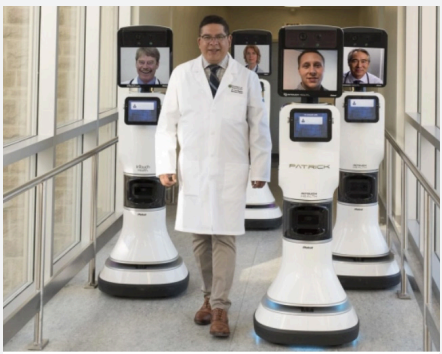

- **High Technology Readiness Level:** 9
- **Unmet need:** Triaging remote patients
- **Capacity to Absorb:** Staff already trained
- **Scalability:** Monthly fee; can order more
- **Efficacy:** 5 years of trials

# TRANSFER / ROBOTICIZE (SHORT TERM)

The transfer and roboticize paths explore opportunities to partner, merge, and adapt existing robotics solutions to pre-existing pandemic technology solutions with well-defined use cases. On the transfer side of the equation, robotics hardware may require some level of re-engineering, interface adaption, software development and testing of newly integrated components. On the roboticize side, operational processes are likely to require testing to validate user needs, evaluate efficacy, and build the business case. The more "commercial-ready" the robotics and underlying technology components are (ie. the higher the TRL levels) and the more well-defined the use case is at the start of the partnership, the shorter the development timeline. A fast development timeline (ie. one that will be ready in time for this pandemic) assumes that:

- The robot is at the top of the Technology Readiness scale (TRL 9) in an adjacent application, even if it requires minor adaptations for this particular use case
- The solution is addressing a pressing unmet need
- End-users have the learning capacity, training and support to adopt the technology
- The robot provider is able to consult, install, deliver, train and/or maintain the system as required

### EXAMPLE
### Augmenting a pre-existing mobile platform with UV disinfection capabilities

An example of a ready-to-transfer/roboticize system that is in development here in Canada is the newly announced partnership between mobile platform OEM OTTO Motors and UV-disinfection provider Prescientx. Funded by the NGen COVID response program, the new partnership aims to rapidly develop an autonomous UV disinfection robot that can serve OTTO Motor's existing global client base of warehouse & logistics customers (see Section 3.2.2). Prescientx already has a number of mature UV-C disinfection solutions, including the portable, tethered UV-C base pictured here. The robot will be built on OTTO's existing mobile robot base and will use the OTTO Fleet Manager to coordinate the movement of the robots and track warehouse sanitation. Testing will be required to ensure efficacy and usability.

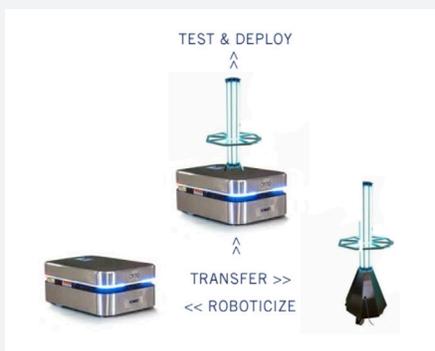

- **High TRL in adjacent application:** 9
- **Unmet need:** Around-the-clock warehouse disinfection
- **Capacity to Absorb:** Warehouse employees are already familiar with the mobile robot platform
- **Scalability:** Large pre-existing customer base that can serve as early adopters
- **Efficacy:** Requires validation & testing

*(mock up for illustration purposes only)*

# DEVELOP (MID-LONG TERM)

The *develop* path is the typical technology maturation process where an existing robot technology at an early TRL could eventually be scaled up to have a real impact in a future pandemic or other disaster scenario. Taking those early stage prototypes and proof-of-concept robots through all TRL involves the entire technology transfer pipeline and substantial investments. As such the Develop Path has a long term time horizon. At the same time radically new and pioneering solutions can be innovated to help us prepare for a world of uncertain futures. The develop path assumes:

- The robot is at the low-mid range of the Technology Readiness scale (TRL2-7)
- The robot offers a potentially transformational innovation with EITHER an order of magnitude efficiency improvement over existing solutions, AND/OR meets a critical need of many people that is currently unmet
- Champions, regulators, and potential early adopters have validated the concept and are involved in the development process

### EXAMPLE
### Validating & testing socially assistive elder care robots for independent living

Long-term care homes in Canada have continuously had staff shortages and high turnover rates, so they were already hard-pressed before the pandemic hit. Some have required military personnel assistance during the COVID-19 outbreak which is only a temporary solution. Long-term care homes are the hardest hit facilities in this country, resulting in a large number of deaths. Assistive robots can complement care staff and address safety and workload issues. The robots need to interact with people and navigate human-centred environments that contain frail and cognitively impaired individuals. As an emerging technology, the research and development of these robots has mainly been in research labs around the world, for example the Casper Socially Assistive Robot designed by the ASBLab at the University of Toronto.

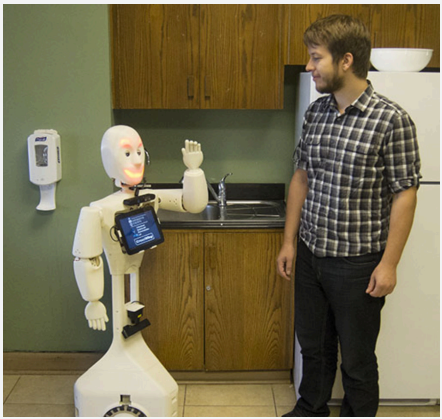

- **Low-Mid Technology Readiness Level**
- **Unmet need:** Age-in place solutions for independent senior living; companionship and support in eldercare home
- **Capacity to Absorb:** Ability to regulate and gain social acceptance has yet to be proven
- **Scalability:** Global aging population
- **Efficacy:** Requires further testing & validation to determine whether these approaches will be effective

# 04

# Deployment Success Factors & How Canada Compares

Why have some countries so successfully deployed robots in response to the pandemic, and why has Canada deployed so few? Robotics is a complex discipline with many success factors and knowledge gaps that are ripe for investigation. We do not attempt to address all these issues here, but instead use this section to identify the key trends that appear to be most relevant to the successful deployment of robots during the pandemic crisis. For each factor discussed, we describe why it matters during the COVID pandemic, we provide examples, and we give insight into how this factor plays out in the Canadian context.

# 4.1 TELEOPERATION

### Why it matters during COVID

COVID-related demand has pushed robots into new operating environments on short notice, leaving developers little time to address the exceptional "edge cases" that make full autonomy a challenge. This does not mean that robots can't be rapidly deployed and immediately useful during the current pandemic however. One solution is to use teleoperation to bridge the autonomy gap, i.e., to use a remote human operator to help guide the robot through unstructured environments or unexpected problems that the robot might encounter. In many settings where robots can make a difference, the technical challenges that remain for teleoperation or semi-autonomous operation can be addressed fairly quickly, making it feasible to deploy such technologies in time for the current pandemic. As autonomy improves slowly over time, teleoperated systems can evolve to become more autonomous and less dependent on human operators, all while remaining largely invisible to end users of the service. This model of semi-autonomous operation requires less development time to bring to market, and allows industry to validate their product concepts and business models before investing in fully autonomous solutions. It also creates new jobs (ie. AV pilots) that can be done remotely (sometimes even from home), helping to protect workers from exposure during the pandemic.

Starship Technologies[149] is a San Francisco-based startup that offers a contactless food delivery service using a fleet of semi-autonomous delivery robots that are small enough to drive on sidewalks. Starship reports seeing significant increased demand for its contactless food delivery service as a result of the pandemic and is now expanding autonomous delivery into five countries[150]. One factor that is enabling Starship to expand so quickly is that their fleet is remotely monitored by a team of operators who are ready to take control in the event that the robot encounters a situation that its autonomy software cannot handle. This approach has also allowed them to validate interest in their service without having to invest in developing full autonomy.

Fetch Robotics[151] is another California-based robotics company that has made significant use of teleoperation during the pandemic. Fetch provides cloud-based autonomous mobile robot solutions to customers in manufacturing and logistics. In a recent interview[152], Fetch CEO Melonee Wise describes how, a few years ago, Fetch took a risk and decided to focus on servicing customers with internet-enabled warehouses. During the pandemic they were able to use the cloud to remotely reconfigure their robots to take on new tasks, rather than having to reprogram them on premises. This provided the flexibility their customers needed to respond to rapidly changing shift and cleaning schedules, while keeping Fetch's own employees safe from exposure.

---

## Long term outlook

- **Growth in cloud-based robotics services:** COVID has created a need for remote work and the cloud services that support it. The cloud services market was growing before COVID and analysts expect the pandemic to accelerate this growth[153]. Such services have given rise to a number of new startups specializing in cloud robotics, online operations, and dashboard systems for fleet management.

- **Demand for teleoperation services:** Demand for services that make use of teleoperation, such as autonomous delivery, was on the rise before the pandemic and COVID has only further accelerated this trend. Several companies that supply or whose business models depend on the use of robotic teleoperation have launched in recent years. Many of these are clustered in the Silicon Valley area, creating a concentration of both supply and demand for robotic teleoperation in this region that is also driving venture capital to invest. For example, Starship Technologies launched in 2015 and has received $82.2M USD in funding to date[154], Phantom Auto launched in 2017 and has received $13.5M in funding[155], and Postmates launched in 2011, has received a total of $903M in funding to date, and was acquired in July 2020 by Uber for $2.65B in stock[156].

- **Viability:** Though many tout teleoperation as the solution to the current reliability gap in certain industries, factors such as interface design and latency mean that teleoperation is likely to be most useful for slow-speed manipulation and driving scenarios in the short term. The ability to teleoperate higher-speed robotic systems will arrive as better interfaces are developed and network speeds improve.

- **Research challenges:** Many fundamental research challenges remain as we move up the autonomy scale from teleoperation to supervised autonomy to full autonomy. One critical research challenge relates to the hand-off of control between the human operator and the robot, an activity that is especially important in teleoperated or semi-supervised driving scenarios where many moving objects add complexity and speed requires fast decision-making. Advances in areas as diverse as sensing, system reliability, failsafe mechanisms, and interface design are needed to address this issue. There is also much interest in having robots learn how to perform a task by watching a human perform the task first, or by having humans teach a manipulation task to a robot by guiding its arm or hand rather than programming. Systems that can learn from humans are an ideal way to bridge the autonomy gap, but advances in machine vision, image recognition, compliance, and deep learning are required to advance the state of the art.

- **Structural challenges:** Given the applied nature of teleoperation, much of the current development in this field is coming from industry, whereas the robotics R&D that takes place in universities is focused on achieving full autonomy because research on full autonomy is more likely to be published and receive funding. As a result, university robotics curricula are set up in ways that largely ignore issues of teleoperation. More research and training programs are needed to support the next generation of robot teleoperators, developers, and designers.

- **Jobs:** While the path to full autonomy comes with the hypothetical risk of job loss, there will be a need to keep humans in the loop for the foreseeable future. As a result, we expect that demand for human teleoperators/pilots, supervisors, developers, user experience (UX) and user interface (UI) designers will continue to grow.

### FOCUS ON CANADA

Despite COVID-related demand for grocery delivery, at present Canada lacks the supply-and-demand activity and regulatory frameworks to support autonomous sidewalk delivery services such as the Starship Technologies example described above.

Yet teleoperation and telepresence are particularly relevant to Canada given its large and sparsely populated country, its long stretches of unsecured borders, and its publicly funded healthcare system. The vast majority of Canada's population and services are located in the southern part of the country near the Canada-US border, and yet northern and remote communities need access to goods and services like healthcare, and remote infrastructure requires monitoring. Teleoperated or semi-autonomous robots can have a measurable impact in remote communities, in applications from telemedicine to delivery of essential supplies to monitoring and surveillance. However Canada's long distances makes the development and maintenance of autonomous robotics solutions in remote communities a challenging proposition for most organizations.

# 4.2 OPEN APIs

### Why it matters during COVID

A major challenge limiting rapid development and large scale deployment of robots is the lack of standardized or "off the shelf" robotics software and hardware components on the market. Even the seemingly simple act of adding a new sensor (for example a new thermal camera for touchless temperature taking) or non-moving hardware component (for example a power-hungry UV light bulb) has significant implications in terms of electrical and software integration. The major engineering companies typically need months of lead time to effectively and reliably integrate a nonstandard sensor into an existing robot system. One way that developers are speeding up robot development during COVID is through the use of an open Application Programming Interface (API), which makes it easy for users to build several products on top of a base robot technology.

One example of how open APIs are being used during the pandemic is France's Kompai Robotics[157] (see Figure 10), which uses a single robot platform to provide multiple solutions to common healthcare use cases. The robot is designed to be quickly and easily reconfigured for new uses, making it adaptable during a pandemic. Boston Dynamics is another example of a company providing an open API, which they call the Open-Source Healthcare Robotics Toolkit[158]. The toolkit offers documentation and CAD files to support development across four use cases: telemedicine, remote vitals inspection, disinfection, and delivery.

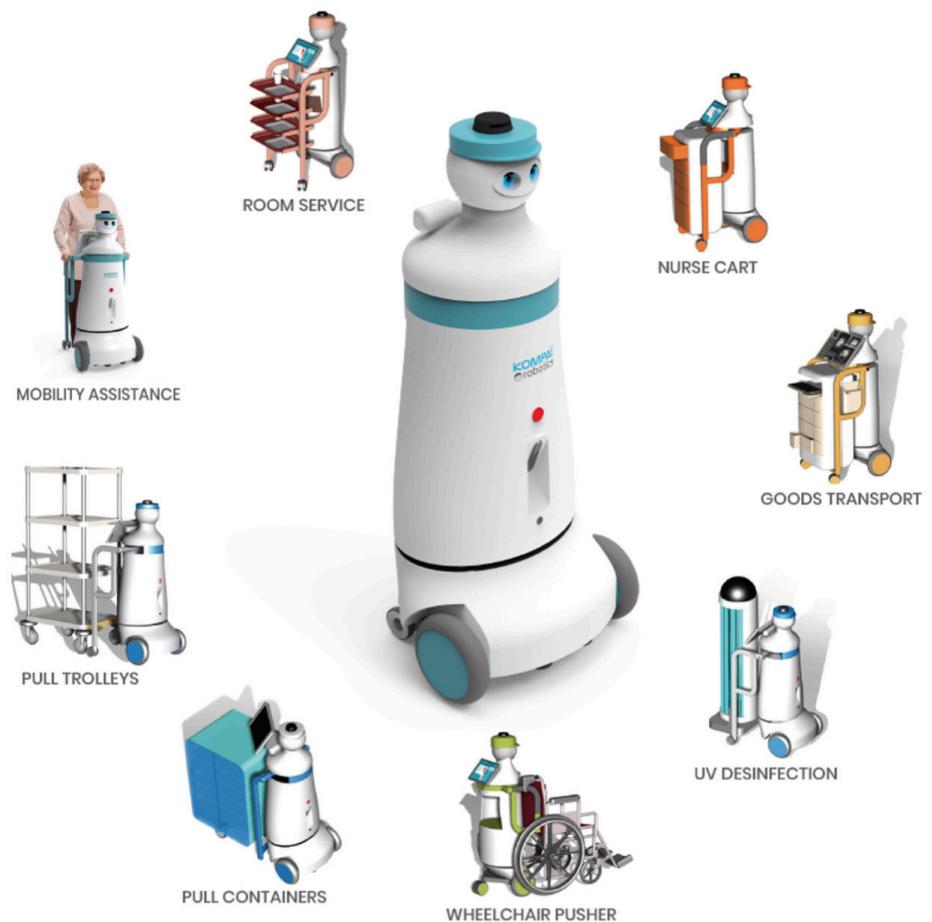

*Figure 10: KOMPAÏ R&D modular robot system. KOMPAÏ Robotics is using an Open API to make it easy for end users to address several different healthcare use cases with their base technology. Image credit: KOMPAÏ Robotics.[159]*

---

157    KOMPAÏ robotics. https://kompairobotics.com/ (accessed Sep. 03, 2020).
158    "Boston Dynamics COVID-19 Response: Using Mobile Robots to Protect Healthcare Workers," Boston Dynamics, Apr. 23, 2020. https://www.bostondynamics.com/COVID-19 (accessed Sep. 03, 2020).
159    "Develop your own healthcare robotic assistant," KOMPAÏ robotics. https://kompairobotics.com/the-robot-kompai/ (accessed Sep. 03, 2020).

## Long term outlook

- **Demand:** Because they facilitate interoperability between different manufacturers, open APIs and other open source approaches are expected by many to be an important driver of development in the robotics industry for the foreseeable future. The Robot Operating System (ROS) is an example of an open source project that has gained significant traction in the community.[160]
- **Structural challenges:** Recognized standards, specifications and protocols for modularity, interoperability, and safety must continue to be developed and formalized.

### FOCUS ON CANADA

In Canada we have (at least) two companies that provide their base robot technologies with an open API that encourages others to build solutions on top. Clearpath Robotics' and Kinova's open API robots were shown previously in Figure 1. Both these companies have seen great success in the international robotics market, in part because their open API allows them to be more easily integrated into their customers' existing hardware, processes, and workflows. To the best of our knowledge, there are few (if any) other countries that are producing both a manipulator and a mobile base with an open API. This presents a major strategic advantage that can be leveraged within Canada (and beyond) to rapidly develop new COVID-related robot products and get these to the top of the TRL ladder.

---

160   https://www.ros.org/ (accessed Sep. 04, 2020)

# 4.3 ROBOT-FRIENDLY ENVIRONMENTS

Building a fully autonomous robot that can function in an environment designed for humans can be time consuming and costly, especially if the robot must be able to open doors or navigate across more than one level of a building. During COVID, retrofitting the environment (e.g. with sensors, markers, and other types of aids) can be a relatively quick and cost effective way to give a simple robot more autonomy. For example, wifi-equipped elevators could enable robots to call the elevator and set the target floor without using the same button interface as humans.

## Long term outlook

- **Accessibility laws:** Wheelchair-friendly environments — which feature ramps, wider spaces to allow movement, and automatic door sensors — are also conducive to robot mobility and are increasingly required under current accessibility laws.

- **Human-centered environments:** Human environments must above all else serve the needs of humans, and should not be compromised by the presence of robots. Smart environments may come at the expense of other liveability factors if privacy and safety concerns are not taken seriously.

**FOCUS ON CANADA**

While cities like Seoul[161] and San Diego[162] are investing in smart city infrastructure, Canada's cities have been more cautious. Toronto was slated to be home to Quayside, Google Sidewalk Lab's first experimental smart city equipped with features like intelligent traffic lights, smart garbage collection, robotic street furniture, and more. However the project was cancelled by Sidewalk Labs this year, citing concerns over economic uncertainty due to COVID-19. Over the two-and-a-half years between when the project was first announced and when it was cancelled, Quayside faced significant controversy over privacy and data protection issues, including questions over data ownership, a legal action over citizen rights, and the resignation of Sidewalk's Privacy Advisor.[163]

While Quayside's cancellation generated a mix of disappointment and relief, it also made clear the significant work yet to be done in Canada (and elsewhere) to clarify expectations and set standards for the governance of publicly-generated data. Canada has a lot of learning to do before technology-enhanced environments become acceptable.

Just weeks after Sidewalk announced the cancellation, the City of Kelowna in BC announced it was claiming the title of "Canada's first real-world 5G smart city". A partnership between the city, Rogers and UBC, the Kelowna project will take on the significantly more modest goal of using the city's new 5G network and wireless sensors and software to collect anonymous traffic pattern data in order to improve road safety.[164]

# 4.4 IN-HOUSE EXPERTISE AND BUY-IN FROM SENIOR LEADERSHIP

### Why it matters during COVID

As Murphy et al. have argued, most organizations aren't willing or able to experiment with complicated new technologies in a crisis. Therefore organizations that had a headstart in robotics before the crisis have been better able to adapt and scale existing solutions. Typically, organizations with significant in-house expertise have a dedicated Chief Technology Officer whose role is to provide strategic direction regarding technology policy and implementation. Small scale pilot projects are typically used to test and validate new solutions before they are scaled.

One example of the importance of expertise lies in how the first COVID patient in the US came to be treated with the assistance of a robot. In January 2020, healthcare staff at Providence St. Joseph Health in Washington used a Vici Robot by Intouch Health to remotely communicate and measure the patient's vital signs, saving on PPE and reducing risk of exposure in the early days of the pandemic when little was known about the virus. Providence began experimenting with telemedicine in 2012, and expects that 10-20% of all its visits will be conducted using telemedicine technologies within 3-5 years. Providence's telemedicine program is overseen by Chief Medical Technology Officer, Dr. Todd Czartosk, a neurologist who first began experimenting with telemedicine as a means to provide timely urgent diagnostic care to stroke patients.[165]

---

165 E. Rosenbaum, "Robotic medicine may be the weapon the world needs to combat the coronavirus," CNBC, Feb. 26, 2020. https://www.cnbc.com/2020/02/26/robotic-medicine-may-be-the-weapon-needed-to-combat-the-coronavirus.html (accessed Sep. 03, 2020).

University of Toronto Robotics Institute | Making Sense of the Robotized Pandemic Response | Success Factors 68

## Long term outlook

- **Fee-for-service models:** Buy-in from leadership is easier to achieve when the cost benefit is clear and the cost of switching from manual processes is low. Robot suppliers have been experimenting with new fee-for-service business models that reduce upfront costs, and free customers from the burden of directly managing ongoing maintenance and repair in exchange for predictable monthly fees. Beyond the business case to customers, managed service models have some upsides for developers as well: the ability to push updates, collect validation data, and call back and replace older robot models makes it easier to test and continue to develop a product while it is live in the field. While managed service models are increasingly common (not only in robotics but in general), robot developers should be cautioned that for fee-for-service to work as a long term business strategy requires a strong economic rationale on the part of the customer, and for robotics this isn't always the case. The robotics market is still young and fluid, and business models for robotics are still evolving and highly dependent on the sector that's being served.

- **Knowledge transfer:** Even if a company does not have its own internal robotics group, it must at least have some qualified, future-thinking, tech-savvy leadership who follow the state of the art and are capable of working with technology or research consultants. Commercially-focused robotics R&D centres such as the 72 Fraunhofer Institutes in Germany, the German Aerospace Centre, or SRI International in the US, offer significant incentive for researchers to collaborate closely with industry and help transfer robotics know-how into new industries and application areas.

- **Workplace acceptance:** Workplace acceptance of robots varies greatly depending on the field of application, the design and function of the robot, and the culture in which the deployment is taking place. Robotic solutions that focus on augmenting the capabilities of human workers and performing dangerous tasks, rather than replacing workers, will no doubt have broader support.

- **Training of end users:** A large number of robotic systems are meant to be deployed in environments that are dominated by users who are not expert roboticists, for instance, in homes, hospitals, factories, and roads. In turn, end-users of robots must be trained to properly and safely operate these machines.

**FOCUS ON CANADA**

Most Canadian organizations lack the experience to understand how advanced technologies such as robotics could improve their processes, as well as the qualified personnel to implement such solutions.

Canada is home to numerous university- and college-level robotics training programs (See Section 2.4), and graduates of these programs are in high demand. However robotics end-user training programs in downstream application areas are less common. In Canada, end user training has focused mainly on operating flying vehicles. To operate drones or Remotely Piloted Aircraft Systems (RPAS) in Canada, one needs a drone pilot licence and vehicle registration for take-off weights of 250 grams (g) up to and including 25 kilograms (kg), via a Special Flights Operations Certificate (SFOC). More specialized end user training exists in surgical robotics, for example at Western University's Canadian Surgical Technologies and Advanced Robotics (CSTAR) facility at the London Health Sciences Centre University Hospital. In Ontario, training for safety drivers for autonomous vehicles is relegated to the company, and the safety protocols have to be reported to the Ministry of Transportation through the Autonomous Vehicle Pilot Program.

More training and knowledge transfer activities are needed in downstream industries with potential for short-term economic impact in Canada during the pandemic. For example, training programs for industrial autonomous mobile robots (AMRs) as well as for new generation manipulators could help our manufacturing and logistics industries keep up with increased operational demands and worker shortages while improving competitiveness over the longer term.

# 4.5 ACCESS TO ROBOT SUPPLY CHAINS

### Why it matters during COVID

The COVID-19 pandemic has brought many economies to a near standstill as workers are sent home to self-isolate. Supply and demand have stopped simultaneously. As a result, procurement is a challenge for everyone, but smaller nations have found themselves lacking in domestic production of medical equipment and other critical supplies, leaving their healthcare systems and economies reliant on imports. As a result many are looking to reinvest in local supply chains and manufacturing, at least for essential goods. The same holds for procuring robots.

Our research shows that the vast majority of ready-to-use robot systems are offered by companies located in a handful of countries. During a pandemic, adapting existing automation solutions to a new challenge is feasible only if the technology is already in the country. Therefore, regions that are home to robot companies with products that can satisfy pandemic needs have been at an advantage when it comes to the variety and scale of their pandemic deployments.

## Long term outlook

- **Reliance on international supply chains for component parts:** While software development for robotics applications is typically completed within the country and not outsourced, scaling up production of robot hardware is critically reliant on manufacturing hubs and well-established international supply chains for electronics and sensors, semiconductors, mechanical components, and batteries.

### FOCUS ON CANADA

UV disinfection and social robots for hospital assistance have turned out to be important front line use cases for the pandemic. However prior to the pandemic, no Canadian robotics OEM was making a UV disinfection robot, and only one was making a social robot (Crosswing). We suspect that this is a major reason for the lack of frontline COVID-related deployments in this country. Nevertheless, with the OEMs Clearpath Robotics and Kinova, Canada has both a homegrown mobile robot platform and manipulator robot. Both companies provide their base robot technologies with an open API to encourage others to build solutions on top. This is a major advantage that can be leveraged within Canada (and beyond) to rapidly develop new COVID-bot prototypes and get these into the maturation process. Indeed, as described in Section 3.2.2, Clearpath is partnering with an established Canadian UV disinfection device maker to develop a mobile disinfection robot built on Clearpath's OTTO Motors' mobile robot platform.

Shenzhen, China, which is currently seen as the electronics capital of the world, is not a critical bottleneck for Canadian manufacturing of intelligent machines. However, it was noted in our interviews that, if faced with an unforeseen need to switch distributors and suppliers or re-engineer subsystems, the majority of Canadian robotics companies would need up to 12 months to adjust.

# 4.6 SUSTAINED ROBOTICS R&D FUNDING

## Why it matters during COVID

While only relatively mature robotic technology is fit for service during a pandemic, mature robotics products would not exist today were it not for investments in early stage research made years or even decades earlier. Robotics R&D funding lays the foundation for an industry ecosystem and supply chain that countries can draw on during a crisis.

According to the International Federation of Robotics Global R&D spending report[166], China launched its Robot Industry Development plan in 2016, and in 2019 alone had allocated $577M USD specifically for robotics R&D; Japan launched its New Robot Strategy in 2016 as well, and in 2019 allocated $351M for this program; South Korea is now on its 3rd five-year Intelligent Robot Basic Plan, having allocated $126M in funding for robotics in 2020 and an additional $150M US towards an Implementation Plan for Intelligent Robotics; the EU launched their Horizon 2020 euRobotics Multi-Annual Roadmap in 2014, with a $780 USD budget for the 6 years of the project.

---

166 "World Robotics R&D Program," International Federation of Robotics (IFR). https://ifr.org/r-and-d#download (accessed Sep. 03, 2020).

These numbers demonstrate both a long term commitment to funding robotics R&D as well as significant current budgets, some of which will no doubt be allocated towards a robotized pandemic response. In our view it is therefore not surprising that these countries would be some of the biggest users of robotics during the pandemic.

## Long term outlook

- **Growth in private investment:** Governments rarely have the appetite to fund research indefinitely without contributions from industry. It's important to note that private investment in robotics has been on the rise ever since the arrival of the smartphone brought down the cost of sensors and electronics. Also, private investment in robotics remains relatively strong despite the economic downturn. The Table below shows a dip in robotics investment in April 2020 at the height of the overall global economic downturn caused by the pandemic lockdown. Unlike many other sectors, however, robotics rebounded to 2019 levels in subsequent months.

**TABLE 11: Q2 GLOBAL VENTURE INVESTMENT IN ROBOTICS (2019 VS 2020)**
*Source: The Robot Report*

|  | 2019 | 2020 |
|---|---|---|
| April | **30** transactions<br>**$6.5** Billion USD | **26** transactions<br>**$0.6** Billion USD |
| May | **27** transactions<br>**$1.5** Billion | **20** transactions<br>**$1.5** Billion USD |
| June | **50** transactions<br>**$1.1** Billion USD | **48** transactions<br>**$1.9** Billion USD |
| July | **49** transactions<br>**$1.1** Billion USD | **47** transactions<br>**$1.9** Billion USD |

- **Overall R&D funding**: A country's overall R&D funding landscape plays a crucial role in its ability to leverage its expertise in robotics. Data from UNESCO shows that many of the top countries in robotics are spending significant portions of their GDP on R&D (see Appendix for details).

## FOCUS ON CANADA

**Robotics funding during COVID**

Many calls for short-term R&D have been created in Canada since the start of the pandemic, and multiple existing programs such as IRAP[167] and NextGEN[168] have been adapted to the current need. However our interviews with Canadian robotics industry members reveal that the business case for pandemic-focused robotics here is weak because demand is expected to subside as soon as a vaccine is released. Many of the most innovative Canadian robotics companies are small in size and are therefore not capable of making large high-risk investments that do not align closely with their existing business. Furthermore, a crisis is the hardest time for small companies to take on new high-risk projects. By focusing funding on TRL7+ technologies, and especially modest adaptations of existing TRL9 robotic platforms with proven in-field operational capabilities, strategic investments could ensure Canadian robotic solutions make it to market in time to support this pressing need.

**Programs for translational robotics research**

Canada has several federal and provincial programs that support the establishment of joint research projects between universities and companies, for instance NSERC Alliance, Mitacs, and others. Many of these programs leverage industrial funding and match it 1:1 or 2:1. We note that while these programs are generally useful for established companies, it is unlikely that startup companies in need of academic expertise would be able to set aside research funding of $50K or $100K/year — an amount that, when matched, would incentivize academics to consider the collaboration. Small startup companies must have lower cash requirements and unfettered access to these programs, especially in the first two to three years.

**History of robotics funding in Canada**

Canada became an early and important investor in robotics technology when it launched the Canada Space Arm (Canadarm) program in 1975. By the time the Canadarm was decommissioned in 2011, the program had invested $1.4B.[169]

---

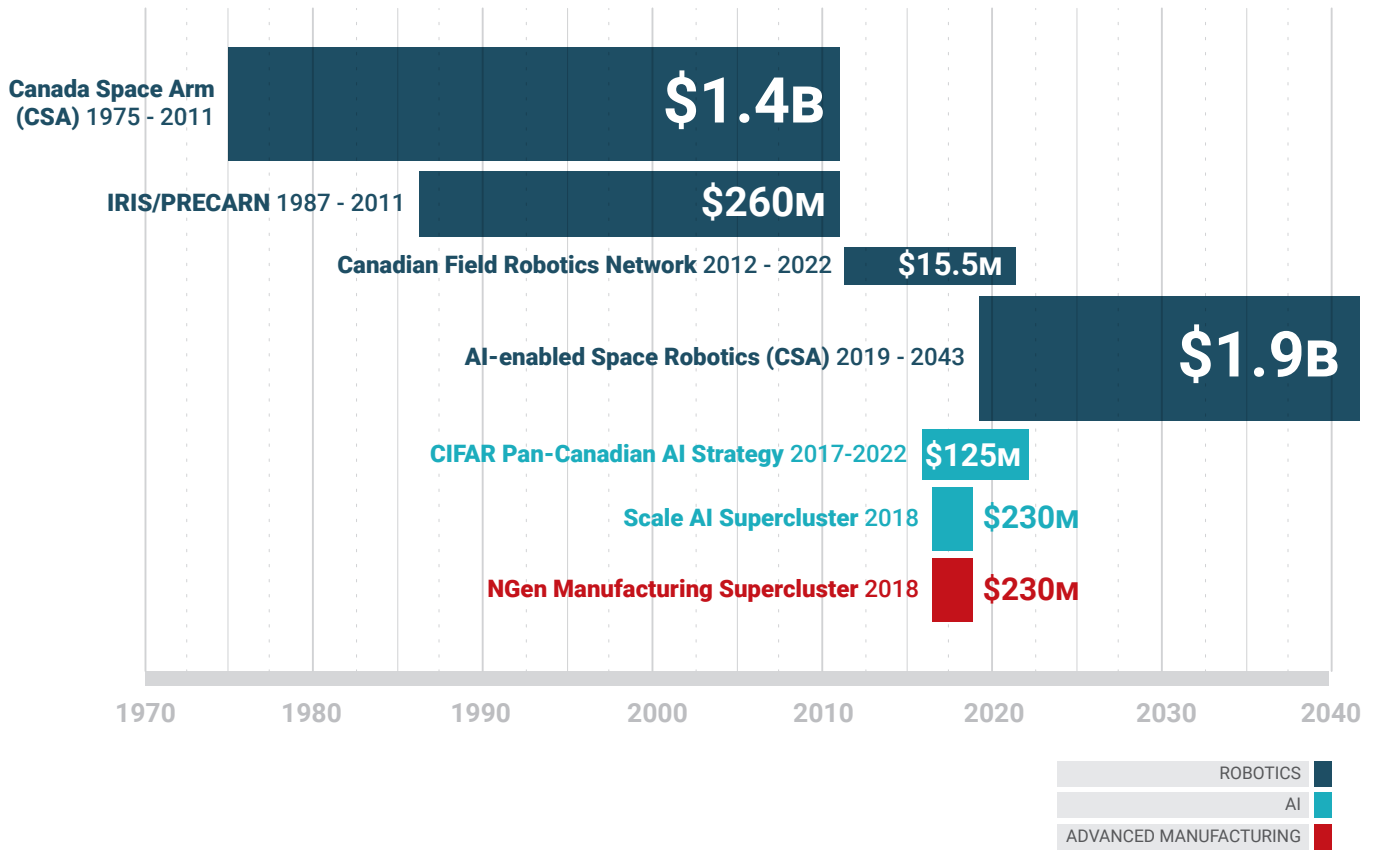

*Figure 11: Canada's national funding programs in robotics & key complementary initiatives*
*Sources: Canadian Space Agency, PRECARN, NSERC Canadian Robotics Network, CIFAR, Scale AI, NGen*

From 1987-2011, Canada contributed an additional $260M over 24 years[170] to PRECARN[171], an industry-led consortium set in motion by the Canadian Institute for Advanced Research (CIFAR), and co-funded by the federal government. PRECARN subsequently launched and led the Institute for Robotics and Intelligent Systems (IRIS), one of Canada's early NSERC Networked Centres of Excellence[172]. After PRECARN/IRIS wound down operations at the end of their respective mandates, Canada's subsequent strategic robotics investments grew significantly smaller. NSERC allocated the Canadian Field Robotics Network with a total of $15.5M in funding for 2012-2022[173][174]. The organization has since broadened its mandate and renamed itself the NSERC Canadian Robotics

---

170   Reference: interview with Paul Johnston, CEO of PRECARN.
171   "Research and Development Funding: Pre-Competitive Advanced Research Network (PRECARN)," Mentor Works Ltd., Feb. 13, 2013. https://www.mentorworks.ca/blog/government-funding/research-and-development-funding-pre-competitive-advanced-research-network/ (accessed Sep. 07, 2020).
172   P. Warrian, "PRECARN Incorporated History," Innovation Policy Lab at the Munk School of Global Affairs, University of Toronto. (unpublished).
173   "NCFRN Fact Sheet." NSERC Canadian Field Robotics Network, 2013, Accessed: Sep. 03, 2020. [Online]. Available: http://ncfrn.mcgill.ca/press-kit/ncfrn-fact-sheet/at_download/file.
174   http://ncfrn.mcgill.ca/press-kit/press-release/view

Network (NCRN), however with limited resources and participants, it falls short of the significant funded strategies of prior years.

Despite this, Canada has invested in some key complementary initiatives. One of these is the $125 million Pan-Canadian Artificial Intelligence Strategy[175], launched in 2017, which established three national AI institutes: the Amii Institute in Alberta, the Mila Institute in Montreal, and the Vector Institute in Toronto. Another is the $950 million Supercluster Initiative, which launched its Scale.ai (artificial intelligence) and NGen (advanced manufacturing) programs[176] in 2018. Most recently, Innovation, Science and Economic Development Canada (ISED) has renewed its space robotics interest with the launch of its new Canadian Space Strategy (2019), which promises to invest

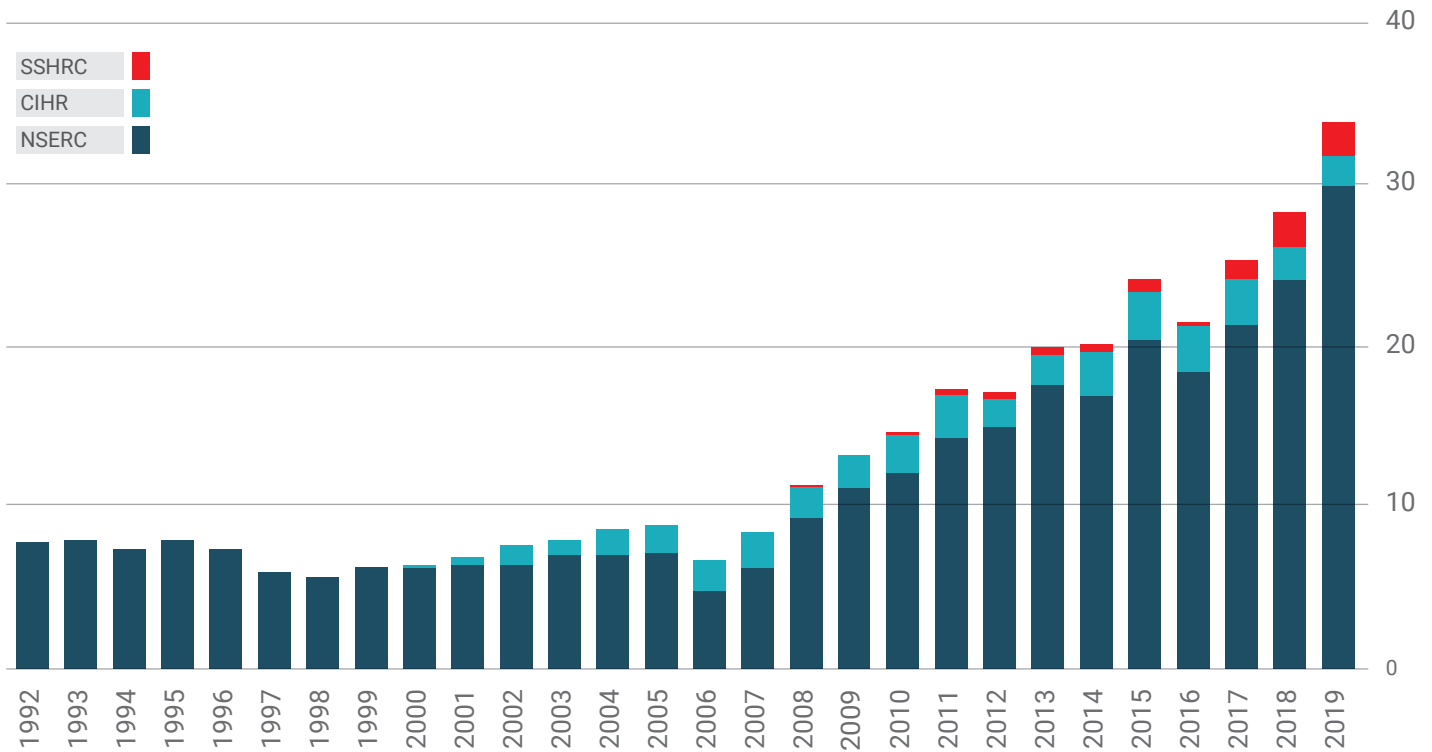

Figure 12: Robotics funding from Canada's Tri-Council Agencies ($Million CDN/year 1992-2019)
Sources: NSERC, CIHR, and SHHRC Awards databases[177]

---

[175] "Pan-Canadian Artificial Intelligence Strategy," CIFAR. https://www.cifar.ca/ai/pan-canadian-artificial-intelligence-strategy (accessed Sep. 01, 2020).
[176] "Canada's new superclusters," Innovation, Science and Economic Development Canada, Jan. 27, 2020. https://www.ic.gc.ca/eic/site/093.nsf/eng/00008.html (accessed Sep. 01, 2020).
[177] Source: Awards databases from Natural Sciences and Engineering Research Council of Canada (NSERC), Canadian Institutes of Health Research (CIHR), and the Social Sciences and Humanities Research Council (SSHRC); search term "robot"

$1.9 billion specifically to develop AI-enabled deep-space robotics systems[178]. In addition, Canada continues to fund robotics research through regular Tri-Council competitive funding mechanisms, and this is steadily growing (see Figure 12). Though Canada now has no formal R&D research program for robotics, a total of $34M CDN was funded for robotics research under Canada's federal research granting agencies in 2019.

**Canada's overall R&D funding**

Canada is known globally for the quality of its R&D. However a recent report on the state of research and development in Canada from the Canadian Council of Academies[179] reports a sustained erosion of Canada's private and public R&D expenditures in recent years. While Canada's R&D expenditures have remained relatively flat when adjusted for inflation, other OECD countries have significantly increased their R&D investments.

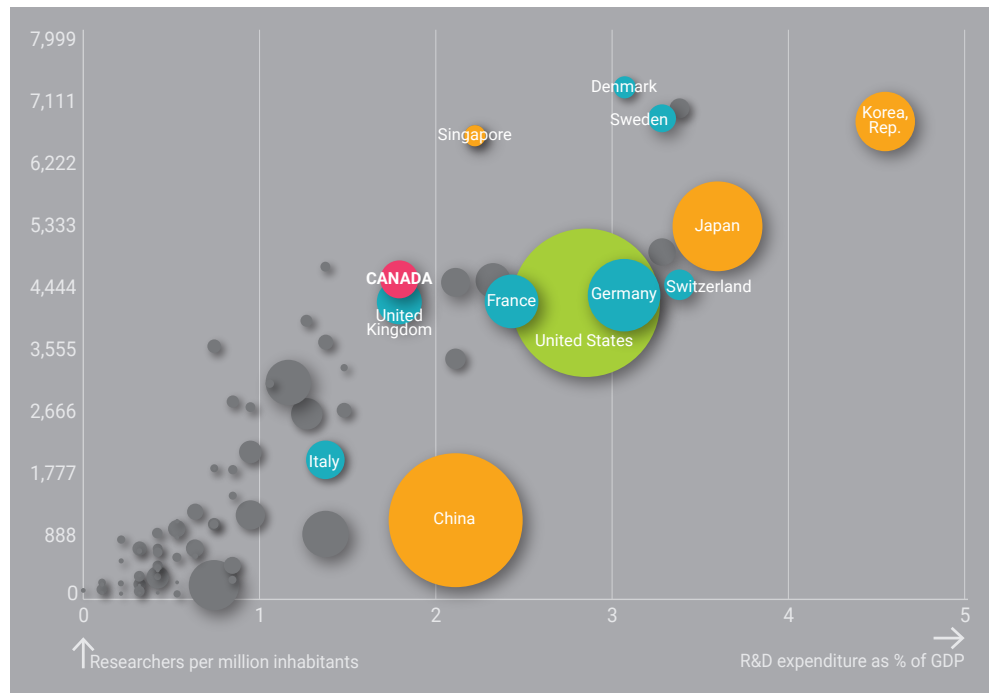

*Figure 13: Global R&D spending as % of GDP. The circles show the amounts countries are spending on total R&D in PPP$. Countries farther to the right are spending relatively more in terms of their GDP. Those closer to the top have higher numbers of researchers per capita. Image credit: Adapted from UNESCO Institute for Statistics.[180]*

# 4.7
# COORDINATION OF STAKEHOLDERS AND EFFORTS

A national pandemic response requires coordination across multiple stakeholders to manage everything from procurement, regulatory fast-tracking, funding and more. To include robotics as part of this response requires additional domain-specific expertise and coordination. Many of the countries where we have seen significant robotics deployments during the COVID crisis have national robotics strategies that have paved the way for significant funding and coordination of efforts prior to the pandemic. A key premise of this paper is that these strategies resulted in relatively mature robotics ecosystems that were better poised than those in other countries to respond with offers of assistance when the pandemic struck.

For example, the EU, which has had a coordinated robotics strategy since 2014, is one of the few regions that has created dedicated funding for the development of COVID-related robotics.

Recognizing the specific challenges of bringing robotics to market, the Digital Innovation Hubs in Healthcare Robotics (DIH-HERO) fund will provide EUR 5M for robotic solutions that can be rapidly deployed in the healthcare sector[181].

Another important example is China, which launched its Robot Industry Development plan in 2016 and has long included automation as an integral part of its economic development strategy. China was able to launch and pilot a robot-equipped Smart Field Hospital within weeks after going into lockdown. This massive coordination effort included liaising with medical staff to determine needs, coordinating robot suppliers and local authorities to procure and ship equipment during lockdown, and telecommunications providers to ensure that the mobile robots would have constant contact with the cloud where technical engineers could provide support[182].

---

181  "Call for funding: robotic tech and solutions immediately deployable in hospitals," Shaping Europe's digital future - European Commission, Apr. 10, 2020. https://ec.europa.eu/digital-single-market/en/news/call-funding-robotic-tech-and-solutions-immediately-deployable-hospitals (accessed Sep. 03, 2020).
182  S. O'Meara, "Meet the engineer behind China's first robot-run coronavirus ward," Nature, vol. 582, no. 7813, Art. no. 7813, Jun. 2020, doi: 10.1038/d41586-020-01794-8.

## Long term outlook

- **Multiple coordination models:** Nations have taken different approaches to coordinating robotics development, with some efforts initiated and sustained by government, and others primarily by industry. Still others involve close partnerships between government, industry and academia alike. National R&D programmes (China), robot strategies (Japan, Germany, USA, Australia), and framework programmes (EU) are the most common tools used to develop leadership in robotics innovation and promote coordination between the various stakeholders.

- **Digital Innovation networks:** It is noteworthy that many of these regions' robotics strategies, though comprehensive in and of themselves, are but one component of a broader digital innovation strategy[183] that involves complementary initiatives in artificial intelligence, wireless infrastructure, sensing technologies, and sometimes even regulatory and ethics frameworks. Such large scale innovation programs are years in the making and involve sustained investment of funding, human resources, and infrastructure.

- **Regulatory fast-tracking:** If robotic devices are to be of benefit during the current pandemic, fast tracking may be needed to bring them more quickly through testing and approvals processes. And yet health and safety standards must in no way be compromised during the fast-tracking process. Several regions have created special programs to fast track approval of medical devices for use during the COVID-19 pandemic, including the US[184], China[185], and Canada, however it remains unclear whether devices that were fast tracked during the pandemic will continue to be approved when the crisis is over. Because robots have many application areas, a "one-size fits all approach" is insufficient; application- and domain-specific regulatory needs must be considered by the appropriate regulatory body.

---

**Regulatory environment**

Due to Canada's relatively small market, a key regulatory issue here is the need to harmonize with international safety standards and the standards of large robotics markets such as the EU. Such an approach helps in three key ways. First, it ensures that Canadian robotics are developed in such a way as to make them suitable for export to international markets. Secondly, it paves the way for robots developed in other countries to more quickly gain approvals for Canadian health and safety standards, making it easier to import robots here when needed. Finally, the participation of Canadian firms and agencies in international standards working groups provides international visibility and experience to Canadian developers.

We describe Canada's regulatory expertise and enumerate some of the relevant Canadian regulations pertaining to robotics in the Appendix, including successes and challenges.

**Coordination of stakeholders**

Canada was one of the early countries to invest in robotics through its Canadarm space robotics program, starting in 1975[186]. According to the Canadian Space Agency, these early contributions helped to generate a Canadian space industry that supports 8,000 skilled jobs and is worth $3.33B each year.[187] In 1984, the Canadian Institute for Advanced Research (CIFAR) launched its first research program in Artificial Intelligence and Robotics, and issued a report that recommended that Canada establish a mechanism to support technology transfer in robotics[188]. Ultimately this report led to the creation in 1987 of a 39-company consortium called the Pre-Competitive Applied Research Network (PRECARN)[189]. Soon after, PRECARN formed a Network Centre of Excellence called the Institute for Robotics and Intelligent Systems to provide basic research to the consortium. In 2010, when IRIS/PRECARN wound down their operations at the end of their respective mandates, Canada became one of the few developed countries to not have a national robotics innovation network. Canada also has no national robotics industry association.

---

The idea for a renewed national robotics strategy for Canada was discussed at a workshop held at the 2017 IEEE Intelligent Robots and Systems (IROS) conference in Vancouver that brought together Canadian stakeholders from across academia, government, and industry. Many of these stakeholders generously provided useful input to this whitepaper. The group identified a number of key opportunities for the Canadian robotics ecosystem, including:

- Aligning stakeholders on a uniquely Canadian vision for robotics that accounts for our robotics strengths, our culture, our economy, and geography. Simply "copying" a robotics roadmap or strategy from another region or country is unlikely to work.

- Clarifying our understanding of the evolving robotics ecosystem in Canada. Many other countries have Robotics Industry Associations and/or national robotics conferences to track growth, impact, and expertise. Canada does not.

- Building visibility for Canadian robotics. As global demand for robotics expertise rises, there is fierce competition to attract and retain top talent and industry funding.

- Strengthening Canada's digital innovation portfolio. Canada already has core strengths in AI, ICT, quantum computing, photonics and sensing. Together with robotics, these form a suite of core digital competencies that are capable of fueling multiple application areas, from healthcare, to mobility, to manufacturing.

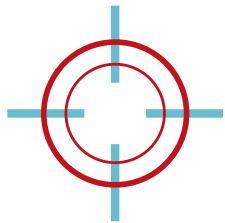

**ALIGN KEY ROBOTICS STAKEHOLDERS**
on a pan-Canadian vision

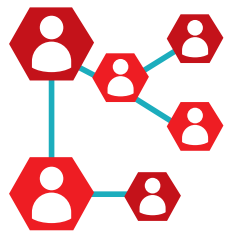

**SUPPORT THE ROBOTICS ECOSYSTEM**
by tracking growth and impact

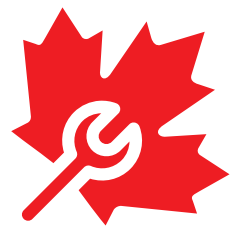

**BUILD VISIBILITY FOR CANADIAN ROBOTICS**
& compete globally to attract $ and talent

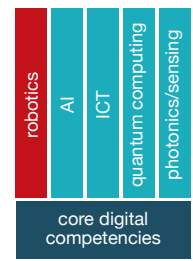

**STRENGTHEN CANADA'S DIGITAL INNOVATION PORTFOLIO**
by complementing existing digital competencies

*Figure 14: Opportunities for a coordinated Canadian robotics ecosystem*

Though Canada has no formal digital innovation strategy, the idea is not new. In 2014, the Innovation Policy Lab at the Munk School of Global Affairs and Public Policy at the University of Toronto launched the Creating Digital Opportunity (CDO) research partnership involving 16 universities and 11 partner organizations[190]. CDO has published several recommendations[191][192][193] for a national digital innovation strategy for Canada.

The Canadian government also appears to be moving in this direction. In 2018, Innovation, Science and Economic Development (ISED) Canada launched a series of national Digital and Data Consultations[194] to invite stakeholders from across the country to share their perspectives on the challenges and opportunities for Canadian digital innovation. Some of the Authors and contributors of this white paper were among those invited to the consultations, which resulted in the publication of Canada's Digital Charter in 2019.[195]

While digital innovation overtures such as these are timely, in Canada they have tended to focus on the virtual and software aspects of digital innovation, and have not given hardware aspects such as robotics and sensing as much attention. Given the unique challenges and opportunities presented by hardware R&D as compared to software (see Section 2.3 for more details), we argue that robotics should not be an afterthought to a digital innovation strategy, but rather a dedicated focus area with its own strategic approach and timeline for development.

---

# 4.8 ADDRESSABLE GAPS: A SUMMARY OF CANADA'S STRENGTHS AND WEAKNESSES

Thus far in Section 4 we have explored the key factors (technical, social and systemic) that are limiting the use of robots during the pandemic, both in general and here in Canada. In this section we summarize Canada's relative strengths and weaknesses.

| TABLE 12: STRENGTHS AND WEAKNESSES IN CANADIAN ROBOTICS | | |
|---|---|---|
| **SUCCESS FACTOR** | **CANADA'S STRENGTHS** | **CANADA'S WEAKNESSES** |
| 4.1 Teleoperation | Strong research capabilities in autonomy and supervised autonomy<br><br>Highly relevant to Canada given the vast distances here; high potential for impact | Lack of incentives for teleoperation-focused research to fill the autonomy gap<br><br>Lack of training programs for teleoperators<br><br>Lack of both supply and demand for teleoperation services |
| 4.2 Open APIs | Two companies (1 mobile, 1 manipulator) with Open API<br><br>Strong research expertise integrating hardware and software systems. We build real robots here. | |
| 4.3 Robot-friendly environments | Introduced barrier-free legislation in 2019 as part of the Accessible Canada Act | Slow to create smart city infrastructure<br><br>Regulatory uncertainty over data ownership and privacy<br><br>Lack of large scale, outdoor robotics test sites |
| 4.4 In-house industry expertise | Multiple centres of graduate training excellence across the country (e.g. Toronto, Montreal, Vancouver, Waterloo)<br><br>Strong college-level programs (Ontario)<br><br>Dedicated training and certification programs for drone pilots and robotic surgery | Lack of robotics experience and knowhow among downstream industry leaders<br><br>Lack of robotics end-user training programs in downstream application areas<br><br>Lack of product-focused, applied research centres to support translational robotics research<br><br>Canadian robotics startups challenged by marketing and selling on a global level due to lack of available, experienced, and relevant talent |
| 4.5 Access to robot supply chains | | Critical electronic components must be imported internationally<br><br>No UV disinfection robot and only one homegrown OEM making social robots prior to COVID |
| 4.6 Sustained robotics funding | Large, early investments in robotics set the stage<br><br>Several new complementary funding mechanisms<br><br>Several new corporate research centres have opened offices recently | No comprehensive and dedicated robotics funding mechanism<br><br>Not currently tracking robotics funding<br><br>Eroding industry R&D investment in Canada as a whole |
| 4.7 Coordination of stakeholders | NSERC Canadian Robotics Network<br><br>Annual Canadian Computer and Robot Vision Conference since 2004<br><br>Pan Canadian AI Strategy<br><br>New Canadian Space Robotics Strategy<br><br>Supercluster programs in Advanced Manufacturing and AI | No national robotics strategy or comprehensive digital innovation strategy that connects researchers to end users<br><br>No Canadian-centred robotics industry association<br><br>Too few policymakers and regulators with expertise in robotics |

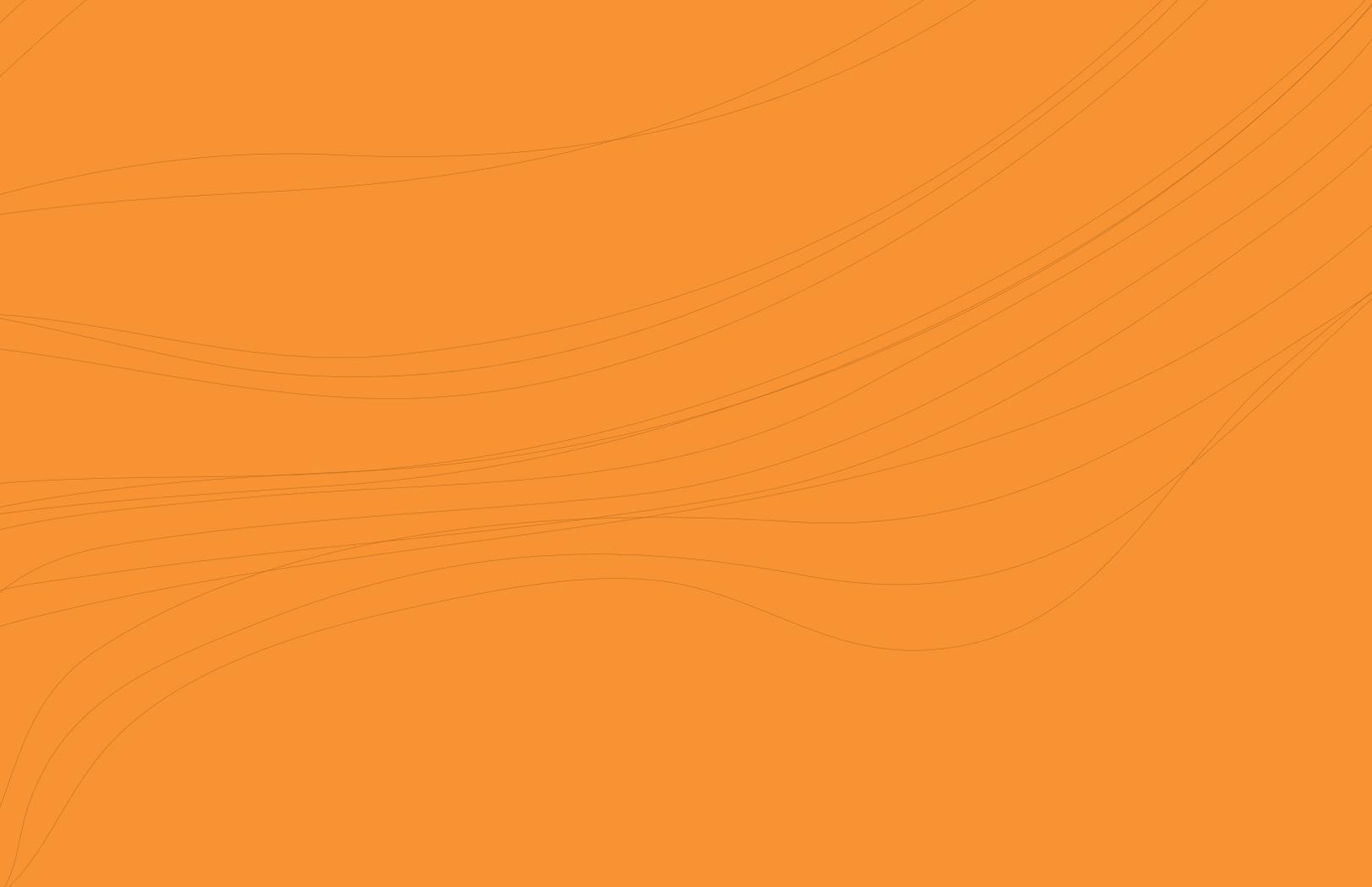

**CONCLUSION**

We started this whitepaper with the question: Where are all the robots fighting COVID-19 in Canada? We interviewed numerous potential and actual end-users — from hospitals to public transit, grocery stores and schools — to understand the most pressing needs where robots can and do play a role, and we conducted a media scan to understand more generally how robots are being put to use in the global pandemic. We also surveyed Canadian robotics companies to understand how the pandemic is impacting their businesses. We then used our experience developing and building robotics systems to evaluate the readiness, robustness, and autonomy of the various robots we were seeing. Ultimately our goal was to gain a general understanding of where robots are working effectively in the pandemic, and what are the factors for successful deployments, both in Canada and more broadly.

We use the remainder of this paper to discuss the implications of our findings in the context of five key questions:

1. Is Canada capable of a robotized future?
2. Does Canada want a robotized future?
3. Does Canada need a robotized future?
4. Is Canada ready for a robotized future?
5. What next steps can Canada take to better exploit its robotics assets?

**Is Canada capable of a robotized future?** This paper has shown that countries and organizations that were invested in robotics before the pandemic are the ones that have most benefited from pandemic demand. It has also shown that Canadian robotics are making some key contributions towards helping the logistics and manufacturing industries navigate staff shortages and increased demand due to COVID-19. And yet Canada has not taken advantage of robotics to the same degree as other nations in its pandemic response, despite its long history of investment, its high academic standing in the field, and the international success of a number of its homegrown robotics companies.

Having examined various factors that influence robotics across a variety of technological, societal and systemic domains, we conclude that Canadian robotics stakeholders do indeed possess the technical know-how to succeed in making impactful contributions to the field of robotics. So what is holding us back?

**Does Canada want a roboticized future?** The news and entertainment media has long presented us with many contradictory views of what a roboticized future might look like, from the inspiring to the downright worrisome. Now the COVID-19 pandemic has accelerated robots' use in the real world, and the pace of change has become mind-boggling. In this context, caution and skepticism are reasonable, as are anticipation and hope. And yet whatever our feelings, we seem to find ourselves at a critical junction: jump in or be left behind. All signs seem to show that global forces are inevitably pulling us towards more automation. In some parts of the world, at least, the roboticized future is already here.

While there is scant data available on Canadians' attitudes towards robots specifically, various studies paint a picture of how Canadians feel about new technologies and disruptive innovation in general. Results from these studies show that Canadians are somewhat more skeptical about technological advances than other nations, but they hold a nuanced view. An HSBC study[196] showed that Canadians are less positive about technology advances than the global average, but are also less overwhelmed by it. And an Ipsos study[197] of Canadians showed that while Canadians tend to recognize the value of new technologies and are generally hopeful that they will bring benefits to society, they are also concerned that the benefits of innovation will not be distributed evenly. Perhaps the reason it's hard to understand whether Canadians actually want a roboticized future is because it's still unclear what in fact this future might look like. As roboticists who are the makers of such technology, we prefer to turn the question around and ask: What is the roboticized future that Canadians want?

Armed with a clear vision of our goals, a mutual understanding of Canadian stakeholders' needs, and a commitment to principles of equity, diversity and inclusion to ensure that our most vulnerable stakeholders are neither adversely impacted nor left behind, Canada has the opportunity to create a roboticized future that reflects the unique cultural, geographic, and economic needs of its citizens. Rather than fear that robots will one day take over our jobs, our freedom, and our privacy, we can engineer and design the robots that we want to ensure a bright future for this country and its residents.

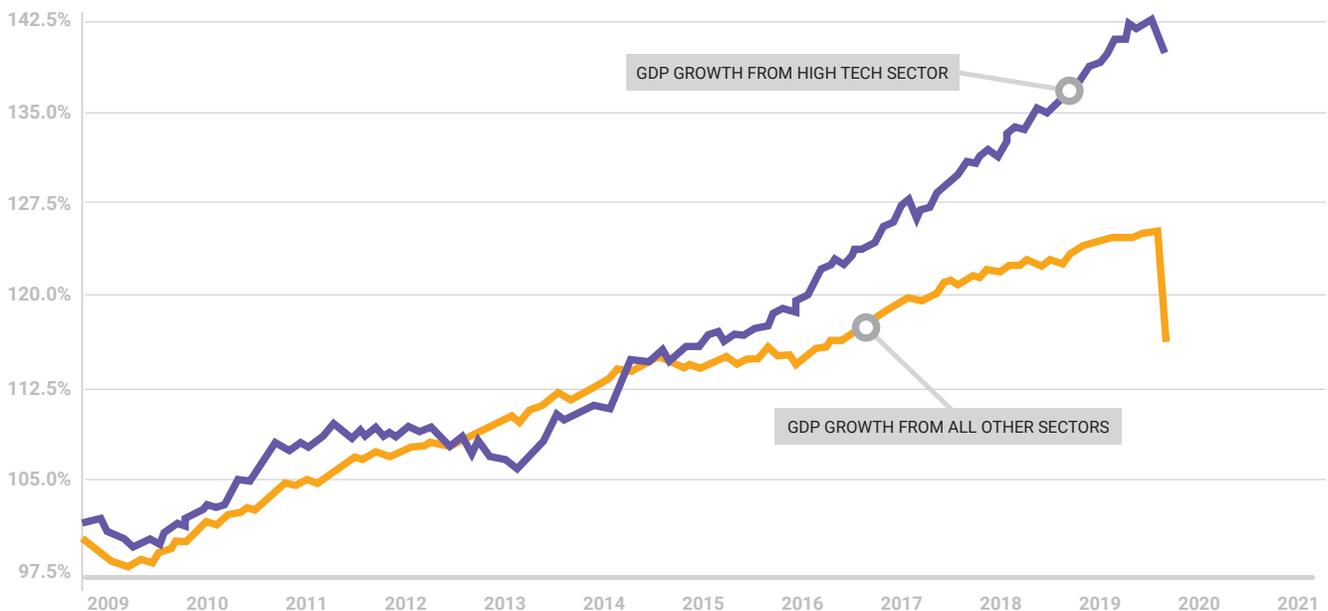

*Figure 15: GDP growth in Canada's high tech sector versus other sectors*
*Sources: BMO, Statistics Canada, Haver Analytics. Image credit: Adapted from CBC News.[195]*

---

**Does Canada <u>need</u> a roboticized future?** When both lives and livelihoods are at risk — as is the case in the current pandemic — it is wise to question whether technology is the right solution. Economies around the world, including Canada's, have taken a beating and will continue to do so for the foreseeable future … is new technology where we should be investing our limited resources right now?

<u>Venture capital seems to think so.</u> In Section 4.6 we saw that despite overall economic slowdowns, the global robotics market continues to attract new investment. Looking at Canada's high tech sector, we see a similar trend. Canada's technology sector is growing faster than any other part of the economy, and now represents about 5% of the total (see Figure 15).[198]

While it's virtually certain that we are headed for some years of economic uncertainty, investment in robotics may prove to be one of the key growth engines for restarting Canada's post-pandemic economy, and ensuring its resilience beyond: even when the pandemic ends, issues like climate change will no doubt drive increased investment in robotics too.

**Is Canada <u>ready</u> for a roboticized future?** Even though on some fronts Canada may appear to be lagging, much of the groundwork is already present for Canada to pull ahead from the middle of the robotics pack. Our robotics research base and startups are strong. We have made significant complementary investments in artificial intelligence, and advanced manufacturing through the CIFAR Pan Canadian Artificial Intelligence Strategy, the SuperCluster initiative, and the new Canadian Space Strategy. And as a politically stable nation with strong public education and healthcare systems, on the whole Canadians will be more immune to the negative effects of automation than many other nations.

But let's look more objectively.

The Economists' Automation Readiness Index[199] is specifically geared toward evaluating whether countries are prepared for a future of automation. The index measures not just technological investment and infrastructure, but also the development of policies (innovation, education, and labour market) designed to prepare for and manage the transition to greater autonomy. Overall, Canada ranks 5th in the world, behind other robotics-forward nations such as South Korea, Germany, Singapore and Japan (See Figure 16).

---

198  P. Evans, "How COVID-19 has changed Canada's economy for the worse — but also for the better," CBC, Jun. 23, 2020. https://www.cbc.ca/news/business/covid-economy-changes-1.5618734 (accessed Sep. 03, 2020).
199  "The Automation Readiness Index 2018," The Economist Intelligence Unit. http://automationreadiness.eiu.com (accessed Sep. 03, 2020).

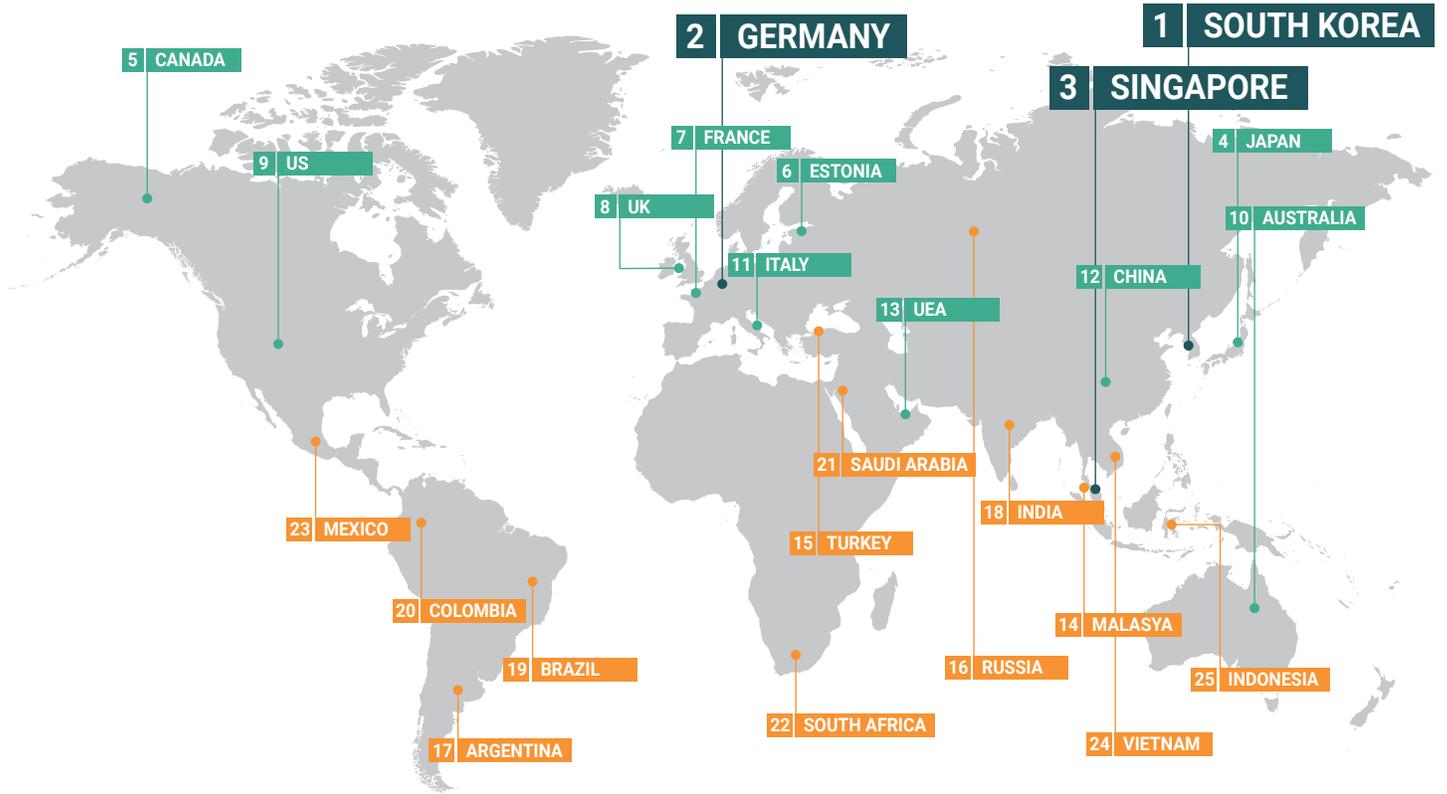

*Figure 16: Automation readiness rankings by country*
*Source: The Economist Intelligence Unit Automation Readiness Index[200]*

A deeper look at the specific factors that were used to assess our automation readiness reveals where Canada is doing well and where more effort is needed. The table 13 summarizes our biggest strengths and weaknesses. It's notable that whereas Canada is a clear frontrunner in most categories, it ranks only 23rd out of the 25 countries examined for its Innovation Policies & Regulations, and 15th for Knowledge Transfer.

In other words, with the right targeted policies in place, Canada can pull ahead.

---

200 "The automation readiness index: Who is ready for the coming wave of automation?," The Economist Intelligence Unit, 2018. Accessed: Sep. 01, 2020. [Online]. Available: https://www.automationreadiness.eiu.com/static/download/PDF.pdf.

**TABLE 13: CANADA'S AUTOMATION READINESS: STRENGTHS AND WEAKNESSES**
*Source: The Economist Intelligence Unit Automation Readiness Index*

### INNOVATION CATEGORY

| Biggest Strengths (Top 3 by ranking) | Biggest weaknesses (bottom half by ranking) |
|---|---|
| 1st in Startup Support | 23rd in Innovation Policies & Regulations |
| 1st in Data Protection | 15th in Knowledge Transfer |
| 1st in Broadband (strategy, usage & speed) | |
| 1st in Citizens Use (data safety awareness) | |
| 3rd in Clusters | |

### EDUCATION CATEGORY

| Biggest Strengths (by ranking) | Biggest weaknesses (bottom half by ranking) |
|---|---|
| 1st in Early Childhood Policies | none |
| 1st in 21st Century Skills and Knowledge | |
| 1st in Career Guidance | |
| 1st in Access to Post Secondary Education | |
| 1st in Continuous Education | |
| 1st for Teacher Training | |
| 1st for Use of Technology and Data | |
| 1st for Social Dialogue | |

### LABOUR MARKET

| Biggest Strengths (by ranking) | Biggest weaknesses (bottom half by ranking) |
|---|---|
| Biggest Strengths (by ranking) | none |
| 1st in Labour Market Research and Policy | |
| 1st in Vocational Training | |
| 1st in Transition from University | |
| 1st in Targeted Retraining | |
| 1st in Public Employment Services | |

What next steps can Canada take to exploit its robotics assets? Our research points to a number of systemic issues that are holding Canada back from fully exploiting its considerable robotics assets. Below are some suggestions to help address these issues and get us moving in the right direction for the future:

## Coordination

- Establish networking mechanism(s) to rapidly share ideas around robotics and COVID-19 in Canada that includes stakeholders from industry, healthcare, robotics and hardware providers (e.g., website, Slack, Canadian Robotics Conference or workshop).
- Develop guidelines for how to pilot robotics solutions in key front-line use cases.

## Training

- Investigate additional vocational workforce training, certification, and regulation for robot operators and technicians to prepare for the growing need for scaled-up robot solutions.

## Funding Schemes

- Develop a funding mechanism to incentivize industry to work on transferring technology to use cases with a 2-3 year runway to circumvent reliance on nascent sales revenue and provide end users with complementary incentives to adopt the technologies.
- Establish sustained and coordinated funding that fosters a healthy robotics ecosystem in Canada across academia, startups, industry, and government labs to ensure a pipeline of technology that is ready for future crises.

## Canadian Robotics Strategy

- <u>Inventory</u> - Create an inventory of Canadian robotics assets (research and test facilities, people, and organizations across the key application areas). Identify existing economies of scale on both the technological and the latent demand fronts, as well as regional areas of application-based expertise.
- <u>Stakeholder consultation</u> - Identify multiple scenarios for Canada's innovation future based on current trends; and create a vision of the future for Canadian robotics that addresses this country's unique geographic, cultural, and economic realities; this should have a clear mandate for specialization that plays to Canada's technological and economic strengths.
- <u>Strategy</u> - Develop a Canadian robotics roadmap on this vision, and establish clear government priorities for the next pandemic and other crises that may lie ahead. Market Canada's strengths and forge international partnerships where 'made in Canada' solutions are not yet viable.

## Digital Innovation Strategy

- Coordinate with digital innovation stakeholders in complementary underlying technologies such as artificial intelligence, photonics and sensing, quantum computing, and ICT, to ensure a comprehensive digital innovation portfolio for Canada.

As a parting message, we urge readers to consider the notion that robotics is the critical (and mostly still missing) link between the information age and "high-touch" sectors of the Canadian economy, from health, to manufacturing, to agriculture and food production. Bringing our physical world into the information age is what is meant by the Fourth Industrial Revolution.

# APPENDIX

# A. CANADIAN ROBOTICS LANDSCAPE

Canada has an expanding robotics ecosystem. In this Appendix, we build on the expertise mentioned in Section 1 to showcase the most relevant expertise and assets across government, and industry.

## A.1 Robotics Companies

Robotics technology is widely used in Canada, primarily for industrial applications. Industrial robots increase capacity and efficiency of Canadian manufacturing, indirectly addressing the needs highlighted above. Over 3,500 industrial robots were installed in Canada in 2018[201], for example in handling parts for assembly, welding, and packaging in factories.

More relevant to this paper are the number of Canadian government organizations and companies that are developing the next generation of robots for potential use in pandemic response. We list a handful of representative examples across six key application areas where robotics can be expected to play a significant role.

### A.1.1 MATERIAL HANDLING

Material handling (MH) involves the simple transport of objects from place to place. By fitting a robot with an appropriate end-of-arm tool (e.g., a gripper), the robot can efficiently and accurately move items from one location to another. Many of these systems are primarily designed for industrial automation (e.g., pick and place of parts in a warehouse) but could also be adapted for manipulating medical supplies and samples, for example. In both cases, the important pandemic response aspect is the elimination of human involvement in transport and laboratory testing, improving our ability to maintain physical distance in production operations.

---

201 "Executive Summary World Robotics 2018 Industrial Robots," International Federation of Robotics, 2018. Accessed: Sep. 03, 2020. [Online]. Available: https://ifr.org/downloads/press2018/Executive_Summary_WR_2018_Industrial_Robots.pdf.

Material handling is one of the areas with heightened private capital interest due to its application in e-commerce. There are a number of enterprises that operate in this technology vertical:

- **Kinova:** Apart from assistive technologies, Kinova is primarily a manufacturer of lightweight robot arms for testing, maintenance and inspection (TRL9), with a wide range of application industries.
- **Kindred Systems:** Kindred focuses on piece picking in warehouse settings with product offerings that include autonomous bin picking and sorting (TRL7 to TRL9). This technology has been validated for soft item handling such as packaged clothing.
- **Robotiq:** Robotiq is a provider of tools and software for cobots. Robotiq's Plug and Play components and end-of-arm tooling (TRL9) come with multiple automation software solutions. Manufacturers choose Robotiq's collaborative robot solutions for applications such as machine tending and pick and place to start production faster.

**A.1.2 MEDICAL ROBOTICS**

Canadian robotics government organizations and companies stand to make significant strides in the medical domain, as they enable physical distancing requirements to be maintained in a variety of high-risk scenarios. Canadian robotic developers are pioneering advanced systems for use in generic medical environments such as operating rooms and doctor's offices.

- **Canadian Space Agency and National Research Council:** The CSA and the NRC have partnered to develop an advanced, semi-automated microfluidic biological analysis system. Although designed originally for rapid sample analysis on board the International Space Station, the system (which is effectively already robotic) could also be deployed in terrestrial settings. The current design is at TRL5, and further development would be needed to enable fully automated robotic sample collection and processing.
- **Defence Research Development Canada:** Although not classified as a 'typical' medical robotic system, DRDC is actively developing a drone-mounted chemical-biological-radiological (CBR) detector (TRL4). This system is based on vacuum filtration of airborne particles and requires substantial air volumes for accurate detection (which is achievable on an aerial platform). Although the CBR system is designed for aerial deployment, it could be repurposed for in situ robotic sampling in enclosed spaces (hospitals, care homes, restaurants, etc.). The sampling rate would be relatively low, requiring on the order of 24 hours to confirm detection of viral particles, but this would still provide an ability to identify 'hot spots' where the virus had been present, potentially enabling mitigation or contact tracing.

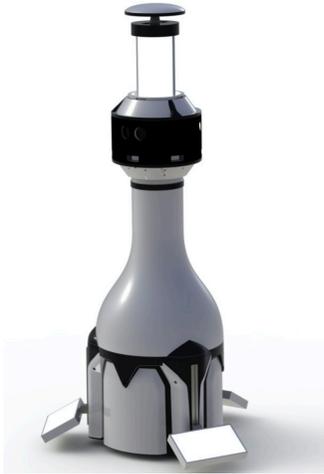

*Figure 17:*
*Bishop robot by Crosswing Inc.*
*Image credit: Crosswing Inc.*

- **CrossWing:** CrossWing is an Ontario company that develops robots for telepresence and for the medical and hospitality industries. They are currently building the Bishop™ Cleanbot™ UV-C Disinfecting Robot (TRL5) specifically for use to fight the COVID-19 pandemic; the robot has an omnidirectional mobile base and an upright 'body' that supports several UV-C LED light panels. Bishop is designed to be capable of autonomously disinfecting public and private space.

- **Synaptive Medical:** Synaptive Medical in Toronto develops advanced surgical imaging systems, including a robotic surgical microscope called Modus, which is TRL9 and deployed in operating rooms in the US. The company specializes in incorporating imaging with surgical planning and visualization to provide detailed information to surgeons. Robot-assisted surgery increases the physical distance between the clinician and the patient and, in certain cases such as neurosurgery, removes the need for surgical apparatus that would otherwise prevent the surgeon from wearing a ventilator or other essential PPE.

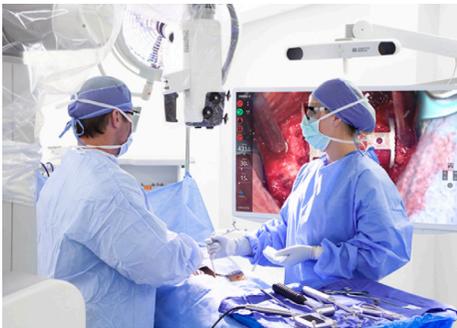

*Figure 18:*
*Modus V robotic digital microscope system by Synaptive Medical*
*Image credit: Synaptive Medical*

- **Titan Medical:** Titan Medical in Toronto is developing a single-port robotic surgical system called SPORT. The system is mounted on a cart and uses disposable tools, leading to lower-cost inclusion into operating rooms than other robotic surgical systems. SPORT also incorporates a high-quality imaging systems for gynecological procedures. Tests have been conducted on live animals and human cadavers, although further results from live-animal testing is needed to secure approval to operate on humans so the system is currently TRL7. The use of robotics in surgery will help with distancing in the operating room.

- **ANVIV Mechatronics in Collaboration with ESI:** ANVIV has developed various robotic systems for inspection, maintenance, and laboratory automation. ANVIV is developing AI-embedded robotic systems for COVID-19 response, including for remote measurement of fever, remote examination of suspected or confirmed carriers of the virus, and disinfection and sterilization of areas in hospitals and public spaces.

- **MDA Robotics:** MDA is well-known internationally for developing the Canadarm for the International Space Station and currently has a division developing image-guided medical robots. As MDA works on a contract basis, they do not produce robots in large quantities. MDA is not developing any robots specifically for response to COVID-19.

**A.1.3 INDOOR MOBILITY**

Indoor mobility includes a range of robotics systems capable of navigating through indoor human environments unaided while constructing maps and avoiding obstacles. These automated robots are the precursors to autonomous driving vehicles and experience unique challenges in each application domain, requiring specialization and customization of their underlying autonomy systems.

- **Avidbots:** Avidbots has deployed fleets of the Neo autonomous floor cleaning robot (TRL9), capable of navigating over flat surfaces scrubbing the floor beneath them. Current operations involve defining a map of the cleaning environment from a human-guided tour, and then executing an autonomous cleaning plan, stopping when dynamic objects appear in the path and navigating around static objects that are new to the mapped area. The robots are restricted to operate in open areas, as they cannot open doors or navigate elevators autonomously. CTO Pablo Molina states that they have seen significantly increased interest from the public sector since the start of the pandemic, due to increased needs to keep employees safe and keep public indoor spaces clean. The company is actively pursuing disinfectant spray adaptations to Neo (TRL8), aiming to avoid some of the pitfalls of UV-based disinfection, and expects to release 3D spraying attachments by Q1 2021.
- **OTTO Motors:** The Clearpath Robotics subsidiary OTTO Motors provides fully autonomous warehouse robotics systems for goods transport in manufacturing and storage operations. The OTTO line of robots (100, 750 and 1500, all TRL9) can transport a wide range of goods over flat surfaces without any human involvement, creating their own maps and navigating seamlessly with human traffic and human-operated vehicles. The vehicles automatically dock at goods loading platforms, enabling end-to-end transport of goods throughout the warehouse. The more recently developed autonomous forklift, OMEGA, can not only navigate and transport goods autonomously, but it can also identify and load pallets on its forklift without human intervention. These capabilities put it at the forefront globally of warehouse automation, and will play an essential role in maintaining physical distancing in logistics hub operations. Clearpath has seen significant acceleration of timelines for robotic deployments across its industrial segments, as the pandemic highlights some of the hidden reliability and safety costs of keeping human operators on the shop and warehouse floor. Although Clearpath is not specifically targeting COVID-related products, its position as a robotic platform company has allowed it to pursue multiple partnerships with UV and spray disinfectant providers for enhanced robotic solutions to the sterilization challenge.

**A.1.4 OUTDOOR MOBILITY**

Numerous ground-based robot platforms are being designed and deployed for outdoor use in human environments. The pressure to maintain social distancing, and the challenge of keeping public transit safe during the pandemic have led to a major acceleration of low-speed and dedicated route operations. The most visible example is autonomous driving, but local delivery and outdoor disinfection applications are also emerging.

- **Gatik:** Point-to-point autonomous driving startup Gatik (HQ in US, research and operations centre in Toronto) is already servicing B2B routes for Walmart in the US, and has seen a major surge in demand to remove reliance on human drivers for regularly repeated routes from warehouses to stores since the start of the pandemic. The ability to limit operations to fixed, established routes has jump-started its deployments and commercialization efforts, eliminating many of the challenges around unlimited edge cases through operational domain constraints. The company is actively integrating disinfection technology for its vehicles, garages and depots, and expects to have a major impact on middle-mile deliveries allowing distinct logistics facilities to better maintain physical separation during goods distribution.

- **Tinymile:** Tinymile are developing user-friendly last-mile delivery robots, which are currently teleoperated and capable of transporting small goods or a modest grocery purchase on pedestrian routes (TRL5). This robotic technology could alleviate some of the challenge around short-range good delivery, particularly in dense urban settings, but needs significant investment to elevate to autonomous operation.

- **Uber ATG, GM Canada, and Huawei:** Autonomous driving fleet development is maturing rapidly around the globe, with strong R&D efforts occurring in Canada at Uber, GM Canada and Huawei (TRL5-7). The pandemic has led to worldwide softening of vehicle demand, but accelerated timelines for autonomous driving fleet deployment due to the pressures of increased delivery traffic and desire for reduced human contact of delivered goods.
- **Clearpath Robotics:** Clearpath Robotics is a Canadian owned and operated firm that is the largest developer of field robotics research products and solutions in the world. Though they do not sell outside of the research and development market, their TRL8-9 offerings have been integral in speeding up various relevant autonomous systems product development efforts around the world.

**A.1.5 AERIAL TRANSPORTATION**

Unmanned aerial vehicles, or drones, are now an established technology for aerial inspection, monitoring and tracking, and continue to improve in their ability to transport small goods efficiently and deploy sprays for agriculture. Within Canada, drone manufacturers are pushing to make their robotic platforms more autonomous, with benefits to delivery of emergency supplies on dedicated routes. This section includes robot platforms designed for outdoor non-built environments (field and air).

- **Drone Delivery Canada:** DDC has recently demonstrated the viability of drone based delivery of goods for point-to-point routes in remote areas, providing timely delivery of medicine, defibrillators and small-sized goods delivery. Point-to-consumer models are under development, which would allow delivery of a large range of goods without the need for human contact. The pandemic has accelerated demand for such solutions, but regulatory hurdles and safety considerations limit the path to large urban markets.
- **FLIR:** Having acquired Waterloo-area startup Aeryon Labs in 2018, FLIR develops long-endurance highly-capable surveillance and reconnaissance drones for military and civilian/industrial markets. The Skyranger series include gimballed IR cameras which could be deployed for mobile crowd temperature monitoring, as well as a Sky Hook package delivery system capable of delivery of goods up to 5 lbs.

### A.1.6 SOCIAL ROBOTICS

Social robotics is a nascent field focused on the design and deployment of robots that are able to interact directly with people. There are myriad applications for social robots in pandemic response, including as interactive information 'kiosks' in hospitals, automated guides in care homes, greeters in banks and even as 'companions' for children and the elderly. Given that the field is just beginning to come into its own, industry activity in Canada at present is limited, but this is likely to change in the near term as there is R&D occurring in several labs.

- **CrossWing:** In addition to the Bishop disinfection robot, CrossWing also sells the VirtualME™ telepresence and guide robot (TRL7), which could enable virtual 'face-to-face' doctor's visits or direct patients in a hospital. The robot is intended as a virtual assistant.

## A.2 Legal & Regulatory Landscape

Canada has been an important international contributor to research on robot law, ethics, and regulation. For example, the pioneering efforts of University of Ottawa Professor Ian Kerr contributed to the development of the standard reference textbook on Robot Law[202] and the establishment of We Robot[203], the leading international conference on robotics law. Other important Canadian contributors to robot law and ethics scholarship include University of Ottawa Professor Jason Millar (Canada Research Chair in the Ethical Engineering of Robotics and Artificial Intelligence), McGill Professor AJung Moon (Founder of the Open Roboethics Institute), and Law Professor Kristen Thomasen from the University of Windsor.

In addition to this scholarly work, Canadian robotics firms and government research agencies have been active contributors to both Canadian and global safety standards for robotics in the application areas that are most relevant to Canadian organizations. For example, Clearpath is contributing to the development of RIA 15.08,[204] a safety standard being developed by the Robotics Industries Association for industrial mobile robots. Another example is Avidbots, which is actively

---

involved in developing IEC 63327[205], the International Electrotechnical Commission's safety standard for the use of motorized cleaning appliances for commercial use. While IEC 63327 is a standard specifically written for cleaning robots, it will be one of the first standards to address the use of robots in everyday environments. The Canadian Space Agency is an active supporter of the development of robotics standards, often representing Canada at international standards associations.

Despite this scholarship and development work, the pace of change in digital technologies has made it difficult for policymakers and regulators to keep up. In this section we enumerate some of the relevant Canadian regulations pertaining to robotics more generally, including successes and challenges:

### UNMANNED AERIAL VEHICLES (UAVS)

Transport Canada's early regulatory clarity on the use of unmanned drones gave its consumer drone sector a multi-year head start[206] over their US counterparts, enabling investment and development in a rich assortment of drone-related businesses in this country. In 2016, the US Federal Aviation Authority (FAA) introduced new rules that provided the clarity needed for the American drone industry to catch up.[207] Commercial drone operations, for crop monitoring in the agricultural sector for example, are now common in both countries, so long as the pilot is certified to fly, the drone is registered, and the drone remains within line of sight of the operator.[208] Drone operations that occur beyond visual line of sight (BVLOS) or over urban areas still require exemptions in both countries and are generally restricted to research and demonstration purposes. Regulatory advances in BVLOS drone flight would open up the commercial drone delivery sector.

---

**AUTONOMOUS VEHICLES**

Autonomous vehicles driving on the road are regulated by Transport Canada[209], which does not specifically regulate fully autonomous vehicles but does require permission for testing. A ranking by the KPMG Autonomous Vehicles Readiness Index places Canada seventh out of 20 countries based on policy, public perception, research activity and industry activity.[210] There remains significant regulatory variation between the provinces in terms of autonomous driving testing and deployment. Further legislation to unify provincial approaches will encourage autonomous vehicle research and industry development in Canada.

**AUTOMATED DECISION-MAKING**

In 2019, the Government of Canada's Directive on Automated Decision-Making (ADM) took effect, which sets out minimum requirements for federal government departments that wish to use ADM technology.

**ASSISTIVE ROBOTICS**

Assistive robots are subject to regulation by a variety of guidelines, but present unique challenges in safety, privacy and even ethics due to their close interactions with people. However, unlike mobile or medical robots, there is not a specific regulatory framework for these devices. While there are international efforts to establish regulatory frameworks for assistive robots[211], Canada's lack of a specific framework in this area creates uncertainty for researchers and industry, and may stifle innovation in this important application area for robotics.

**MEDICAL ROBOTICS**

New medical devices must be approved by Health Canada for clinical use in Canada. These regulations are designed to ensure the safety of patients and protect them from harmful devices and treatments. The approval process typically requires significant animal and human trial data to demonstrate the safety of the device or treatment. The approval process is highly dependent on the class of device or treatment, and can

---

take years to complete. As such it is a major barrier to rapid deployment of new technologies in a crisis. For this reason, Canada must ensure the continual development and approval of new technologies if it wants to have new innovations ready for future crises.

**MEDICAL DEVICES**

In March 2020 Health Canada approved an Interim Order (IO)[212] to speed the review of medical devices used to diagnose, treat or prevent COVID-19. In addition, access to COVID-19 related medical devices not yet licensed in Canada can be requested by healthcare professionals through Health Canada's Special Access Program (SAP).[213] Health Canada established a new Digital Health Division[214] in 2018 with the express goals of developing regulatory pathways that improve access to health data, facilitating more timely diagnosis, and improving access to care for patients in care centres, at home, and in rural and remote communities. As part of this initiative, Health Canada created a division within the Therapeutic Products Directorate's Medical Devices Bureau[215] to improve regulatory pathways to allow for targeted pre-market review of some of the many digital health technologies that play a role in healthcare robotics, including Wireless Medical Devices, Mobile Medical Apps, Telemedicine, Software as a Medical Device (SaMD), Artificial Intelligence, and Cybersecurity, and Medical Device Interoperability.

---

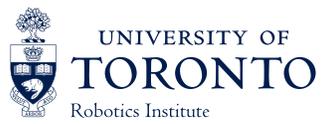

robotics.utoronto.ca | @UofTRobotics